\documentclass{jfm}

\usepackage{graphicx}
\usepackage{epstopdf,epsfig}
\usepackage{natbib}
\usepackage{hyperref}
\usepackage{bm}
\usepackage{subcaption}

\hypersetup{
    colorlinks = true,
    urlcolor   = blue,
    citecolor  = black,
}

\newcommand{\RomanNumeralCaps}[1]
\linenumbers

\title{Inertial and gravity wave transmissions near radiative-convective boundaries}

\author{Tao Cai\aff{1}
  \corresp{\email{tcai@must.edu.mo}},
  Cong Yu\aff{2,1}
  \corresp{\email{yucong@mail.sysu.edu.cn}}
 \and Xing Wei\aff{3}}

\affiliation{\aff{1}State Key Laboratory of Lunar and Planetary Sciences, Macau University of Science and Technology, Macau, People's Republic of China
\aff{2}School of Physics and Astronomy, Sun Yat-sen University, Zhuhai, 519082, People's Republic of China
\aff{3}Department of Astronomy, Beijing Normal University, Beijing, People's Republic of China}

\begin{document}
\maketitle

\begin{abstract}
In this paper, we study inertial and gravity wave transmissions near radiative-convective boundaries on the {\it f}-plane. Two configurations have been considered: waves propagate from the convective layer to the radiative stratified stable layer, or the other way around. It has been found that waves prefer to survive at low latitudes when the stable layer is strongly stratified ($N^2/(2\Omega)^2>1$). When the stable layer is weakly stratified ($N^2/(2\Omega)^2<1$), however, waves can survive at any latitude if the meridional wavenumber is large. Then we have discussed transmission ratios for two buoyancy frequency structures: the uniform stratification, and the continuously varying stratification. For the uniform stratification, we have found that the transmission is efficient when the rotation is rapid, or when the wave is near the critical colatitude. For the continuously varying stratification, we have discussed the transmission ratio when the square of buoyancy frequency is an algebraic function $N^2\propto z^{\nu} (\nu >0)$. We have found that the transmission can be efficient when the rotation is rapid, or when the wave is near the critical colatitude, or when the thickness of the stratification layer is far greater than the horizontal wave length. The transmission ratio does not depend on the configurations (radiative layer sits above convective layer, or vice versa; wave propagates outward or inward), but only on characteristics of the wave (frequency and wavenumber) and the fluid (degree of stratification).
\end{abstract}



\section{Introduction}
\label{sec:intro}
The interior of a star or planet can be divided into convective and radiative layers. The convective layer is unstable to convection, and thus, the energy from the interior of star can be efficiently transported by convection. The radiative layer, on the other hand, is stable to convection, and the energy is mainly transported by radiation. Many stars and planets have adjacent convective and radiative layers. For example, intermediate-mass main-sequence stars (mass is within 0.5-1.5 solar masses) have an outer convective layer adjacent to an inner radiative layer, while massive main-sequence stars (mass is above 1.5 solar masses) have an outer radiative layer adjacent to an inner convective layer. Gas giant planets, such as hot Jupiters, are also believed to have adjacent radiative and convective layers. The convective stability of a convective or radiative layer can be measured by the square of buoyancy frequency $N^2$. Convective layer has $N^2\leq 0$ ($N^2=0$ means neutrally unstable), while radiative layer has $N^2>0$. As convection can transport energy efficiently, the thermal structure in the convective layer is almost adiabatic, which yields $N^2\approx 0$. Wave propagation is an important physical process in stars and planets. In the radiative layer, gravity waves can be excited, because the buoyancy force can act as a restoring force of perturbations from the thermal equilibrium state. In contrast, gravity waves cannot survive in the convective layer. When a gravity wave propagates from a radiative layer into its adjacent convective layer, the wave amplitude will decay quickly. In the presence of rotation, inertial waves can be excited because Coriolis force can act as a restoring force. Unlike gravity waves, inertial waves can survive in both radiative and convective layers. Gravity waves in a rotating system are sometimes called inertia-gravity waves or gravito-inertial waves. In a rotating system, inertial waves can propagate from convective layer into radiative layer as gravito-inertial waves, and vice versa. The purpose of this paper is to investigate wave propagations near radiative-convective boundaries in a rotating system.

It is well known that internal gravity waves (IGWs) are likely to be generated near the radiative-convective boundary in stars by the shear stress \citep{kumar99} or stochastic overshooting plumes \citep{rogers05,mathis14}. Various numerical simulations have verified that IGWs can be excited efficiently near radiative-convective boundaries in turbulent convection \citep{meakin2007turbulent,brun2011modeling,alvan2014theoretical}. When IGWs propagate in the radiative stable layer, it could induce material mixing \citep{rogers17} and transport angular momentum \citep{rogers13,fuller14}. For a stable layer above a convective layer, gravity waves are important because they propagate upward into a decreasing density region, which yields increasing wave amplitudes and wave breaking \citep{rogers13}. On Earth's atmosphere, it has been confirmed that the breaking of gravity wave is one of major factors for the transport of water from troposphere into the stratosphere \citep{qu20}. Using the analogy of Earth's atmosphere, \citet{rogers13} proposed that IGWs are dynamically and chemically important in massive stars. \citet{bowman19} detected low-frequency variability in massive stars by asteroseismology, and they interpreted it as a signal of IGWs. \citet{lecoanet2019low} argued that this observed low-frequency variability might be produced by the subsurface convection, rather than IGWs excited by core convection.
Wave propagation also plays an essential role on angular momentum transport. \citet{zahn97} developed a theoretical model on the angular momentum transported by waves. For stars, the model of \citet{fuller14} predicts that IGWs tend to reduce differential rotation (different parts of a star have different angular velocities) on short timescales in low mass stars. Thus it helps to explain the observed small amounts of internal differential rotation in low-mass main-sequence stars. For planets, it has been shown that IGWs possibly play important roles in planetary formations and evolutions \citep{rogers12,yu20}.

Apart from gravity waves, in rotating stars or planets, inertial waves could also be generated because of the presence of Coriolis force \citep{ogilvie04,wu05a,goodman09}. Gravity waves can only survive in convectively stable layers, while inertial waves can survive in both convectively stable and unstable layers. \citet{wu05b} found that resonantly exited inertial modes have significant impact on the tidal dissipation of a coreless Jupiter \citep{wu05b}. \citet{goodman09} argued that a rigid core should be included in the model. In the discussion of \citet{goodman09}, inertial waves were assumed to be fully reflected at the surface of the rigid core. However, recent study \citep{liu19} revealed that Jupiter probably has a diluted core formed by a giant impact. If the rigid core is replaced by a diluted core, inertial waves are expected to be partially reflected at the surface. An analogy can be readily drawn between this problem and that of wave propagations near radiative-convective boundaries.

A question that remains unclear is how waves reflect and transmit near radiative-convective boundaries in stars and planets. \citet{wei20,wei2020erratum} has discussed the reflection and transmission of an incident wave at the radiative-convective boundary on the {\it f}-plane, with the assumption that the buoyancy frequency abruptly changes from zero at the convective layer to a constant value at the radiative stable layer. He discovered that waves can efficiently transmit across the boundary in a rapidly rotating fluid. In real stars or planets, transitions of thermal structures are likely to be continuous. For example, the transition of static stability profile in Venus's atmosphere is abrupt at middle and polar latitudes, but smooth at equatorial latitudes \citep{tellmann2009structure}.
\citet{borovikov1990t} have considered the trajectory rays of internal waves in a two-layer vertically infinite medium with a variable $N^2$ layer above and a constant $N^2$ layer below. 
\citet{lecoanet2013internal} investigated internal gravity waves excited by turbulent convection in discontinuous and smooth transitions near radiative-convective boundaries. They found that the flux of IGWs in the smooth transition case is larger than that calculated in the discontinuous transition case.
Therefore it is necessary to consider the wave reflection and transmission near a continuously transited radiative-convective boundary. In this paper, we extend the work of \citet{wei20}, by considering two different stratification structures in the stable layer: a uniform stratified layer, or a continuously varying stratified layer. Efficiencies of wave transmissions were estimated and compared.

\section{Method}
For a Boussinesq flow on a rotating {\it f}-plane, the linearized hydrodynamic equations of mass and momentum conservations \citep{phillips1966dynamics} are
\begin{eqnarray}
&& \bnabla \bcdot \bm{u} = 0~,\label{eq1}\\
&&\bm{u}_{t}  + \bm{f} \bm{\times} \bm{u}  +\bnabla p-b \hat{\bm{z}} = 0~,\label{eq2}\\
&&b_{t}+ N^2 \bm{u} \bcdot \hat{\bm{z}} = 0~.\label{eq3}
\end{eqnarray}
where $\bm{u}=(u,v,w)$ is the velocity; $\bm{f}=(0,\tilde{f},f)$ is a vector form of horizontal and vertical Coriolis parameters, with $\tilde{f}=2\Omega\sin\theta$ and $f=2\Omega\cos\theta$ ($f$ and $\tilde{f}$ are constants at a given $\theta$); $\Omega$ is the rotation rate, and $\theta$ is the colatitude of the {\it f}-plane; $p$ is the modified pressure perturbation (pressure perturbation scaled by constant background density); $b$ is the buoyancy; $N^2$ is the square of buoyancy frequency; $\hat{\bm{x}}$, $\hat{\bm{y}}$, and $\hat{\bm{z}}$ are the unit vectors in the west-east direction, the south-north direction, and the vertical direction, respectively. We consider a two-layer thermal structure, with an upper density-stratified stable layer adjacent to a lower adiabatically convective layer (fig.\ref{fig:f1}a). By this setting, $N^2$ is positive in the stable layer, and zero in the convective layer, respectively.

\subsection{Wave dispersion relation and frequency range}
\begin{figure}
\centering
\begin{subfigure}{0.45\textwidth}
\includegraphics[width=0.8\linewidth]{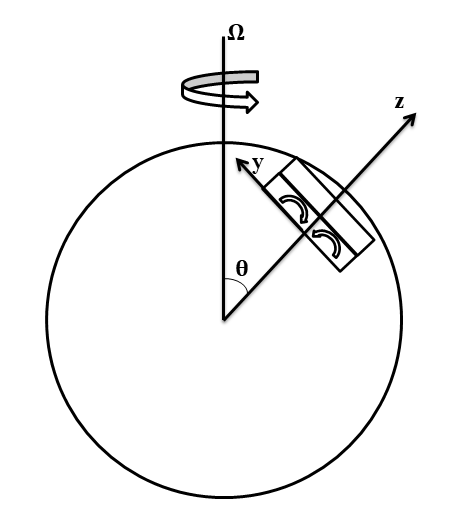}
\caption{}
\end{subfigure}
\begin{subfigure}{0.45\textwidth}
\includegraphics[width=\linewidth]{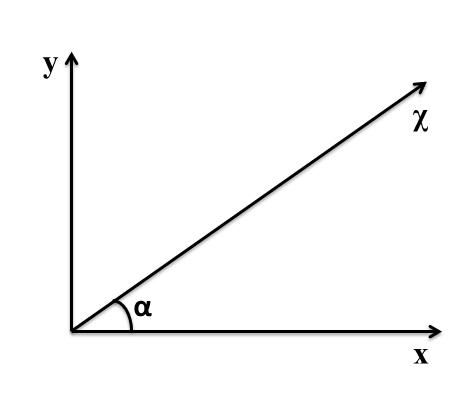}
\caption{}
\end{subfigure}

\medskip

\begin{subfigure}{0.45\textwidth}
\includegraphics[width=\linewidth]{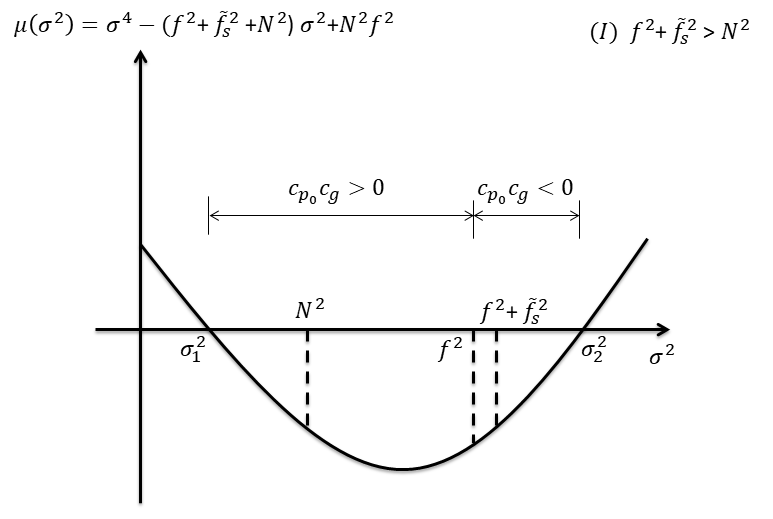}
\caption{}
\end{subfigure}
\begin{subfigure}{0.45\textwidth}
\includegraphics[width=\linewidth]{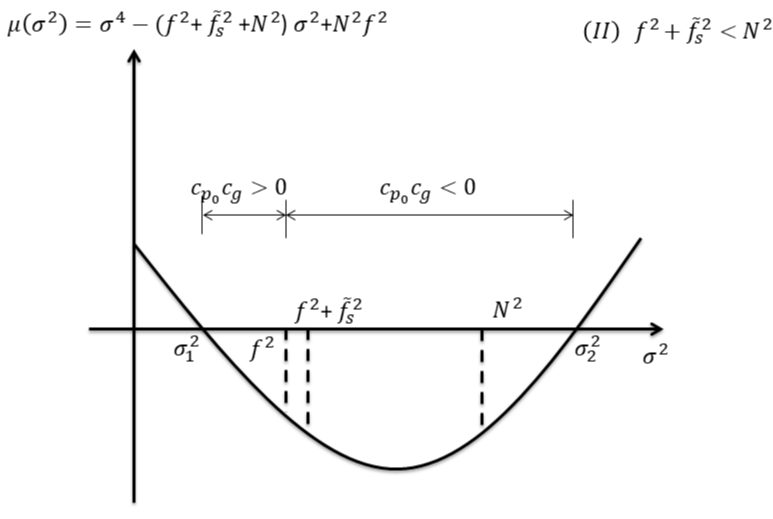}
\caption{}
\end{subfigure}
\caption{Sketch plots of {\it f}-plane structure and wave direction. (a)Illustration of a two-layer setting {\it f}-plane. The stable layer sits above the convective layer. The {\it f}-plane is inclined an angle $\theta$ with the north pole. $\hat{\bm{y}}$ is the south-north direction. $\hat{\bm{z}}$ is the vertical direction. $\hat{\bm{x}}$ is the west-east direction pointing into the plane of the paper. (b)The plane wave propagates in a direction at an angle $\alpha$ to $x$-axis. (c) and (d) The sign of $c_{p_{0}}c_{g}$ in weakly (case I) and strongly (case II) stratified cases, respectively. Note that $f^2$ can be smaller than $N^2$ in the weakly stratified case. \label{fig:f1}}
\end{figure}

After some manipulations, the above equations can be reduced into the following equation on vertical velocity \citep{gerkema05}:
\begin{eqnarray}
\nabla^2 w_{tt}+(\bm{f} \bcdot {\bnabla})^2 w +N^2 \nabla_{h}^2 w=0~, \label{eq4}
\end{eqnarray}
where $\nabla_{h}^2=\partial_{x}^2+\partial_{y}^2$ is the horizontal Laplacian operator, and $\nabla^2=\nabla_{h}^2+\partial_{z}^2$ is the Laplacian operator.  For a plane wave propagating in the direction $(\cos\alpha,\sin\alpha)$, the problem can be further simplified by introducing a new variable $\chi$, which satisfies $x=\chi \cos\alpha$ and $y=\chi\sin\alpha$ (fig.\ref{fig:f1}b).
Substituting $\partial_{x}=\cos\alpha\partial_{\chi}$, $\partial_{y}=\sin\alpha\partial_{\chi}$, and $w=W(\chi,z)\exp(-i\sigma t)$ into the above equation, we obtain
\begin{eqnarray}
AW_{\chi\chi}+2BW_{\chi z}+CW_{zz}=0~,\label{eq5}
\end{eqnarray}
where $\sigma$ is the time frequency, $A=\tilde{f}_{s}^2-\sigma^2+N^2$, $B=f\tilde{f}_{s}$, $C=f^2-\sigma^2$, and $\tilde{f}_{s}=\tilde{f}\sin\alpha$. Wave solution exists when this partial differential equation is hyperbolic, thus it requires $\Delta=B^2-AC>0$ \citep{gerkema05}, or equivalently
\begin{eqnarray}
\sigma^4-(f^2+\tilde{f}_{s}^2+N^2)\sigma^2+N^2 f^2<0~. \label{eq6}
\end{eqnarray}
Then the frequency range for wave solutions is
\begin{eqnarray}
\sigma_{1}^2<\sigma^2<\sigma_{2}^2~,
\end{eqnarray}
where $\sigma_{1,2}^2=\left[(f^2+\tilde{f}_{s}^2+N^2)\mp \sqrt{(f^2+\tilde{f}_{s}^2+N^2)^2-4N^2 f^2}\right]/2$. For a wave to propagate through the whole domain, this inequality should be satisfied for all the possible values of $N^2(z)$. Let $N_{min}^2$ and $N_{max}^2$ be the minimum and maximum values of $N^2(z)$, respectively. For a wave to survive in both convective and stable layers, it requires
\begin{eqnarray}
\sigma_{min}^2<\sigma^2<\sigma_{max}^2~,
\end{eqnarray}
where $\sigma_{min}^2=\left[(f^2+\tilde{f}_{s}^2+N_{max}^2)- \sqrt{(f^2+\tilde{f}_{s}^2+N_{max}^2)^2-4N_{max}^2 f^2}\right]/2$ and $\sigma_{max}^2=\left[(f^2+\tilde{f}_{s}^2+N_{min}^2)+ \sqrt{(f^2+\tilde{f}_{s}^2+N_{min}^2)^2-4N_{min}^2 f^2}\right]/2$. In our problem $N_{min}^2=0$, thus $\sigma_{max}^2=f^2+\tilde{f}_{s}^2$. The width of the frequency range is
\begin{eqnarray}
\sigma_{max}^2-\sigma_{min}^2=\frac{1}{2}\left[(f^2+\tilde{f}_{s}^2-N_{max}^2)+ \sqrt{(f^2+\tilde{f}_{s}^2-N_{max}^2)^2+4N_{max}^2 \tilde{f}_{s}^2}\right]~. \label{eq9}
\end{eqnarray}
This width is always positive unless $\tilde{f}_{s}^2=0$ and $f^2+\tilde{f}_{s}^2\leq N_{max}^2$. The first condition $\tilde{f}_{s}^2=0$ can be achieved only when $\sin\theta=0$ or $\sin\alpha=0$. Thus, in polar regions, waves cannot survive in both convective and stable layers in a slowly rotating fluid ($N^2/4\Omega^2 \gg 1$, or in other words, the stable layer is strongly stratifed) \citep{wei20}. However, the conclusion at other latitudes is completely different.

At other latitudes, if the meridional wavenumber is nonzero, waves can survive in both convective and stable layers even in a slowly rotating fluid. From (\ref{eq9}), we see that the width of frequency range varies with the degree of stratification $N_{max}^2/(4\Omega^2)$, the colatitude $\theta$, and the plane wave propagating angle $\alpha$. The dependence of the width of frequency range on these parameters can be studied by analyzing the monotonicity of (\ref{eq9}). After some trivial calculations (see Appendix \ref{appendixa}), we have the following conclusions. First, the width of the frequency range decreases with $N_{max}^2/(4\Omega^2)$ , with an upper limit value $f^2+\tilde{f}_{s}^2$ achieved at $N_{max}^2/(4\Omega^2)\rightarrow 0$ and a lower limit value $\tilde{f}_{s}^2$ achieved at $N_{max}^2/(4\Omega^2)\rightarrow \infty$. As a result, waves are more likely to survive at low latitudes in a slowly rotating fluid, where $\tilde{f}_{s}^2$ is larger. In a rapidly rotating fluid ($N^2/4\Omega^2 \ll 1$, or in other words, the stable layer is weakly stratifed), waves are more likely to survive at high latitudes, but the chance at low latitudes is still moderate if the meridional wavenumber dominates the zonal wavenumber. This conclusion shows agreement with the result obtained in the spherical geometry by \citet{friedlander1982internal} (see their figures 1 and 3). Second, the width of frequency range increases with $\sin^2\alpha$,
which indicates that waves with higher meridional wavenumbers are more likely to survive. Third, the variation of the width of frequency range on $\theta$ is not always monotonic. It depends on the degree of stratification. If $N_{max}^2/4\Omega^2>1$, then the width increases with $\theta$. If $N_{max}^2/4\Omega^2<1$, however, there exists a critical value $(\sin^2\alpha)_{c}$, and the width increases with $\theta$ in the range $\sin^2\alpha>(\sin^2\alpha)_{c}$, and decreases with $\theta$ in the range $\sin^2\alpha<(\sin^2\alpha)_{c}$. Again, we have verified that the frequency range is wider at low latitudes in a slowly rotating fluid. In a rapidly rotating fluid, on the other hand, waves can survive at any latitude when the meridional wavenumber is large.

In summary, we have the following findings:
\begin{itemize}
\setlength{\itemsep}{0pt}
\setlength{\parsep}{0pt}
\setlength{\parskip}{0pt}
\item[(1)] Waves are more likely to survive at low latitudes for strongly stratified fluid, while they can survive at any latitude for weakly stratified fluid when the meridional wavenumber is large.
\item[(2)] Waves with higher meridional wavenumber are more likely to survive.
\item[(3)] The wave frequency width is wider at lower latitudes for strongly stratified fluid. For weakly stratified fluid, the width achieves its maximum value at a certain latitude.
\end{itemize}

\subsection{Phase and group velocities}
Now we move back to find the solution of (\ref{eq5}). Given $\sigma$ in the frequency range $(\sigma_{min},\sigma_{max})$, the wave solution can be obtained if (\ref{eq5}) can be solved. The general solution of (\ref{eq5}) can be written in a form \citep{gerkema05} of
\begin{eqnarray}
W=\Psi(z)\exp(ik\chi)=\psi(z)\exp{i(k\chi+\delta z)}~,
\end{eqnarray}
where $\delta=-kB/C$ and $\Psi(z)=\psi(z)\exp(i\delta z)$. Making a substitution, we obtain
\begin{eqnarray}
\psi_{zz}+k^2 (\frac{B^2-AC}{C^2})\psi=0. \label{eq12}
\end{eqnarray}
or
\begin{eqnarray}
\psi_{zz}+\frac{k^2}{C}(\frac{B^2-A_{0}C}{C}-N^2)\psi=0~, \label{eq13}
\end{eqnarray}
where $A_{0}=\tilde{f}_{s}^2-\sigma^2$. From this equation, we know that the wave solution depends on the structure of $N^2(z)$.

Before the discussion of wave transmission, it is necessary to consider wave directions in detail. Here we determine the wave direction by its group velocity, because it represents the direction of energy propagation. For the convenience of discussion, let us first consider a simple case with constant $N^2$. In such case, the basis functions of wave solution can be written as $e^{i(\pm r_{0}+\delta)z}$, where $r=\pm r_{0}+\delta $ is the wavenumber along the vertical direction. For the convenience of discussion, we call $\pm r_{0}$ as modified vertical wavenumbers. Substituting $e^{\pm ir_{0}z}$ into (\ref{eq12}), we obtain the dispersion relation
\begin{eqnarray}
r_{0}^2=k^2\frac{B^2-AC}{C^2}~.
\end{eqnarray}
and its derivative
\begin{eqnarray}
\frac{dr_{0}}{d\sigma}=\frac{k^2\sigma}{r_{0}}\frac{C^2-AC+2B^2}{C^3}~.
\end{eqnarray}
Here we define $c_{g_{0}}=d\sigma/dr_{0}$ as the modified vertical group velocity, and $c_{p_{0}}=\sigma/r_{0}$ as the modified vertical phase velocity.

Now we consider the vertical group velocity of the wave $e^{irz}$. It can be shown that
\begin{eqnarray}
\frac{dr}{d\sigma}=\frac{k \sigma }{C^2} [\frac{\pm(C^2-AC+2B^2)-2B (B^2-AC)^{1/2}}{(B^2-AC)^{1/2}}]~.
\end{eqnarray}
Correspondingly, we define $c_{g}=d\sigma/dr$ as the vertical group velocity, and $c_{p}$ as the vertical phase velocity. Note that $c_{g}$ has the same sign as $dr/d\sigma$. Compared to group
velocity, the judgement of phase velocity direction is much easier. So here we try to build up some relations between the vertical group velocity $c_{g}$ and the modified phase velocity $c_{p_{0}}$. It can be shown that
\begin{eqnarray}
\frac{r_{0}}{\sigma}\frac{dr}{d\sigma}=\frac{k^2  }{C^3} [(C^2-AC+2B^2)\mp 2B (B^2-AC)^{1/2}]>0~.
\end{eqnarray}
Thus we conclude that
\begin{eqnarray}
Sgn(c_{g}c_{p_{0}})=Sgn(\frac{r_{0}}{\sigma}\frac{dr}{d\sigma})=Sgn(C)~,
\end{eqnarray}
where the operator $Sgn$ is a sign function. As a result, the sign of $c_{p_{0}}c_{g}$ is only determined by the sign of $C$. In other words, the group velocity and the modified phase velocity have same directions when the frequency is sub-inertial $\sigma^2< f^2$, and opposite directions when the frequency is super-inertial $\sigma^2>f^2$. Singularity appears when $\sigma^2=f^2$ and it defines the critical colatitudes $\theta_{c}=\cos^{-1}\pm\sigma/(2\Omega)$ \citep{rieutord01,goodman09}.

Our conclusion is general for the wave propagation on a tilted $\it f$-plane. The special cases of the pure gravity wave ($f^2=0$) and the pure inertial wave ($N^2=0$) have been well studied in a non-tilted plane ($\tilde{f}^2=0$ and $c_{p_{0}}=c_{p}$) \citep{rieutord15}. For the pure gravity wave case ($f^2=0$ and $\tilde{f}^2=0$), we have $f^2=0<\sigma^2<N^2$. Thus $c_{g}$ always points in the opposite direction to $c_{p}$. For the pure inertial wave case ($N^2=0$ and $\tilde{f}^2=0$), we have $0<\sigma^2<f^2$. Thus $c_{g}$ always propagates in the same direction as $c_{p}$. These two results are implied in the calculations of \citet{rieutord15} (see equations (5.48) and (8.20) in the book).

Now let us consider the gravito-inertia wave on a non-tilted {\it f}-plane ($\tilde{f}=0$ and $f=2\Omega$). For such case, the lower and upper limits of wave frequency are $\sigma^2_{1,2}=(f^2+N^2\mp|f^2-N^2|)/2$. When the density structure is weakly stratified $N^2/f^2<1$, we have $\sigma^2<\sigma_{2}^2=f^2$, therefore $c_{g}$ has the same direction as $c_{p}$. When the density structure is strongly stratified $N^2/f^2>1$, we have $f^2=\sigma_{1}^2<\sigma^2$, therefore $c_{g}$ has the opposite direction to $c_{p}$. When $N^2/f^2=1$, the group velocity $c_{g}=0$ and wave energy does not propagate along the vertical direction. Clearly, the direction of $c_{g}$ depends on the density stratification. In summary, $c_{g}$ in weakly (strongly) stratified density structure prefers the same (opposite) direction as (to) $c_{p}$.

Now we consider the more general case for the gravito-inertial wave on a tilted {\it f}-plane ($\tilde{f}\neq 0$). The discussion is similar to that of the non-tilted case. If $\tilde{f}_{s}\neq 0$, then the frequency domain $\sigma^2\in(\sigma_{1}^2,\sigma_{2}^2)$ can be separated into two regions by $f^2$. In the sub-inertial region $\sigma_{1}^2\in(\sigma_{1}^2,f^2)$, the group velocity $c_{g}$ has the same direction as the modified phase velocity $c_{p_{0}}$ (fig.\ref{fig:f1}(c)). By contrast, in the super-inertial region $\sigma^2\in(f^2,\sigma_{2}^2)$, the group velocity $c_{g}$ has the opposite direction to the modified phase velocity $c_{p_{0}}$ (fig.~\ref{fig:f1}(d)). In the strongly stratified (slowly rotating) case, the sub-inertial range is thinner than the super-inertial range. So $c_{g}$ and $c_{p_{0}}$ prefer the opposite direction in a strongly stratified density structure. In a weakly stratified (rapidly rotating) case, the conclusion depends on latitudes. $c_{g}$ and $c_{p_{0}}$ prefer the same direction at high latitudes, but the opposite direction at low latitudes.

Our findings are summarised as follows:
\begin{itemize}
\setlength{\itemsep}{0pt}
\setlength{\parsep}{0pt}
\setlength{\parskip}{0pt}
\item[(1)] The vertical group velocity and the modified vertical phase velocity have same directions for a sub-inertial wave, and opposite directions for a super-inertial wave.
\item[(2)] On a non-tilted plane, the vertical group and phase velocities are always in opposite directions for a pure gravity wave, and in same directions for a pure inertial wave.
\item[(3)] On a non-tilted plane, the vertical group and phase velocities of a gravito-inertia wave are always in opposite directions in strongly stratified fluid, and same directions in weakly stratified fluid.
\end{itemize}

\subsection{Wave energy fluxes and transmission cofficients}
\begin{figure}
\centering
\begin{subfigure}{0.45\textwidth}
\includegraphics[width=0.9\linewidth]{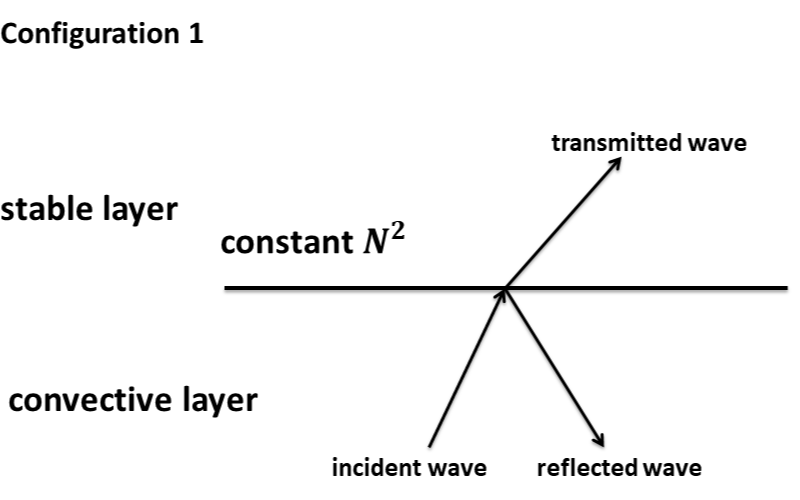}
\caption{}
\end{subfigure}
\begin{subfigure}{0.45\textwidth}
\includegraphics[width=0.9\linewidth]{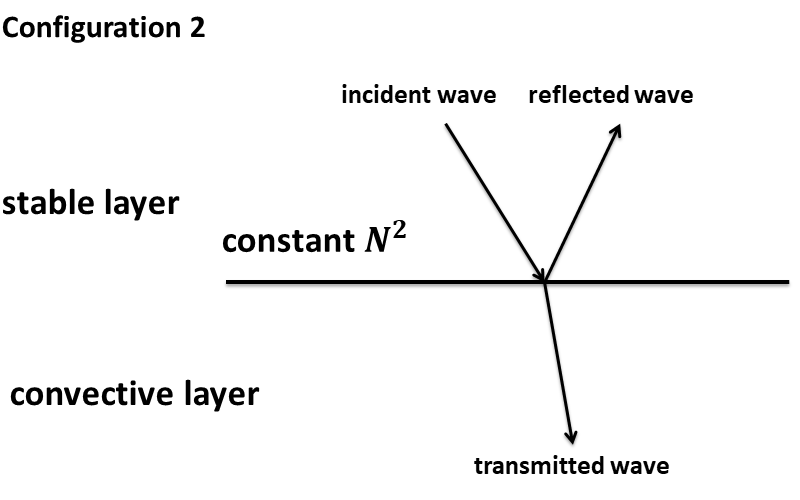}
\caption{}
\end{subfigure}

\medskip

\begin{subfigure}{0.45\textwidth}
\includegraphics[width=0.9\linewidth]{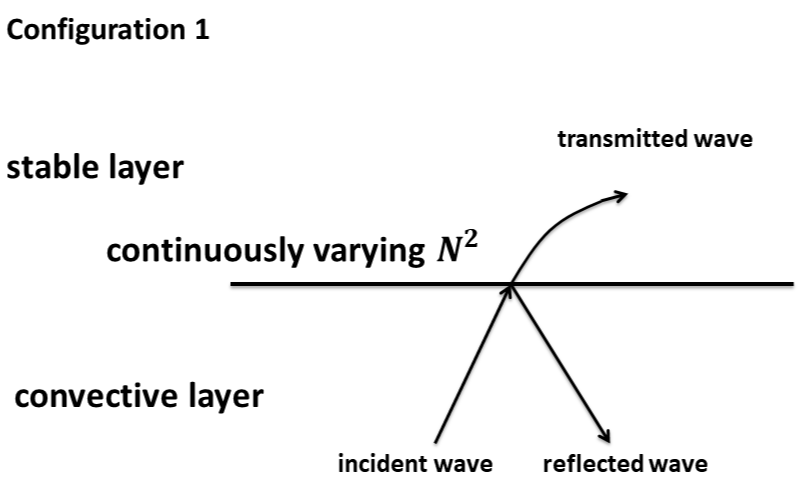}
\caption{}
\end{subfigure}
\begin{subfigure}{0.45\textwidth}
\includegraphics[width=0.9\linewidth]{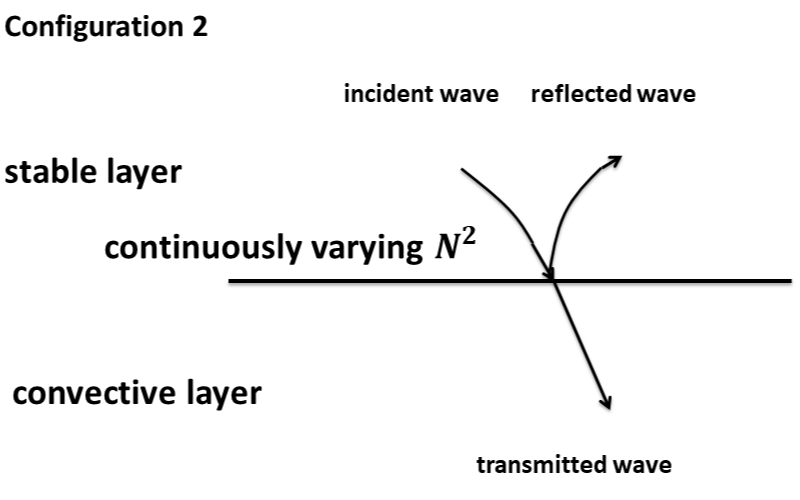}
\caption{}
\end{subfigure}
\caption{Sketch plots of wave propagation in different configurations. (a and c)Configuration 1: the wave propagates from the convective layer to the stable layer. (b and d)Configuration 2: the wave propagates from the stable layer to the convective layer. (a and b)The stable layer is uniform stratification. (c and d)The stable layer is continuously varying stratification.\label{fig:f2}}
\end{figure}

Now we discuss the wave reflection and transmission at the interface between the convective layer and the stratified stable layer. \citet{wei20} has discussed four configurations on wave reflection and transmission: from convective (stable) to stable (convective) layers, and stable layer sitting above (below) convective layer. It is easy to verify that when the stable layer is switched with the convective layer, the Boussinesq equations on the {\it f}-plane are invariant under the transformation $z \leftrightarrow -z$ and $w\leftrightarrow -w$, where the location of the interface is set at $z=0$. Thus it is sufficient to discuss one of the cases. The other case can be inferred from the symmetric property. The up/down symmetry also holds under the transformation $\theta\leftrightarrow \pi-\theta$, so hereafter we only focus on the {\it f}-plane in the northern atmosphere. In this paper, we consider two configurations for the stable layer sitting above the convective layer: the wave propagates from the convective layer to the stable layer (the left panel of fig.~\ref{fig:f2}), and the wave propagates from the stable layer to the convective layer (the right panel of fig.\ref{fig:f2}). In either case, we consider two different stratification structures in the stable layer: uniform stratification (figs.~\ref{fig:f2}(a and b)), and continuously varying stratification (figs~\ref{fig:f2}(c and d)).

To obtain the wave solution, we firstly discuss the general solutions in the convective layer and the stable layer, and then determine the coefficients by matching boundary conditions at the interface. We have two boundary conditions at the interface. The first is the condition for continuous vertical velocity $w(0^{+})=w(0^{-})$. The second is the condition that Lagrangian perturbation of pressure is continuous at the interface, which requires that the first derivative of vertical velocity is continuous $w_{z}(0^{+})=w_{z}(0^{-})$ (see \citet{wei20} and Appendix \ref{appendixb} for explanation).

In the adiabatically convective layer with $N^2=0$, it is easy to obtain the following wave solution:
\begin{eqnarray}
\psi=a_{1}\exp(iqz)+a_{2}\exp(-iqz)~,
\end{eqnarray}
where $q>0$ and $q^2=k^2 ({B^2-A_{0}C})/{C^2}$ is the square of wavenumber in the vertical direction. The coefficients $a_{1}$ and $a_{2}$ would be determined from matching boundary conditions at the interface. The wave solution in the stable layer is more complicated. It depends on the stratification structure of $N^2(z)$. In the following, we discuss the wave solutions of two different stratification structures: uniform stratification and continuously varying stratification. Before the discussion, we introduce the vertical component of the averaged internal wave energy flux (see Appendix \ref{appendixb} for the derivation)
\begin{eqnarray}
\left<F\right>=\frac{C\Imag(\psi_{z}/\psi)}{2k^2\sigma}|\psi|^2~,
\end{eqnarray}
where the bracket $\langle\rangle$ represents that the average is taken in a wave period, and the operator $\Real$ and $\Imag$ denote the real and imaginary parts of a complex number, respectively. Given the averaged kinetic energy fluxes of the incident wave $\left<F_{i}\right>$, the reflective wave $\left<F_{r}\right>$, and the transmitted wave $\left<F_{t}\right>$, we define the reflection ratio
\begin{eqnarray}
\zeta=\frac{|\left<F_{r}\right>|}{|\left<F_{i}\right>|}~,
\end{eqnarray}
and the transmission ratio
\begin{eqnarray}
\eta=\frac{|\left<F_{t}\right>|}{|\left<F_{i}\right>|}~.
\end{eqnarray}
Note that here the definitions of the reflection ratio and the transmission ratio are calculated by energy flux (\citet{wei20} used kinetic energy but in his corrigendum \citep{wei2020erratum} he used energy flux). A detailed discussion on the definitions of transmission and reflection ratios can be found in \citet{sutherland2010internal}.
In the followings, we discuss wave solutions in the two different stratifications.

\section{Result}
\subsection{Uniform stratification in the stable layer}
The case of uniform stratification has been discussed in \citet{wei20}. Here we discuss it again in a general form for the convenience of comparison. For a uniform stratification $N_{max}^2=\gamma_{1}$ ($\gamma_{1}>0$), the wave solution can be written as
\begin{eqnarray}
\psi=b_{1}\exp(isz)+b_{2}\exp(-isz)~,
\end{eqnarray}
where $s>0$ and $s^2=k^2 ({B^2-A_{0}C-\gamma_{1} C})/{C^2}$ is the square of wavenumber in the vertical direction. The wave solution contains two separated branches with opposite vertical propagating directions. The selection of wave directions depends on configurations. We consider configurations 1 and 2 in the followings, respectively.
\subsubsection{Configuration 1}
The vertical component of the group velocity ($c_{g}$) of the transmitted wave should be in the same direction as that of the incident wave. For configuration 1, $c_{g}$ of the transmitted wave is outward. The direction of $c_{g}$ is determined by the sign of $C$. Here we discuss the two cases $C>0$ and $C<0$, separately.

For the case $C>0$ (or $\sigma^2<f^2$), it requires $b_{2}=0$. Waves with wavenumber $+q$ is incident wave, the wave with wavenumber $-q$ is reflective wave, and the wave with wavenumber $+s$ is transmitted wave. Matching boundary conditions at the interface gives
\begin{eqnarray}
&& a_{1}+a_{2}=b_{1}~,\\
&& q a_{1}-q a_{2}=s b_{1}~.
\end{eqnarray}
Note that terms containing $\delta$ can be cancelled on both sides in the second matching boundary condition.
Solving these equations, we obtain the wave amplitude ratios
\begin{eqnarray}
\frac{a_{2}}{a_{1}}=\frac{q-s}{q+s}~,\\
\frac{b_{1}}{a_{1}}=\frac{2q}{q+s}~.
\end{eqnarray}
With the amplitude ratios, the reflection ratio can be calculated as
\begin{eqnarray}
\zeta=(\frac{a_{2}}{a_{1}})^2=(\frac{q-s}{q+s})^2~,
\end{eqnarray}
and the transmission ratio can be calculated as
\begin{eqnarray}
\eta=\frac{s}{q}(\frac{b_{1}}{a_{1}})^2=\frac{4sq}{(q+s)^2}~.
\end{eqnarray}
It can be seen that
\begin{eqnarray}
\zeta+\eta=1~,
\end{eqnarray}
which means that the incident wave energy is either reflected or transmitted. We are interested in whether waves could penetrate across the interface efficiently. The efficiency can be estimated by the transmission ratio (or the reflection ratio). The higher transmission ratio (or the lower reflection ratio), the more wave energy is transmitted.

Noting $q=k\sqrt{\Delta_{0}}/C$ and $s=k\sqrt{\Delta_{m}}/C$, we can rewrite the transmission ratio as
\begin{eqnarray}
\eta&&=1-(\frac{\sqrt{\Delta_{0}}-\sqrt{\Delta_{m}}}{\sqrt{\Delta_{0}}+\sqrt{\Delta_{m}}})^2=1-(1-\frac{2}{\sqrt{\Delta_{0}/\Delta_{m}}+1})^2~, \label{eq39}
\end{eqnarray}
where $\Delta_{0}=B^2-A_{0}C$ and $\Delta_{m}=B^2-A_{0}C-N_{max}^2C$. It can be seen that $\eta$ decreases with $\Delta_{0}/\Delta_{m}$. By taking the first derivatives, it is easy to prove that $\Delta_{0}/\Delta_{m}$ increases with $N_{max}^2/(2\Omega)^2$, and decreases with $\sigma^2$. As a result, we find that $\eta$ decreases with $N_{max}^2/(2\Omega)^2$ and increases with $\sigma^2$. Specifically, we have $\eta \rightarrow 1$ when $N_{max}^2/(2\Omega)^2\rightarrow 0$ or $\sigma^2\rightarrow f^2$. Therefore, the transmission is efficient when the rotation is fast, or when the absolute value of the frequency is close to the vertical Coriolis parameter.

For the case $C<0$ (or $\sigma^2>f^2$), it requires $b_{1}=0$. The wave with wavenumber $-q$ is incident wave, the wave with wavenumber $+q$ is reflective wave, and the wave with wavenumber $-s$ is the transmitted wave. Following the same procedure, we find that the transmission ratio has the same as (\ref{eq39}).
Noting $\Delta_{0}/\Delta_{m}<1$, we see that $\eta$ increases with $\Delta_{0}/\Delta_{m}$. By taking derivatives, we find that $\Delta_{0}/\Delta_{m}$ decreases with $N_{max}^2$ and $\sigma^2$. As a result, we find that $\eta$ decreases with $N_{max}^2/(2\Omega)^2$ and $\sigma^2$. Again, we conclude that the transmission ratio is efficient when the stable layer is weakly stratified, or when the wave is near the critical colatitude.

\begin{figure}
\centering

\begin{subfigure}{0.32\textwidth}
\includegraphics[width=\linewidth]{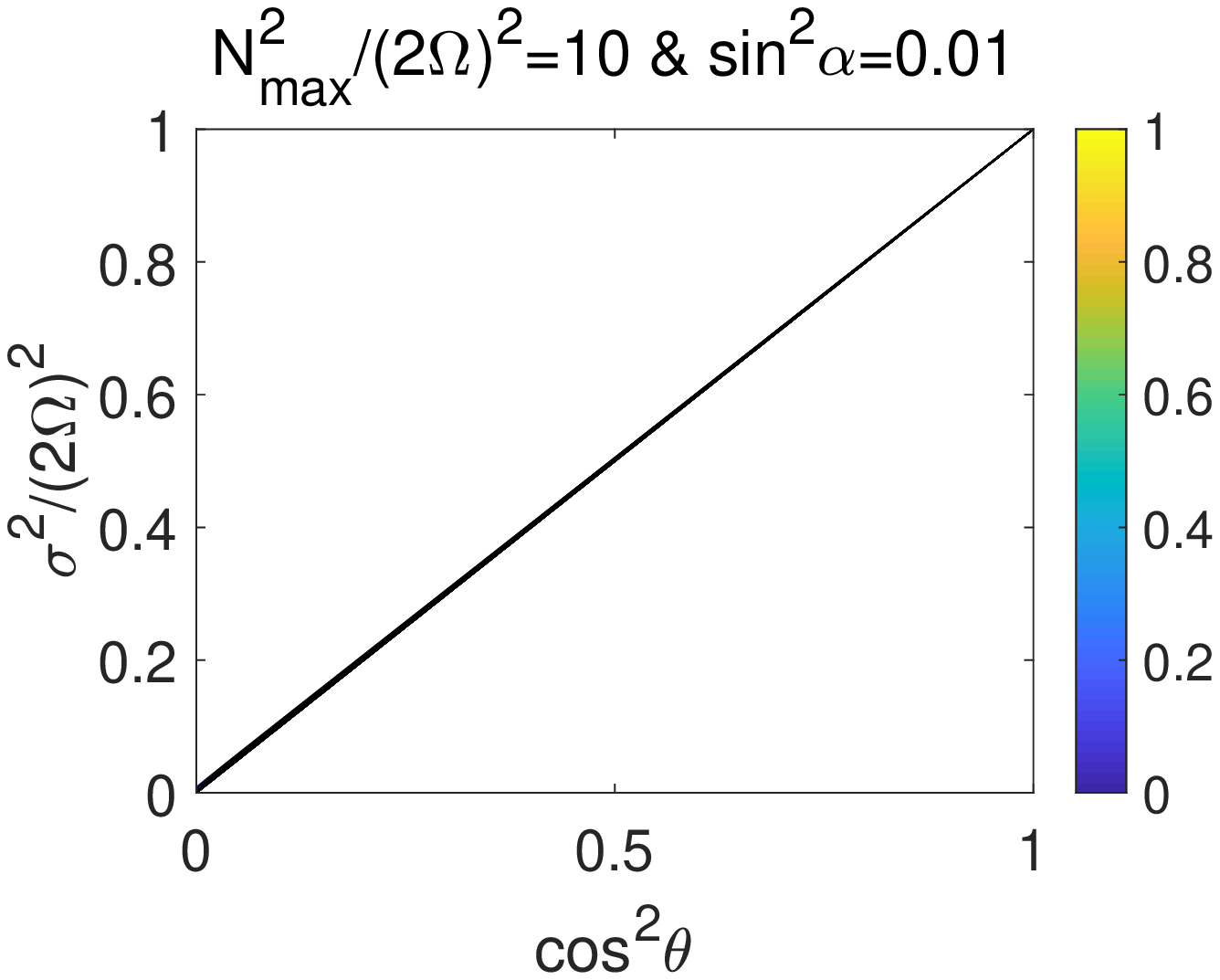}
\caption{}
\end{subfigure}
\begin{subfigure}{0.32\textwidth}
\includegraphics[width=\linewidth]{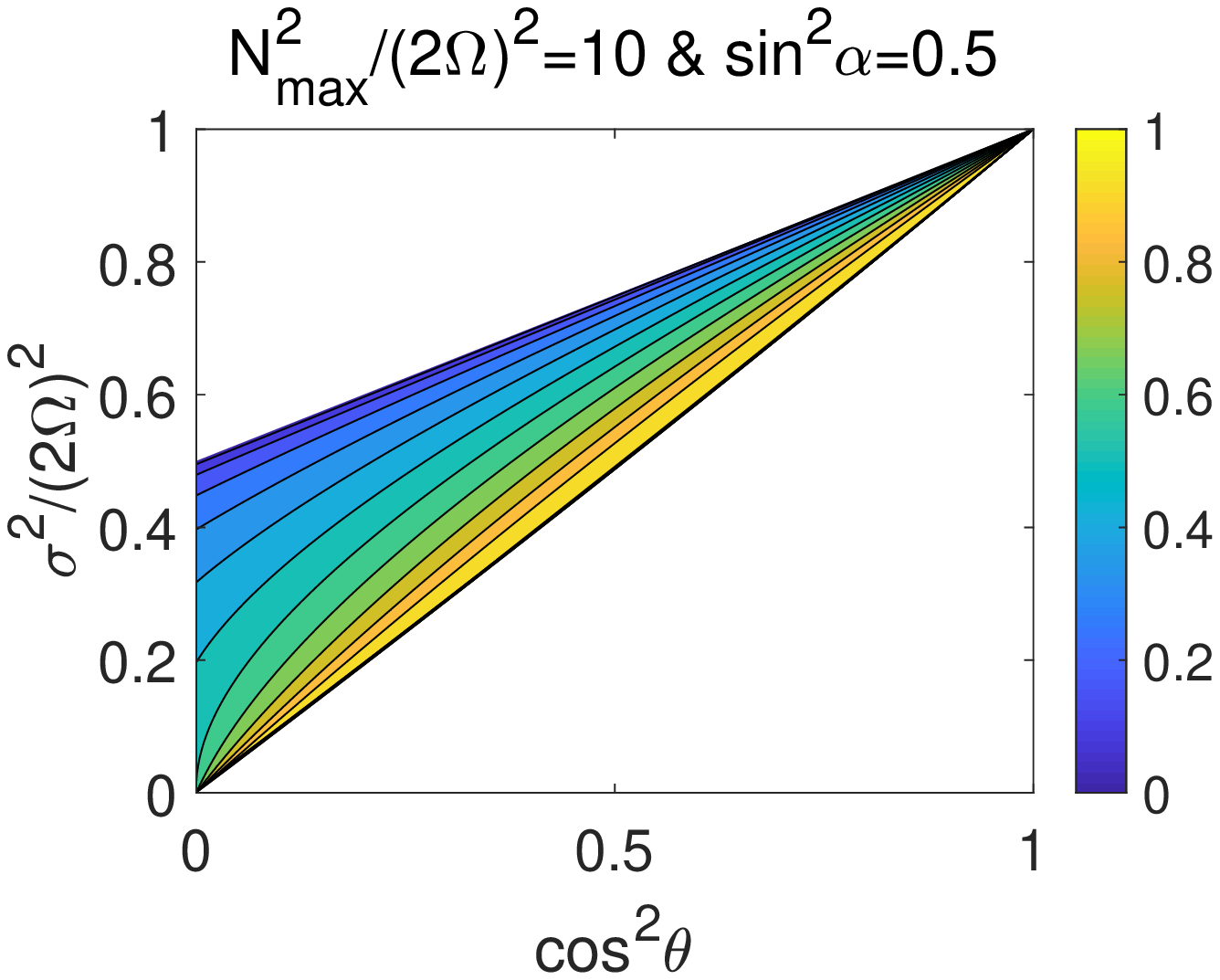}
\caption{}
\end{subfigure}
\begin{subfigure}{0.32\textwidth}
\includegraphics[width=\linewidth]{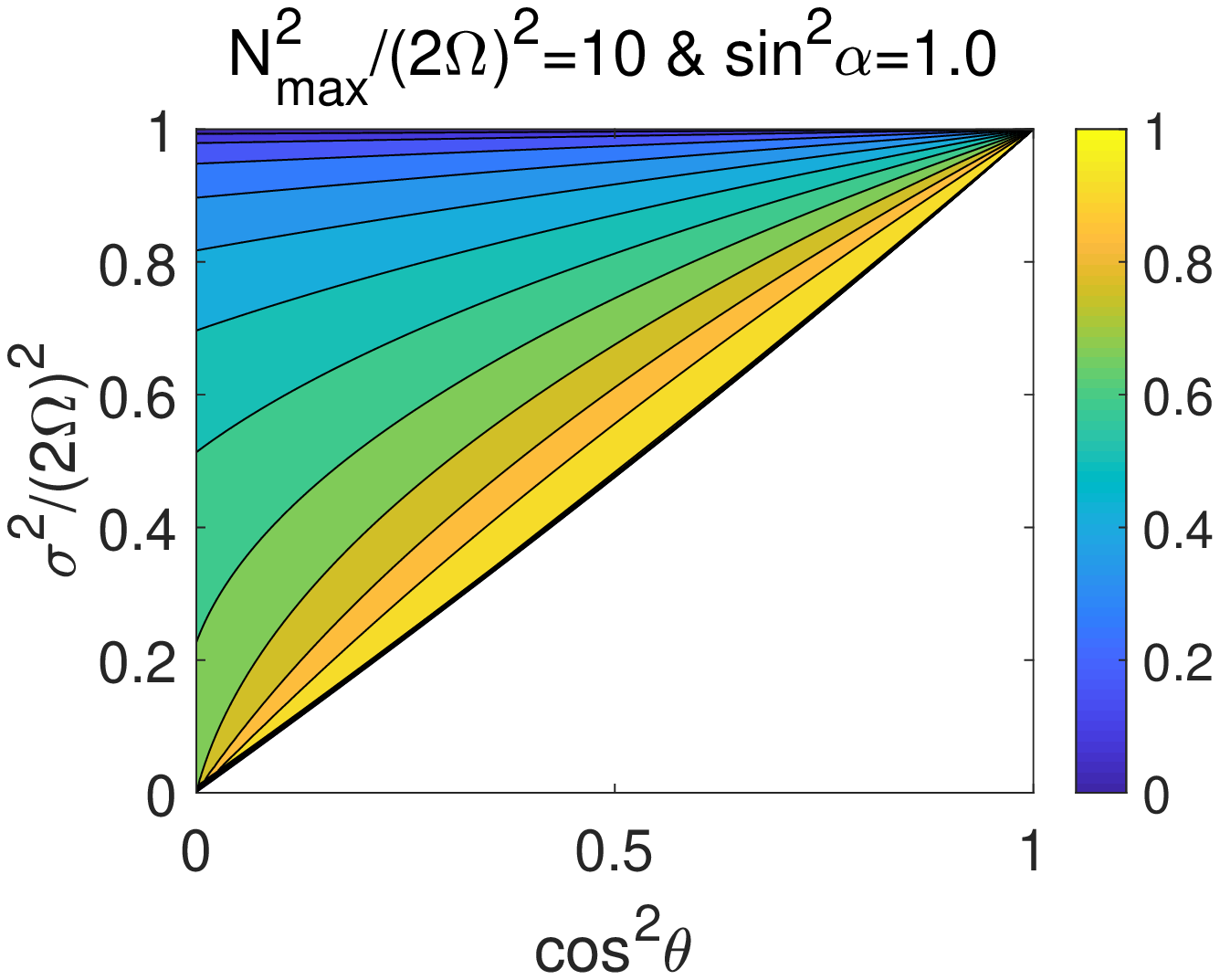}
\caption{}
\end{subfigure}

\medskip

\begin{subfigure}{0.32\textwidth}
\includegraphics[width=\linewidth]{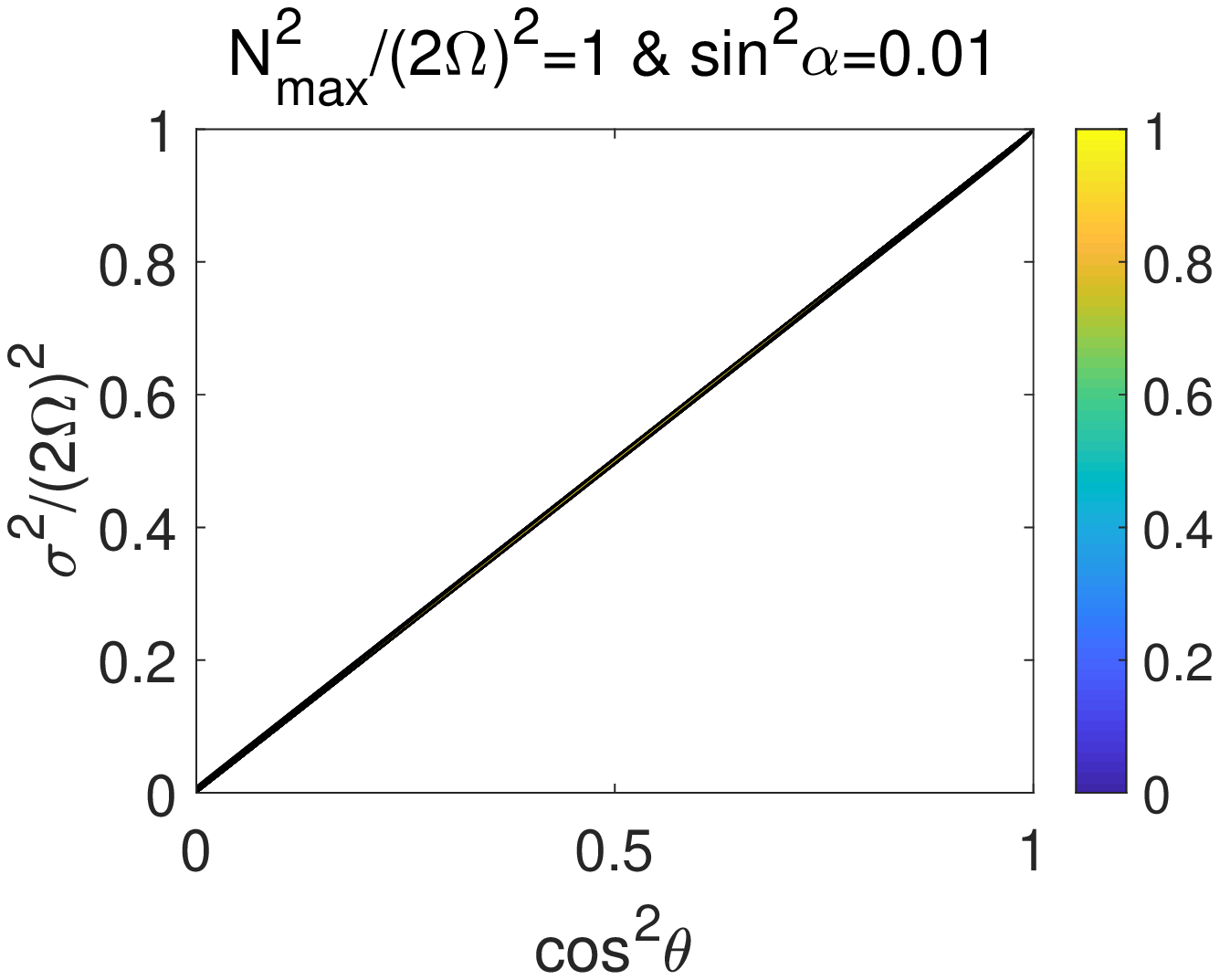}
\caption{}
\end{subfigure}
\begin{subfigure}{0.32\textwidth}
\includegraphics[width=\linewidth]{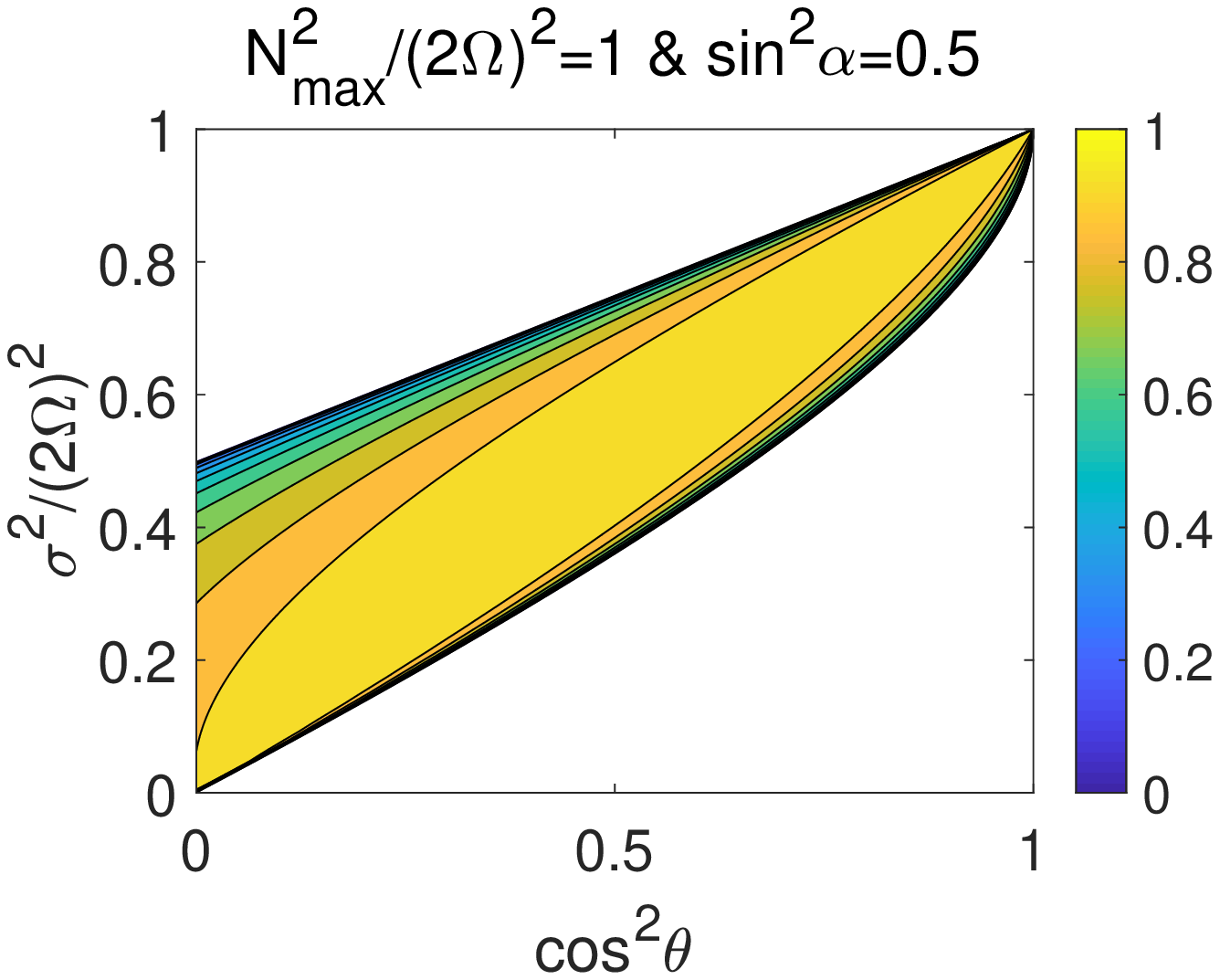}
\caption{}
\end{subfigure}
\begin{subfigure}{0.32\textwidth}
\includegraphics[width=\linewidth]{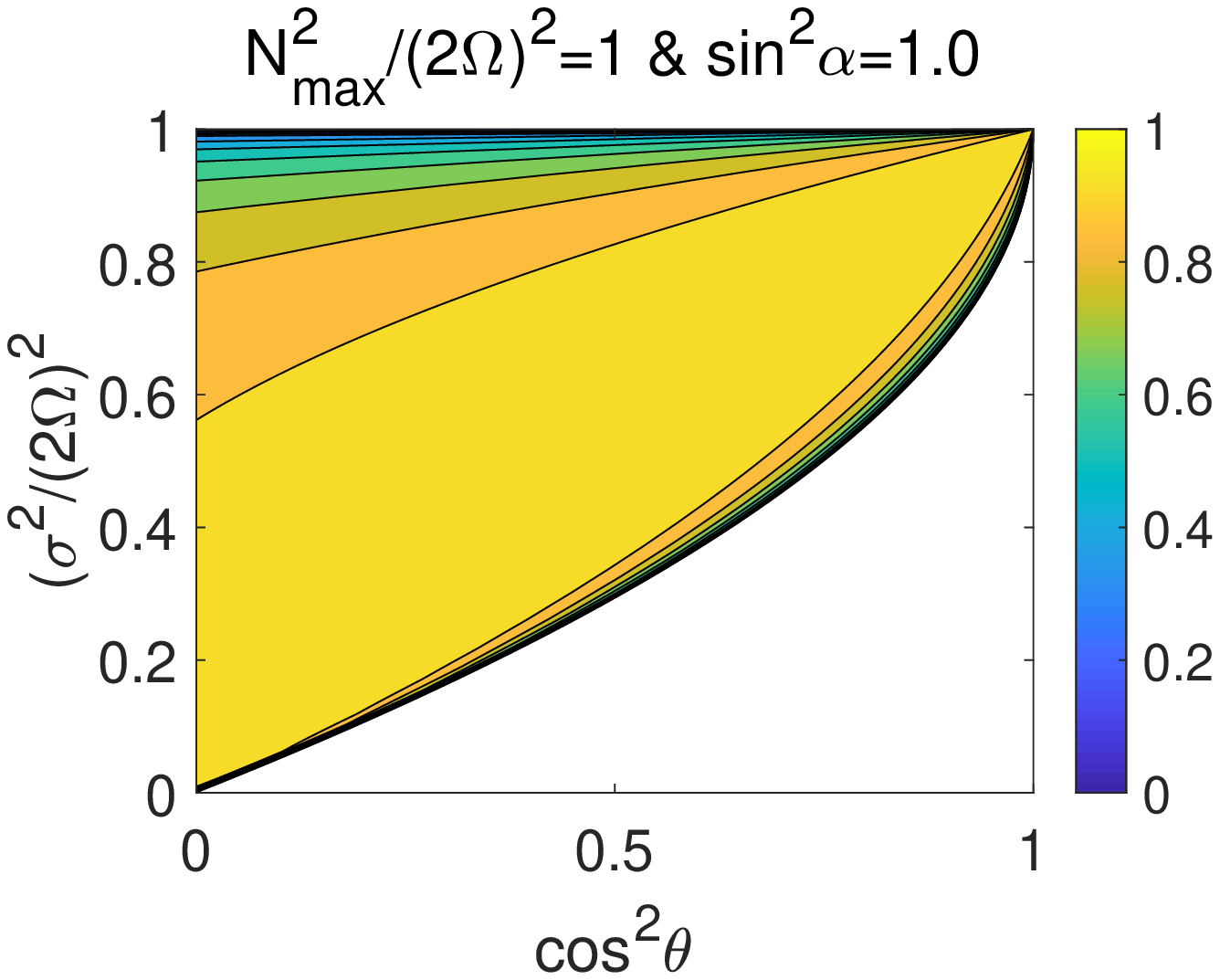}
\caption{}
\end{subfigure}

\medskip

\begin{subfigure}{0.32\textwidth}
\includegraphics[width=\linewidth]{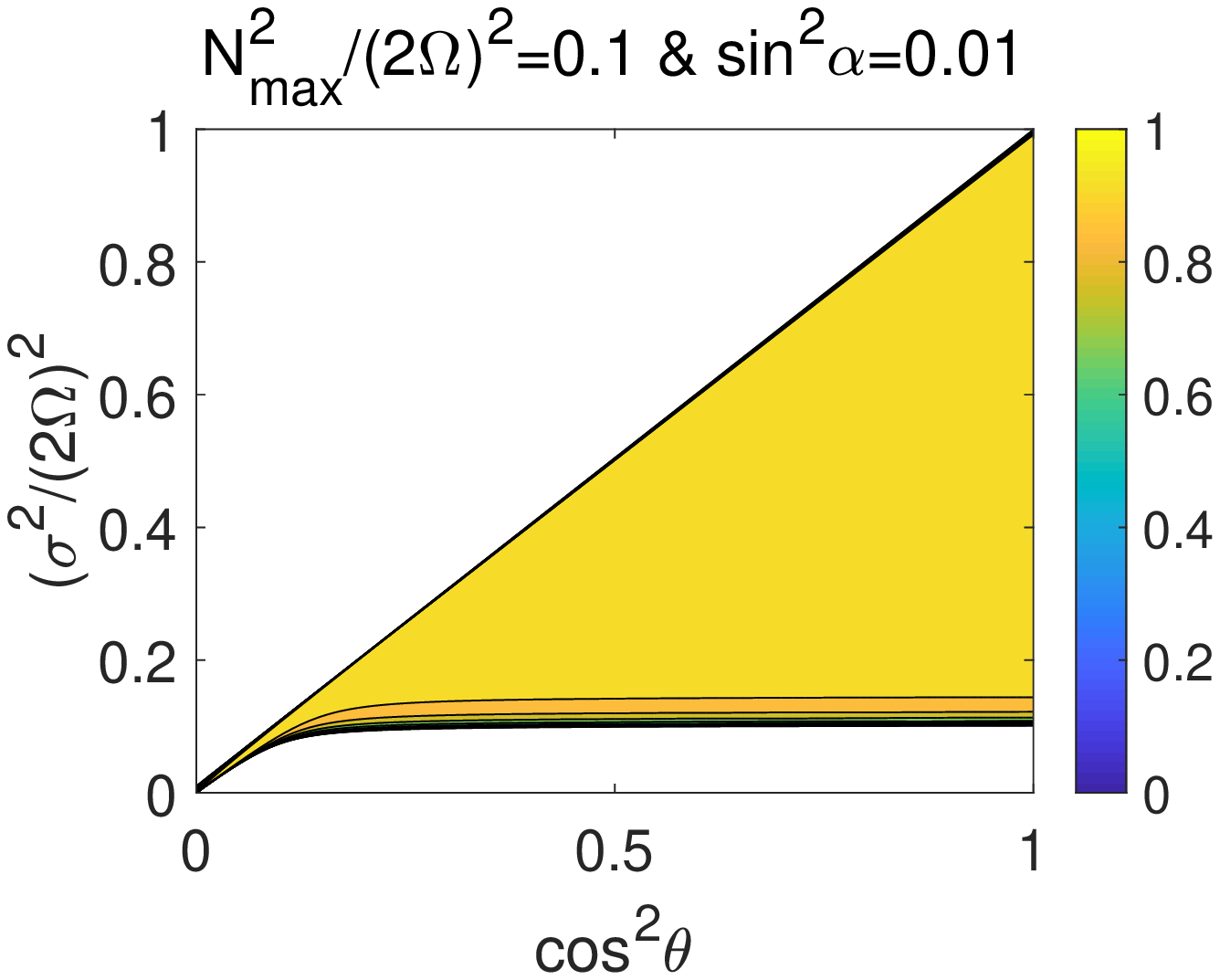}
\caption{}
\end{subfigure}
\begin{subfigure}{0.32\textwidth}
\includegraphics[width=\linewidth]{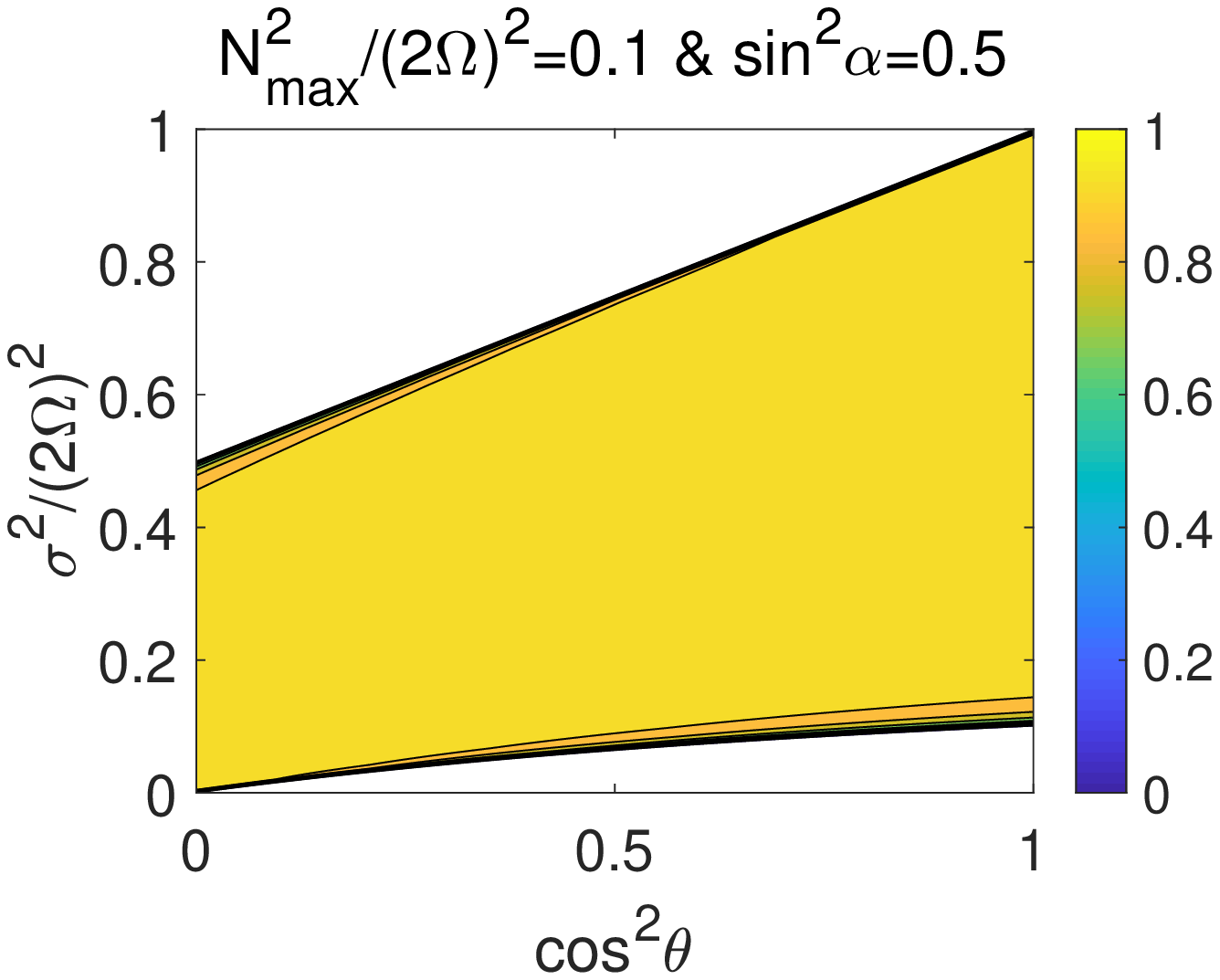}
\caption{}
\end{subfigure}
\begin{subfigure}{0.32\textwidth}
\includegraphics[width=\linewidth]{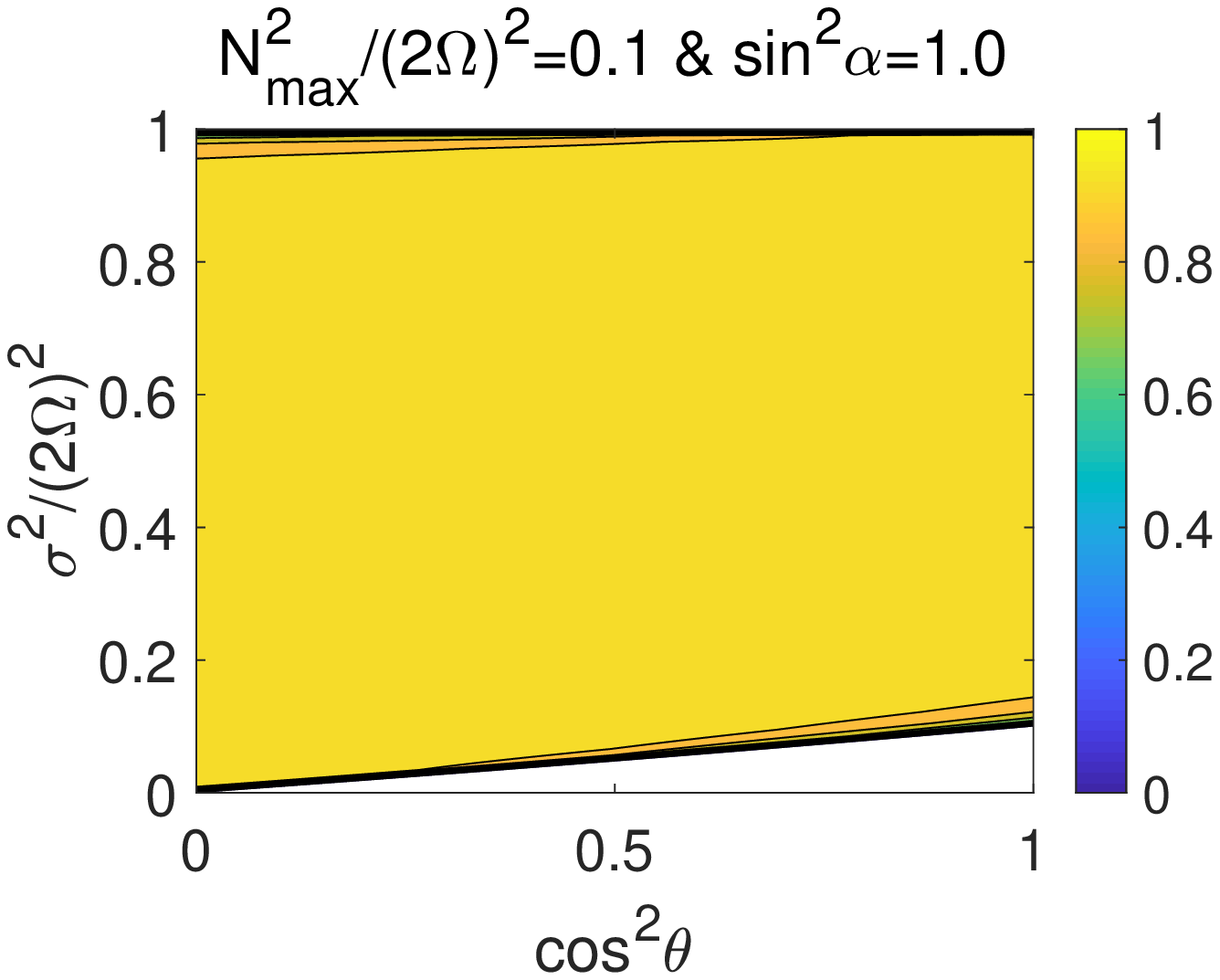}
\caption{}
\end{subfigure}

\caption{Transmission ratios for different $N_{max}^2/(2\Omega)^2$ and $\sin^2\alpha$. The horizontal axis is $\cos^2\theta$, and the vertical axis is $\sigma^2/(2\Omega)^2$. (a)-(c) The slowly rotating group ($N_{max}^2/(2\Omega)^2=10$) with $\sin^2\alpha \in \{0.01,0.5,1.0\}$ from the left to right. (d)-(f) The moderately rotating group ($N_{max}^2/(2\Omega)^2=1$) with $\sin^2\alpha \in \{0.01,0.5,1.0\}$ from the left to right. (g)-(i) The rapidly rotating group ($N_{max}^2/(2\Omega)^2=0.1$) with $\sin^2\alpha \in \{0.01,0.5,1.0\}$ from the left to right. Waves can only survive in the colored regions. The regions are left white in the contour plots if waves cannot survive \label{fig:f3}}
\end{figure}

For the configuration 1, we have plotted the transmission ratios for different $N_{max}^2/(2\Omega)^2$ and $\sin^2\alpha$ in fig.~\ref{fig:f3}. The effect of $N_{max}^2/(2\Omega)^2$ is shown from the top to bottom ($N_{max}^2/(2\Omega)^2=10,1,0.1$), and the effect of $\sin^2\alpha$ is shown from the left to right ($\sin^2\alpha=0.01,0.5,1.0$). First, we observe that the degree of stratification ($N_{max}^2/(2\Omega)^2$) has great impact on the frequency domain. The frequency domain is much wider at low (high) latitudes when the stable layer is strongly (weakly) stratified. Second, we notice that the wavenumber also has important effect on the width of frequency domain. The frequency domain is much wider when the meridional wavenumber dominates the zonal wavenumber. These two points have already been confirmed previously in the frequency domain analysis. Third, we notice that the transmission ratio largely depends on the degree of stratification and wave frequency. High transmission ratios are achieved when $N_{max}^2/(2\Omega^2)$ is small (the stable layer is weakly stratified), or when $\sigma^2-f^2\simeq 0$ (the wave is near the critical colatitude). This result is consistent with our theoretical analysis. In the study of wave transmission in a multi-layered structure, \citet{andre2017layered} also found that low-frequency waves of all wavelengths are able to perfectly transmit near the critical colatitude. The reason is that only one non-zero solution of $c_{g}$ exists at the critical colatitude, and thus the amplitude of incident wave must be equal to that of transmitted wave at the boundary \citep{andre2017layered}.

\subsubsection{Configuration 2}
For configuration 2, wave propagates from the stable layer to the convective layer. We discuss it in two cases $C>0$ and $C<0$, respectively. For the first case $C>0$, it requires $a_{1}=0$. Waves with wavenumber $-s$, $+s$, and $-q$ are the incident wave, reflective wave, and transmitted wave, respectively. From boundary conditions, we obtain
\begin{eqnarray}
&& b_{1}+b_{2}=a_{2}~,\\
&& sb_{1}-sb_{2}=-qa_{2}~,
\end{eqnarray}
from which we obtain the transmission ratio
\begin{eqnarray}
\eta=\frac{q}{s}(\frac{a_{2}}{b_{2}})^2=\frac{4sq}{(s+q)^2}=1-(\frac{\sqrt{\Delta_{m}}-\sqrt{\Delta_{0}}}{\sqrt{\Delta_{m}}+\sqrt{\Delta_{0}}})^2~.
\end{eqnarray}
The analysis is similar to configuration 1, and the conclusion is the same.

For the case $C<0$, we can show that the conclusion remains unchanged, and here we will not repeat it.

\subsection{Continuously varying stratification in the stable layer}
\subsubsection{Transmission at the interface}
For a continuously varying stratification in the stable layer, we assume that $N^2(z)$ is a function of z in the region $z\in(0,z_{m})$. For the existence of wave solution, it requires $B^2-A_{0}C-N^2(z)C>0$ for $z\in(0,z_{m})$. It also requires $N^2(0)=0$ because $N^2(z)$ is assumed to be continuous at the interface. In the stable layer, we write (\ref{eq13}) as
\begin{eqnarray}
\psi_{zz}+s^2 \psi=0~,\label{eq53}
\end{eqnarray}
where $s^2=k^2[(B^2-A_{0}C-N^2C)/{C^2}]$ and $s$ is a function of $z$. We perform WKB analysis to solve this equation. Assuming $\psi \propto e^{iX(z)}$ and substituting it into the above equation, we obtain
\begin{eqnarray}
 iX_{zz}-X_{z}^2+s^2=0~. \label{eq55}
\end{eqnarray}
It is a nonlinear equation which is generally difficult to solve. However, approximate solution can be obtained if $X(z)$ is assumed to be a slowly varying function. If $X(z)$ is a slowly varying function, then it is reasonable to assume that $X_{zz}$ is small compared to $X_{z}^2$. Neglecting $X_{zz}$ in (\ref{eq55}) gives the first approximation
\begin{eqnarray}
-X_{z}^2+s^2=0~,
\end{eqnarray}
which has the solution
\begin{eqnarray}
X(z)=\pm \int_{0}^{z} s(z') dz'~. \label{eq57}
\end{eqnarray}
From a comparison of $X_{zz}$ and $X_{z}^2$, we find that the assumption on slowly varying function is valid as long as $|s_{z}/s^2| \ll 1$. In other words, it requires that the wavelength ($1/s$) is much smaller than the variation scale length of the wavenumber ($\partial z/\partial{\log s}$).
Substitution of (\ref{eq57}) into (\ref{eq55}) yields the following equation
\begin{eqnarray}
X_{z}^2=\pm i s_{z}+s^2~,
\end{eqnarray}
from which we obtain the second approximation
\begin{eqnarray}
X_{z}=\pm(\pm i s_{z}+s^2)^{1/2}\approx \pm s +  i s_{z}/(2s)~.
\end{eqnarray}
An integration of the above expression gives
\begin{eqnarray}
X(z)\approx \frac{i}{2}[\ln s(z)-\ln s(0)] \pm \int_{0}^{z} s(z') dz'~.
\end{eqnarray}
Substituting $X$ back into $\psi$, we obtain an approximate solution of (\ref{eq53})
\begin{eqnarray}
\psi \propto [s(0)/s(z)]^{1/2} e^{\pm i\int_{0}^{z} s(z') dz'}~.
\end{eqnarray}
Thus the general solution of $\psi$ can be written as
\begin{eqnarray}
\psi=[s(0)/s(z)]^{1/2} \left[ b_{1}e^{i\int_{0}^{z}s(z') dz'}+b_{2}e^{-i\int_{0}^{z} s(z') dz'} \right]~,
\end{eqnarray}
and its derivative is
\begin{eqnarray}
\psi_{z}=-2^{-1} s_{z}(z) s(z)^{-1} \psi + i[s(z)s(0)]^{1/2}\left[ b_{1} e^{i\int_{0}^{z} s(z') dz'} - b_{2} e^{-i\int_{0}^{z} s(z') dz'}\right]~.
\end{eqnarray}
At $z=0$, we have
\begin{eqnarray}
&&\psi(0)=b_{1}+b_{2}~,\\
&&\psi_{z}(0)=-2q\beta\psi(0)+iq(b_{1}-b_{2})~,
\end{eqnarray}
where $\beta=(2s(0))^{-2}s_{z}(0)$ and $s(0)=q$ have been used. Here the explicit form of $\beta$ is
\begin{eqnarray}
|\beta|=\frac{|dN^2(0)/dz| C^2}{8k(B^2-A_{0}C)^{3/2}}~,
\end{eqnarray}
which depends on the slope of $N^2$. As $\beta$ can be expressed as $(1/s)/(\partial {z}/\partial \log s)|_{z=0}$, it measures the importance of the wavelength to the variation scale length of the wavenumber at the interface. $\beta$ is small when the wavenumber varies slowly near the interface. Later, we will show that $\beta$ plays a critical role in wave transmissions at the interface. Now we deduce wave transmission ratios at the interface for the case $C>0$ in configuration 1.
For $C>0$, the outward transmitted wave requires $b_{2}=0$, and thus, the following equations can be deduced from the boundary conditions
\begin{eqnarray}
&&a_{1}+a_{2}=b_{1}~,\\
&&a_{1}-a_{2}=2\beta ib_{1}+b_{1}~.
\end{eqnarray}
The solution of the above equations is
\begin{eqnarray}
&&\frac{a_{2}}{a_{1}}=\frac{-\beta i}{1+\beta i}~, \\
&&\frac{b_{1}}{a_{1}}=\frac{1}{1+\beta i}~,
\end{eqnarray}
from which we can deduce the transmission ratio
\begin{eqnarray}
\eta=\frac{q}{q}\frac{|b_{1}|^2}{|a_{1}|^2}=\frac{1}{1+\beta^2}=1-\zeta~. \label{eq70}
\end{eqnarray}
Obviously, the efficiency of transmission mainly depends on the value of $|\beta|$.
If $|\beta| \rightarrow \infty$, then $\eta \rightarrow 0$ and the incident wave is totally reflected. If $|\beta|\rightarrow 0$, then $\eta \rightarrow 1$ and the incident wave is totally transmitted. If $\infty>|\beta|>0$, then $1>\eta>0$ and the incident wave is partially reflected.

Let us further assume that $N^2$ varies as an algebraic function $N^2(z) =\gamma_{2} z^{\nu}$, where $\gamma_{2}>0$ and $\nu\geq 1$ are constants. Making a substitution, we obtain
\begin{eqnarray}
|\beta|=\frac{\gamma_{2} \nu z^{\nu-1} C^2}{8k(B^2-A_{0}C)^{3/2}}|_{z=0}~.
\end{eqnarray}
From the above equation, we have the following conclusions:
\begin{itemize}
\setlength{\itemsep}{0pt}
\setlength{\parsep}{0pt}
\setlength{\parskip}{0pt}
\item[(1)] When $\nu>1$, we have $|\beta| =0$, thus the incident wave is totally transmitted at the interface.
\item[(2)] When $\nu=1$, we have $\infty>|\beta|>0$, thus the incident wave is partially transmitted at the interface.
\end{itemize}
It has to be mentioned that above conclusions are drawn based on the WKB analysis, which generally requires $|\beta|=|(2s(0))^{-2}s_{z}(0)|\ll 1$. Under this condition, wave transmission is expected to be efficient. Although (\ref{eq70}) is obtained under the WKB approximations, in the appendix~\ref{appendixc} we have shown in a non-WKB analysis that it probably still holds for large $|\beta|$.

\begin{figure}
\centering

\begin{subfigure}{0.32\textwidth}
\includegraphics[width=\linewidth]{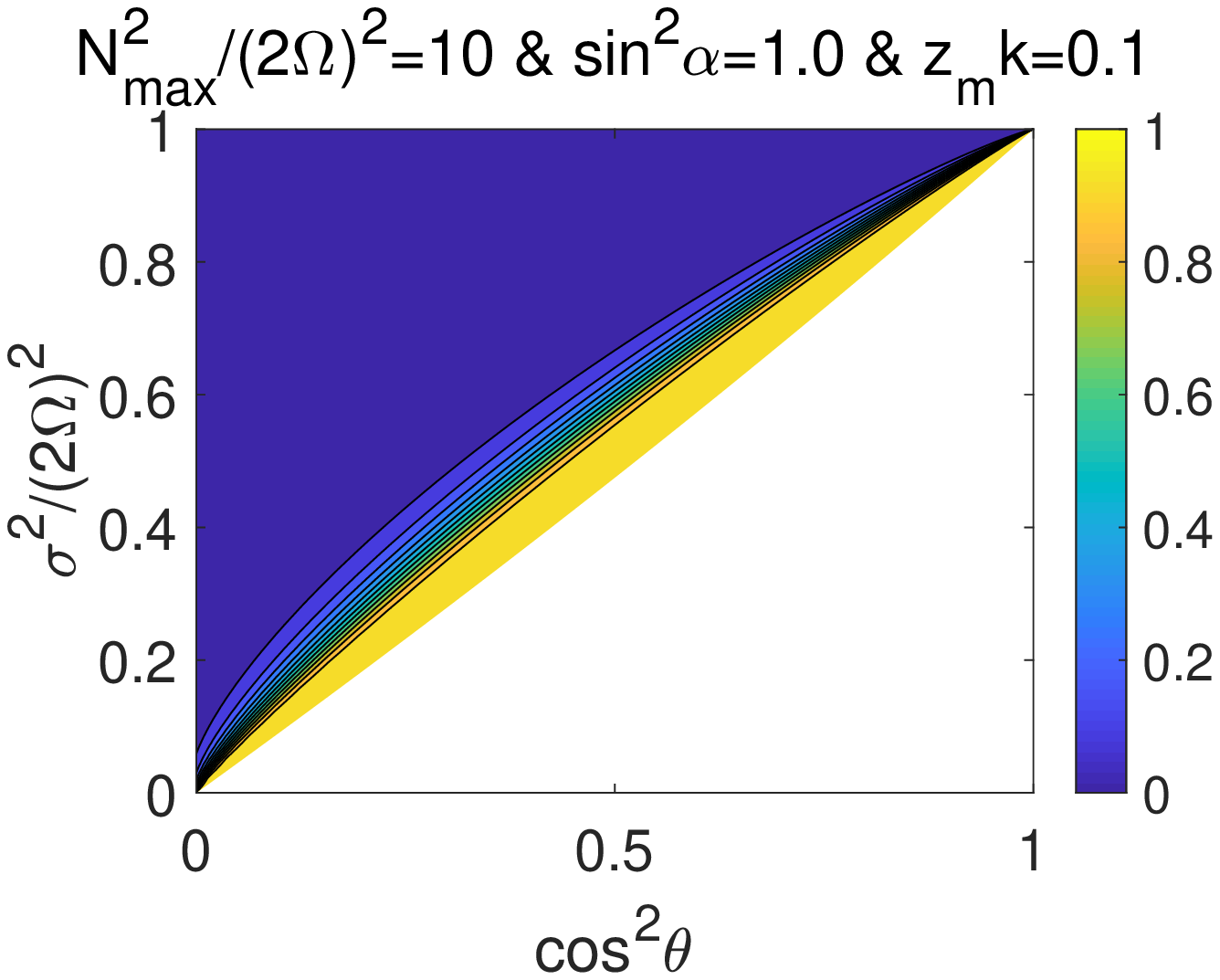}
\caption{}
\end{subfigure}
\begin{subfigure}{0.32\textwidth}
\includegraphics[width=\linewidth]{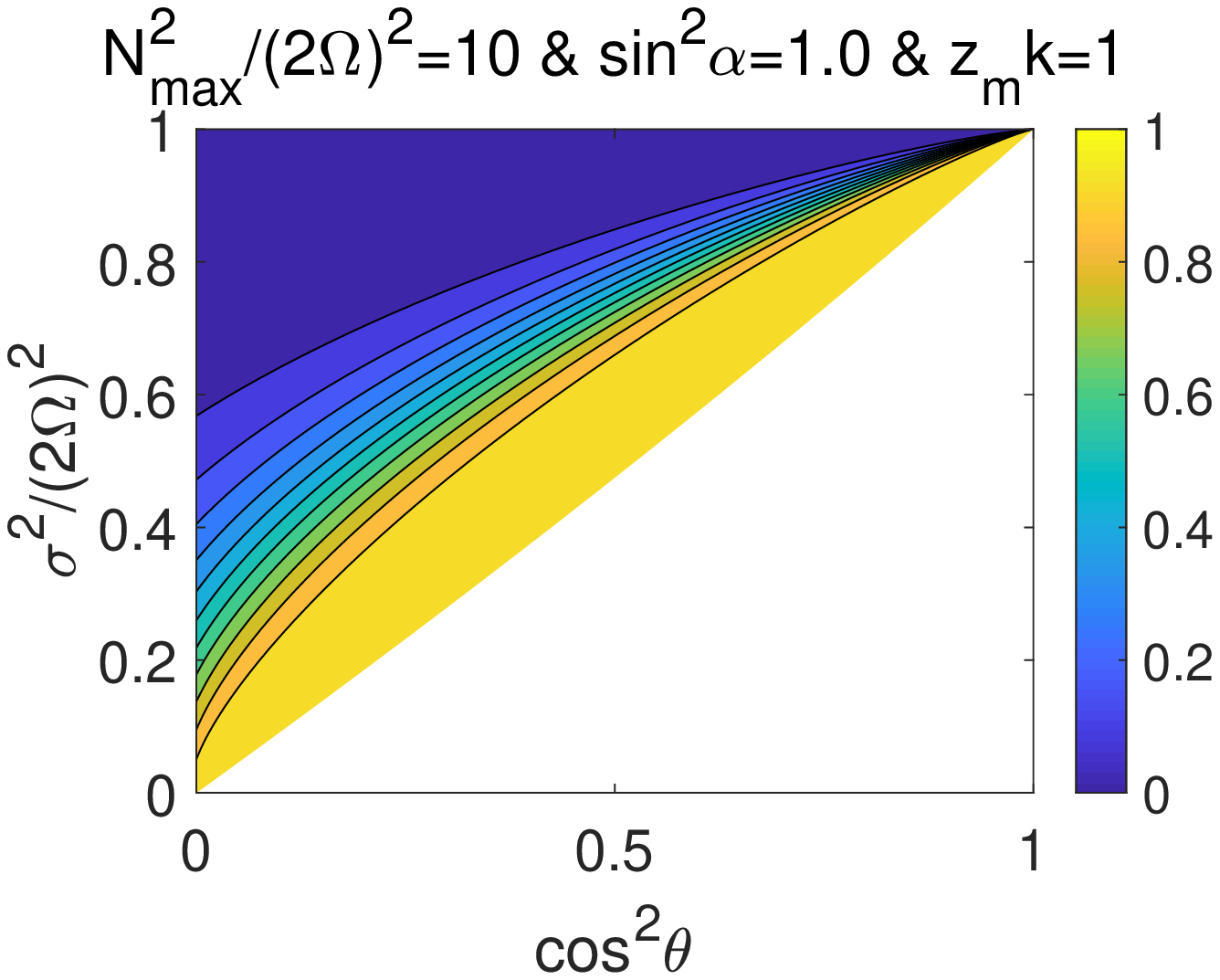}
\caption{}
\end{subfigure}
\begin{subfigure}{0.32\textwidth}
\includegraphics[width=\linewidth]{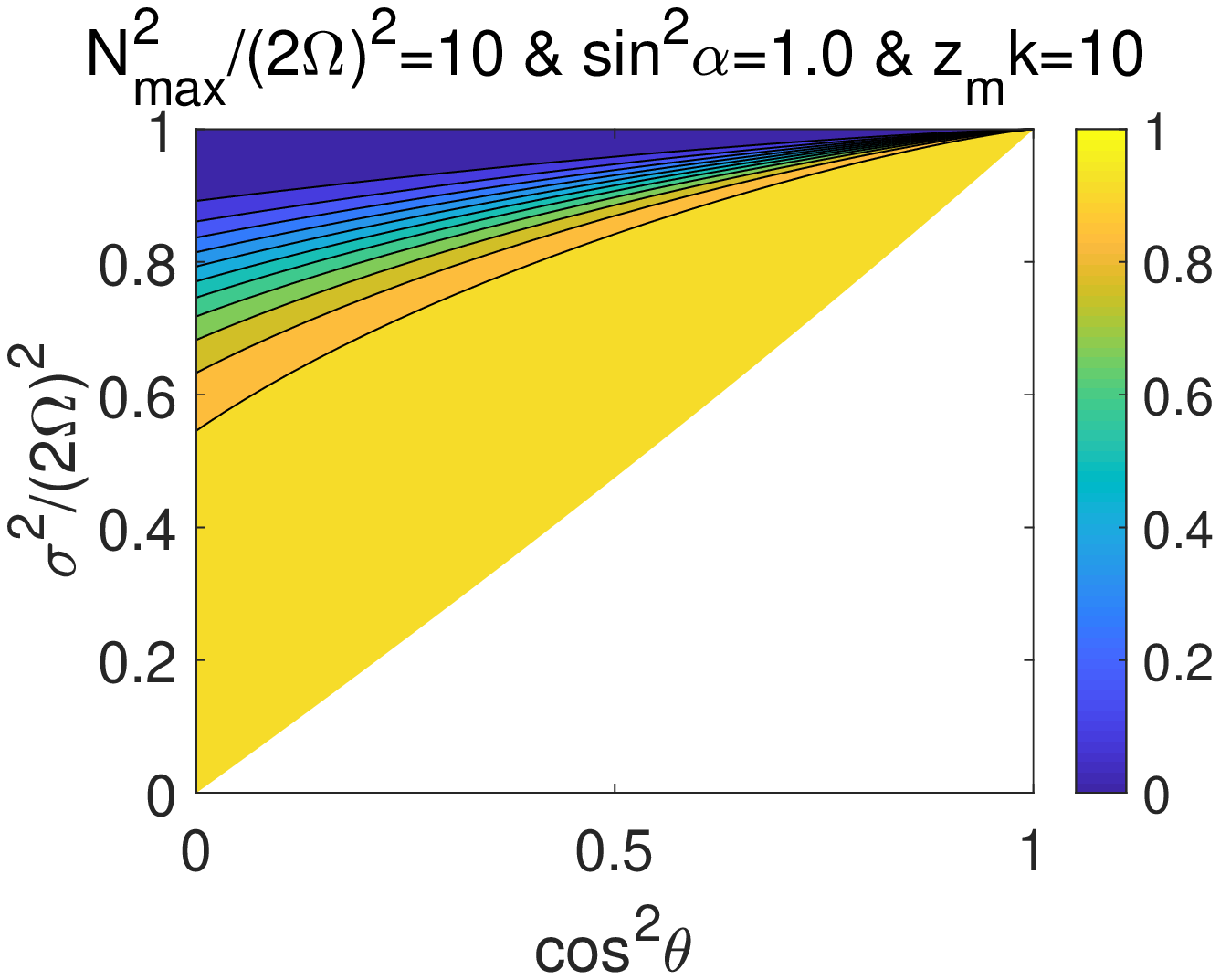}
\caption{}
\end{subfigure}

\medskip

\begin{subfigure}{0.32\textwidth}
\includegraphics[width=\linewidth]{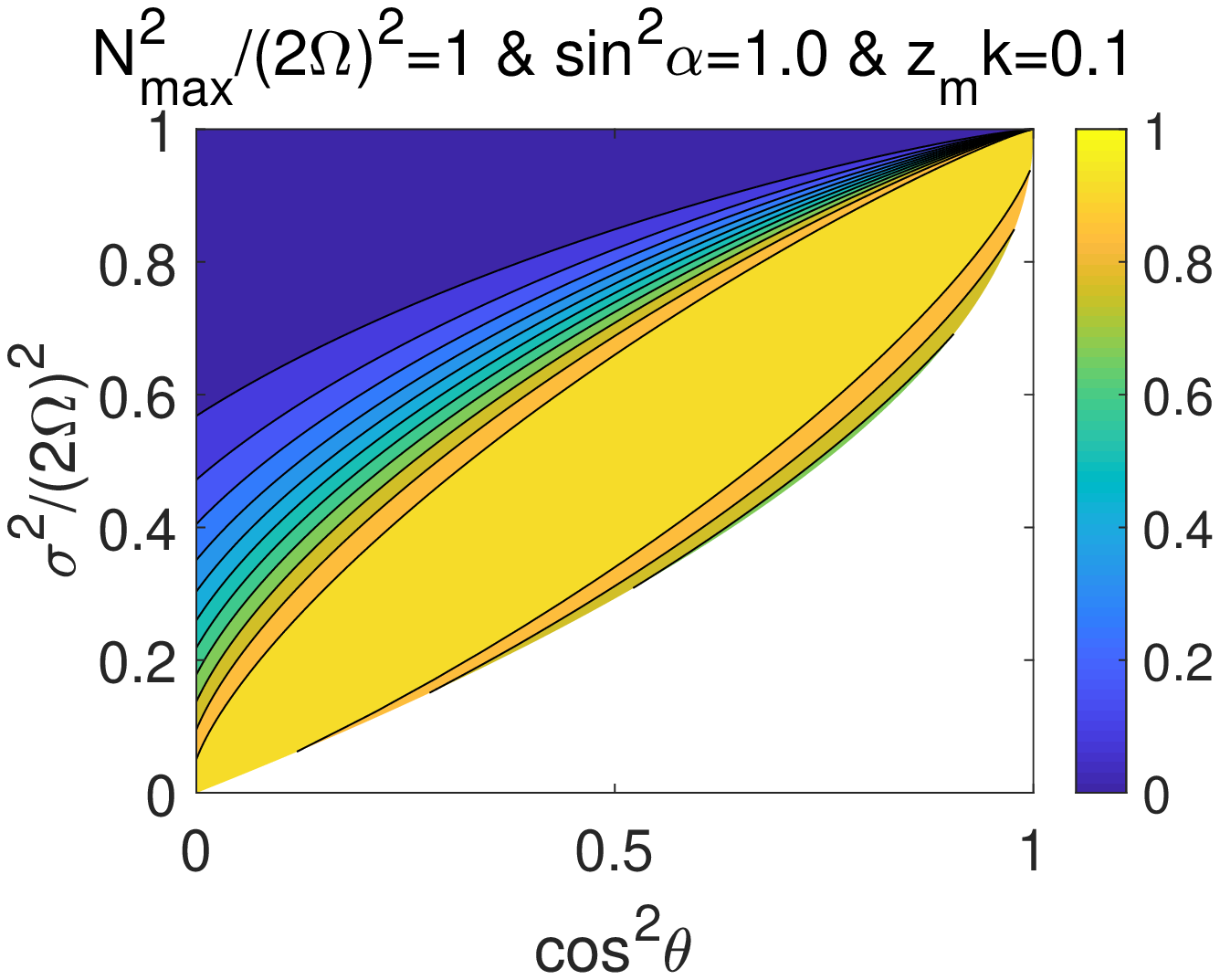}
\caption{}
\end{subfigure}
\begin{subfigure}{0.32\textwidth}
\includegraphics[width=\linewidth]{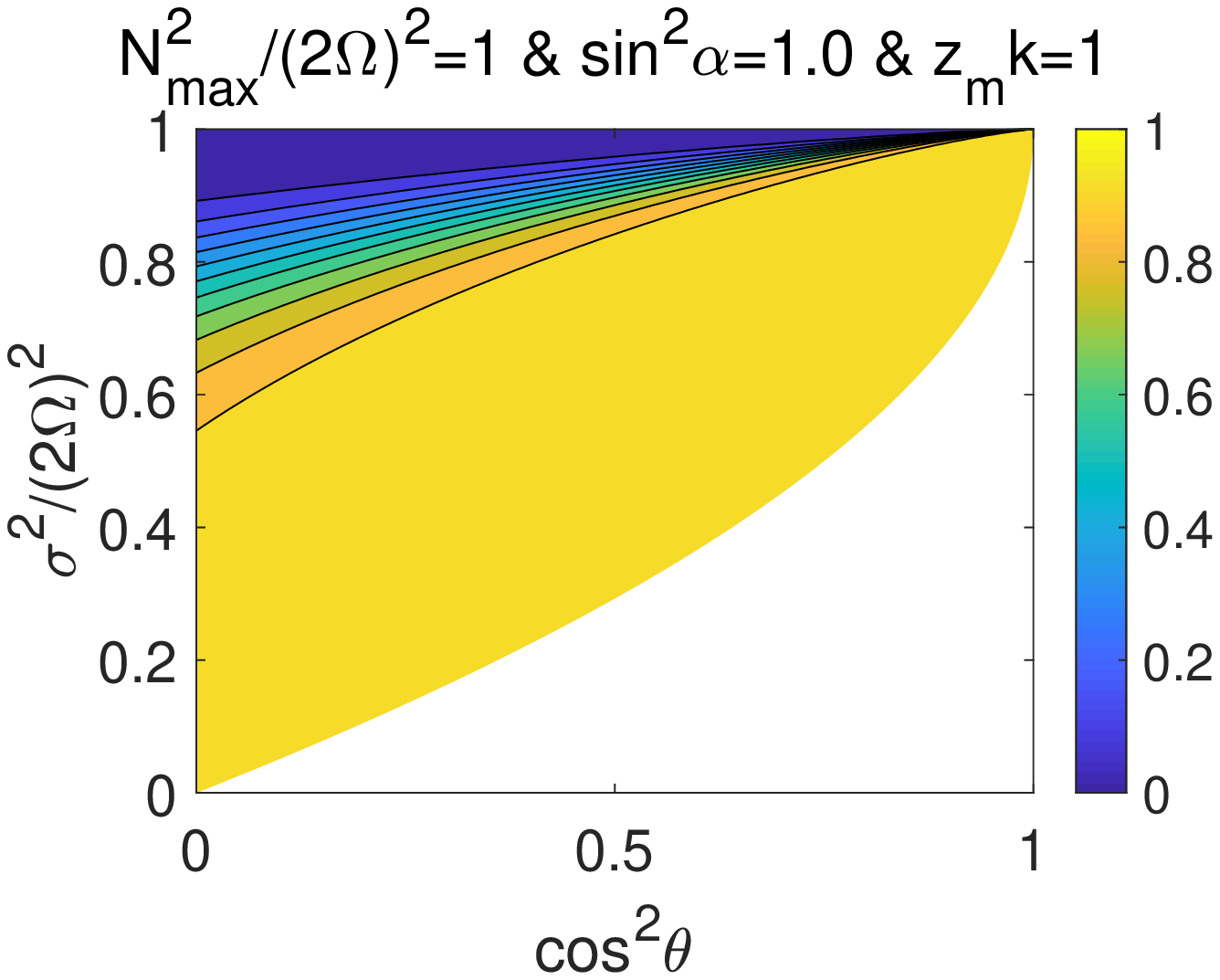}
\caption{}
\end{subfigure}
\begin{subfigure}{0.32\textwidth}
\includegraphics[width=\linewidth]{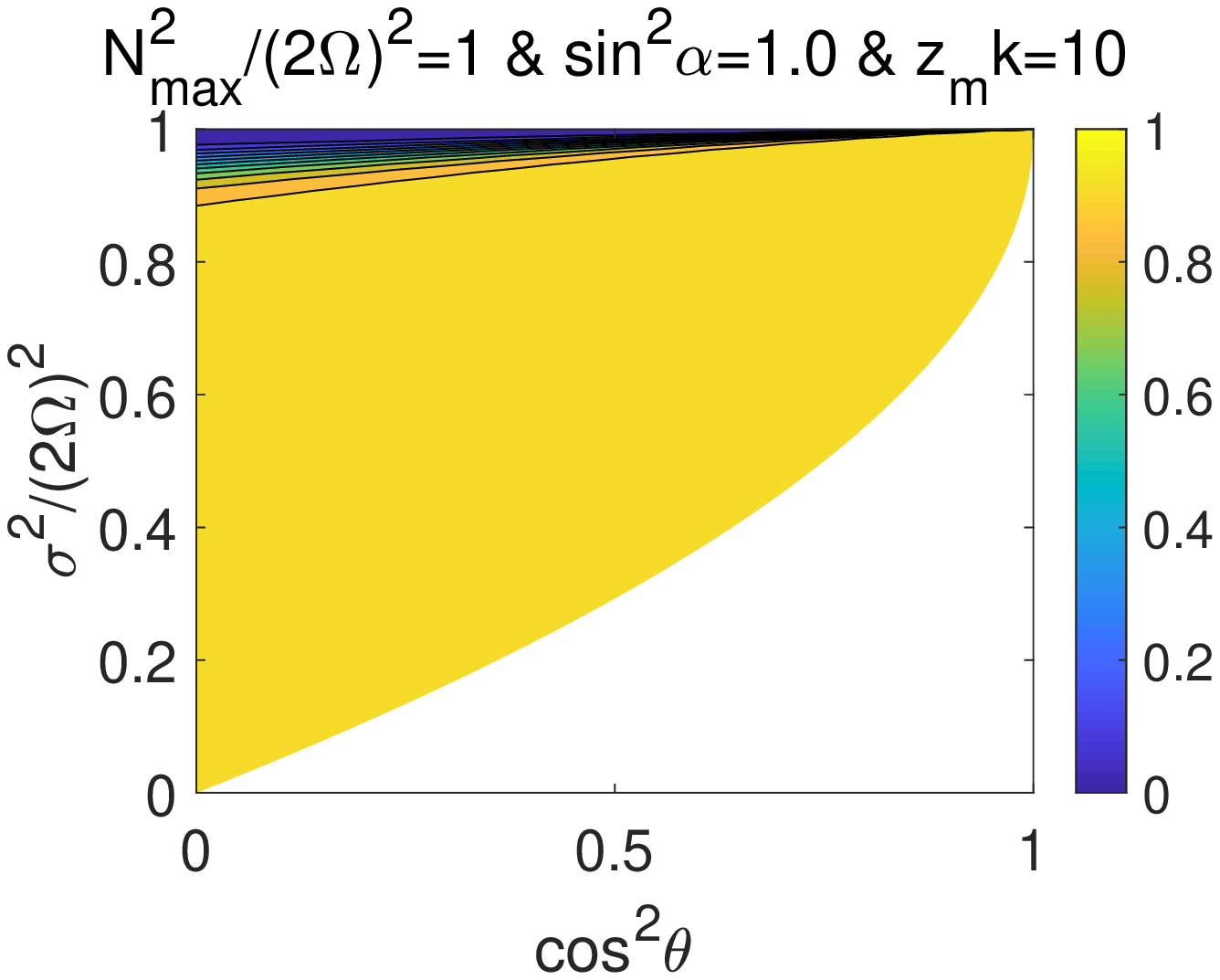}
\caption{}
\end{subfigure}

\medskip

\begin{subfigure}{0.32\textwidth}
\includegraphics[width=\linewidth]{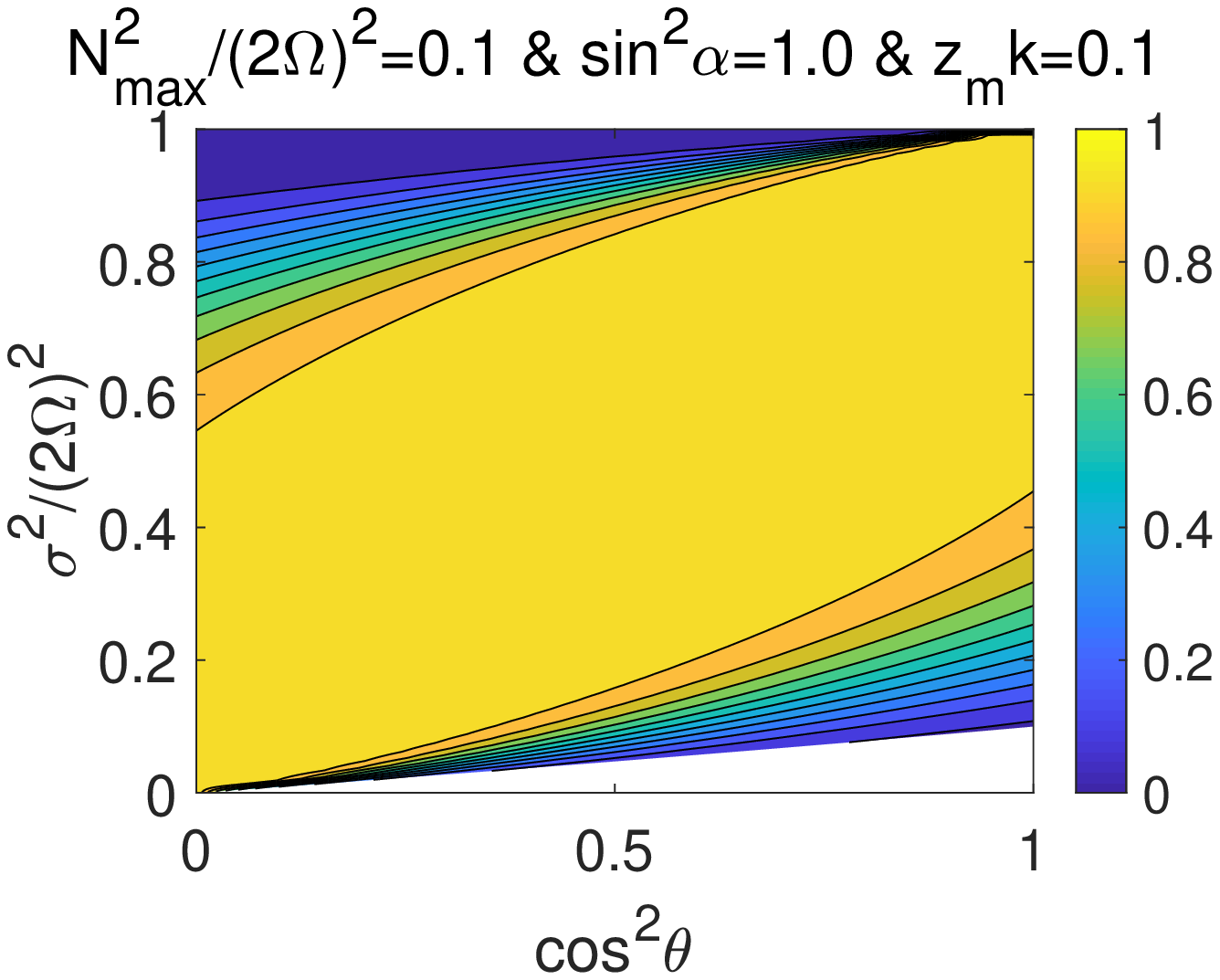}
\caption{}
\end{subfigure}
\begin{subfigure}{0.32\textwidth}
\includegraphics[width=\linewidth]{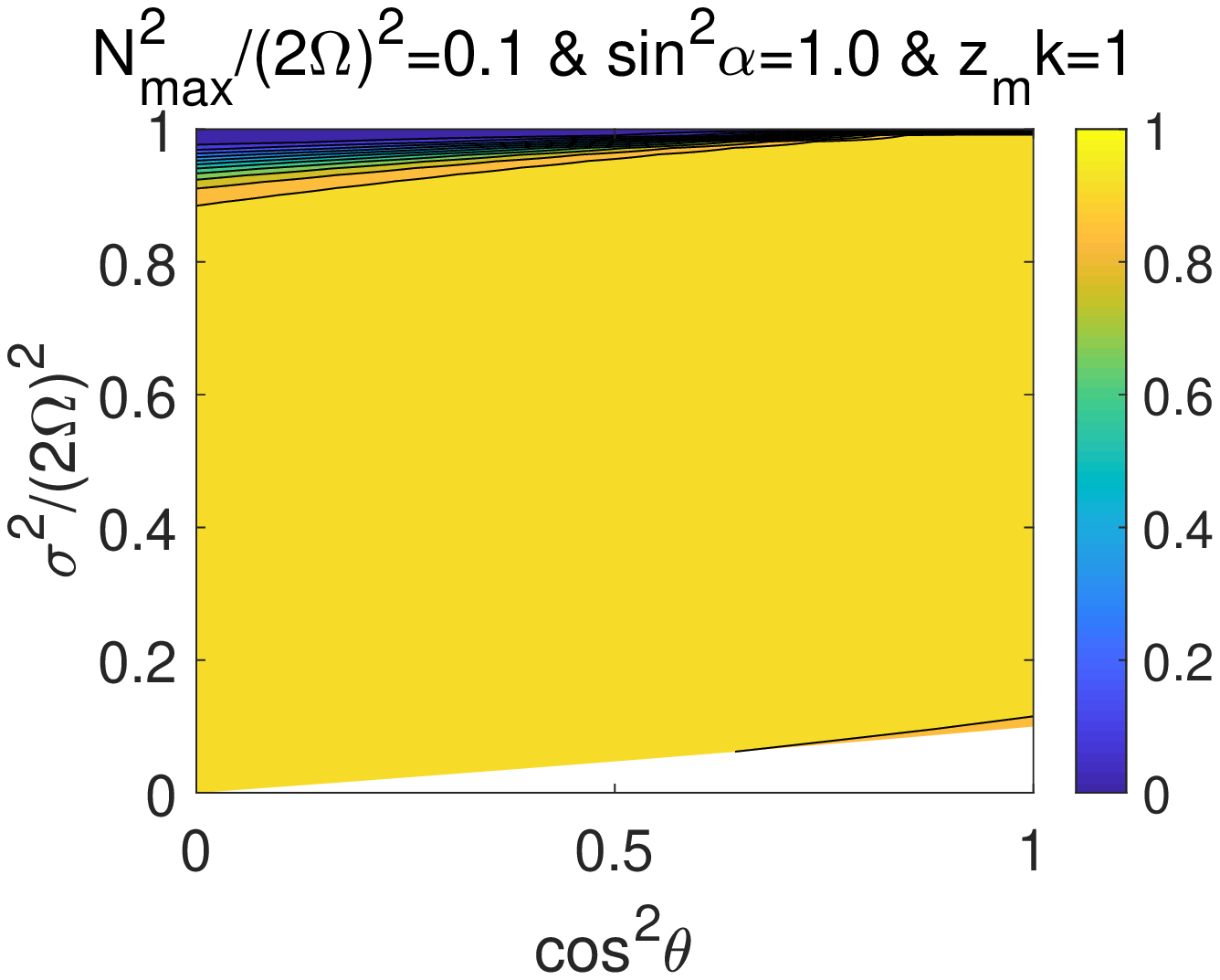}
\caption{}
\end{subfigure}
\begin{subfigure}{0.32\textwidth}
\includegraphics[width=\linewidth]{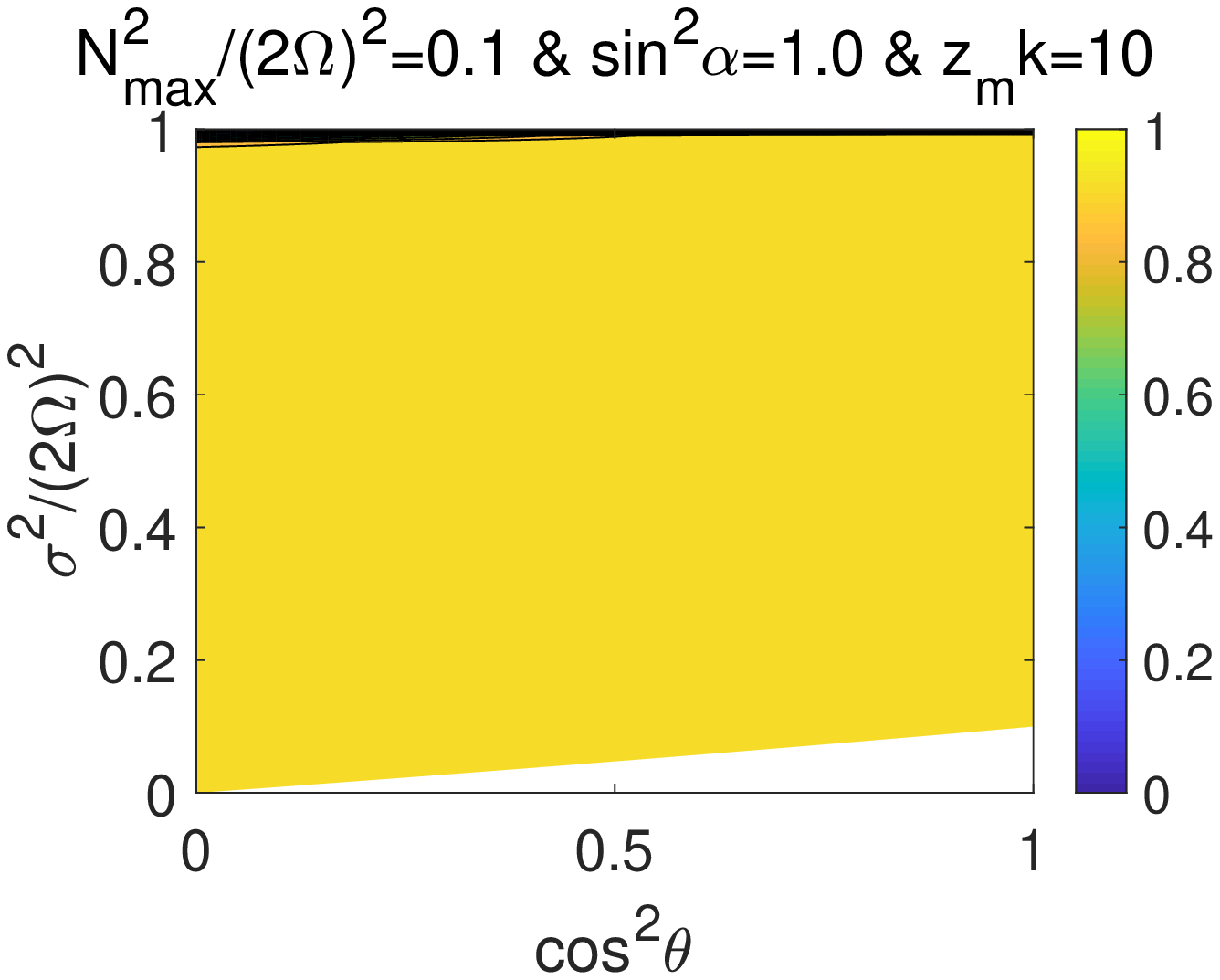}
\caption{}
\end{subfigure}

\medskip

\caption{Transmission ratios at different $z_{m}k$ (=0.1,1,10 from the left to right) and $N_{max}^2/(2\Omega)^2$ (=10,1,0.1 from the top to bottom) when the stratification is linearly varying $N^2=\gamma_{2}z$. The parameter $\sin^2\alpha=1.0$ for all the shown cases. Waves can only survive in the colored regions. The regions are left white in the contour plots if waves cannot survive. \label{fig:f4}}
\end{figure}

Let us now consider the special case $\nu=1$. For this case, we have
\begin{eqnarray}
|\beta|&&=\frac{\gamma_{2} C^2}{8k(B^2-A_{0}C)^{3/2}}\\
&&=\frac{N_{max}^2 C^2}{8 z_{m}k(B^2-A_{0}C)^{3/2}} \label{eq73}
\end{eqnarray}
where the relation $N_{max}^2=\gamma_{2}z_{m}$ has been applied.
It can be seen that $|\beta|$ decreases with $z_{m}k$ and increases with $N_{max}^2/(2\Omega)^2$. Also, it can be proved that $|\beta|$ decreases with $\sigma^2$ by taking derivatives. As a result, $\eta$ increases with $z_{m}k$ and $\sigma^2$, and decreases with $N_{max}^2/(2\Omega)^2$. Specifically, we have $\eta\rightarrow 1$ when $N_{max}^2/(2\Omega)^2 \rightarrow 0$, or $\sigma^2 \rightarrow f^2$, or $z_{m}\gg k^{-1}$. Therefore, the transmission is efficient when the stable layer is weakly stratified, or when the wave is near the critical latitude, or when the thickness of the linearly varying stratification layer is far greater than the horizontal wavelength. Fig.\ref{fig:f4} illustrates the transmission contours at different $z_{m}k$ and $N_{max}^2/(2\Omega)^2$. It clearly shows that the transmission ratio increases with $z_{m}k$ even when the stable layer is strongly stratified. A large value of $z_{m}k$ can be easily realized for short waves. It indicates that short waves are more likely to transmit in a strongly stratified fluid when the stratification in the stable layer is linearly varying.

Following similar procedures, we find that the transmission ratios remain unchanged for $C<0$, and for configuration 2. Here we will not repeat the discussion for these cases.

\subsubsection{Far-field transmission away from the interface}
In previous sections, we mainly consider wave transmissions at the interface. By the WKB analysis, we have found that the incident wave can be totally transmitted at the interface for a convexly varying stratification. However, wave reflection still can occur in the far-field away from the interface. In this section, we will discuss the far-field wave transmission for linearly and convexly varying buoyancy frequency profiles ($\nu\geq 1$) through WKB analysis. We consider a three-layer structure, with a bottom convective layer, a middle transition stable layer with a algebraic function of $N^2=\gamma_{2} z^{\nu}$, and a top stable layer with constant $N^2=N_{max}^2$. The bottom, middle, and top layers are separated at $z=0$ and $z=z_{m}$, respectively. Here we only consider the case with $C>0$. The discussion on the case with $C<0$ is similar, and the conclusion remains unchanged. Let $a_{1}$ and $a_{2}$ be the amplitudes of incident and reflective waves at the bottom layer, $b_{1}$ and $b_{2}$ be the amplitudes of upward and downward propagating waves in the middle layer, and $b_{3}$ be the transmitted wave at the top layer. Matching the boundary conditions at the two interfaces, we can obtain the relations among these amplitudes. After some trivial calculations, we find that the far-field transmission can be written as
\begin{eqnarray}
\eta_{z_{m}}\approx\frac{1}{\beta_{m}^2+\beta^2+1+2\beta^2\beta_{m}^2-2\beta\beta_{m}[(\beta_{m}-\beta)\sin\chi+(\beta\beta_{m}+1)\cos\chi]}~,
\end{eqnarray}
where $\beta_{m}=(2s(z_{m}))^{-2}s_{z}(z_{m})$, and $\chi$ is the argument of $\exp(-i\int_{0}^{z_{m}}2s(z')dz')$. When $\nu>1$, it is easy to obtain
\begin{eqnarray}
\eta_{z_{m}}\approx\frac{1}{\beta_{m}^2+1}~,
\end{eqnarray}
with
\begin{eqnarray}
|\beta_{m}|=\frac{\nu N_{max}^2 C^2}{8 z_{m}k(B^2-A_{0}C-N_{max}^2C)^{3/2}}~.
\end{eqnarray}
Interestingly, it can be shown that the far-field transmission ratio for the cases $\nu>1$ also increases with $z_{m}k$ and decreases with $N_{max}^2/(2\Omega)^2$, which is quite similar to the result obtained at the interface for the case $\nu=1$.

WKB analysis is not valid for the cases $0<\nu<1$. We will discuss these cases in the following section by numerical calculations. In summary, for far-field transmission, we mainly have the following conclusion:
\begin{itemize}
\setlength{\itemsep}{0pt}
\setlength{\parsep}{0pt}
\setlength{\parskip}{0pt}
\item[(1)] For $\nu>1$, although the incident wave is totally transmitted at the interface, it is partially transmitted at the far-field away the interface.
\item[(2)] For $\nu>1$, the far-field transmission can be efficient when the thickness of the stratification layer is far greater than the horizontal wavelength, but less efficient the other way around.
\end{itemize}

\subsection{Numerical solutions of transmission ratios for different $\nu$}\label{sectionNum}
In the above WKB and non-WKB analysis, we have considered the transmission ratio for $N^2=\gamma_{2}z^{\nu}$ with $\nu\geq 1$. The case $0<\nu<1$ has not been discussed yet. When $0<\nu<1$, WKB approximation is not valid. For this case, we try to solve (\ref{eq13}) numerically. However, solving (\ref{eq13}) directly is not easy for the current configurations, because the wavenumber of outgoing wave is not well defined. As a result, the wavenumber of the transmitted wave has to be estimated. This estimation may affect the transmission ratio at the radiative-convective boundary. To gain some insights, we estimate the wavenumber of the transmitted wave at an arbitrary points $0<z_{b}<z_{m}$, and then investigate the effects on the transmission ratio. At the location $z=z_{b}$, we estimate the wavenumber $s_{z_{b}}$ of the transmitted wave at $z=z_{b}$ as
\begin{eqnarray}
s_{z_{b}}^2=k^2\frac{N_{max}^2}{C}\left( \frac{B^2-A_{0}C}{N_{max}^2C}-\left(\frac{z_{b}}{z_{m}}\right)^{\nu}\right)=c_{1}\left(c_{2}-\left(\frac{z_{b}}{z_{m}}\right)^{\nu}\right)~,
\end{eqnarray}
where the coefficients $c_{1}$ and $c_{2}$ are
\begin{eqnarray}
&&c_{1}=k^2\frac{N_{max}^2}{f^2-\sigma^2}~,\label{eq45}\\
&&c_{2}=\frac{\sigma^2(f^2+\tilde{f}_{s}^2-\sigma^2)}{N_{max}^2(f^2-\sigma^2)}~.\label{eq46}
\end{eqnarray}
As $s_{z_{b}}^2$ is a constant, wave directions can be easily separated, and the proper transmitted wave can be selected. Given boundary conditions at $z=z_{b}$, (\ref{eq13}) can be integrated backward to the interface. Once $\psi(0)$ and $\psi_{z}(0)$ are found numerically, we can calculate the amplitude ratio of the reflected wave to the incident wave, from which the reflection and transmission ratios can be deduced as well.

\begin{figure}
\centering

\begin{subfigure}{0.32\textwidth}
\includegraphics[width=\linewidth]{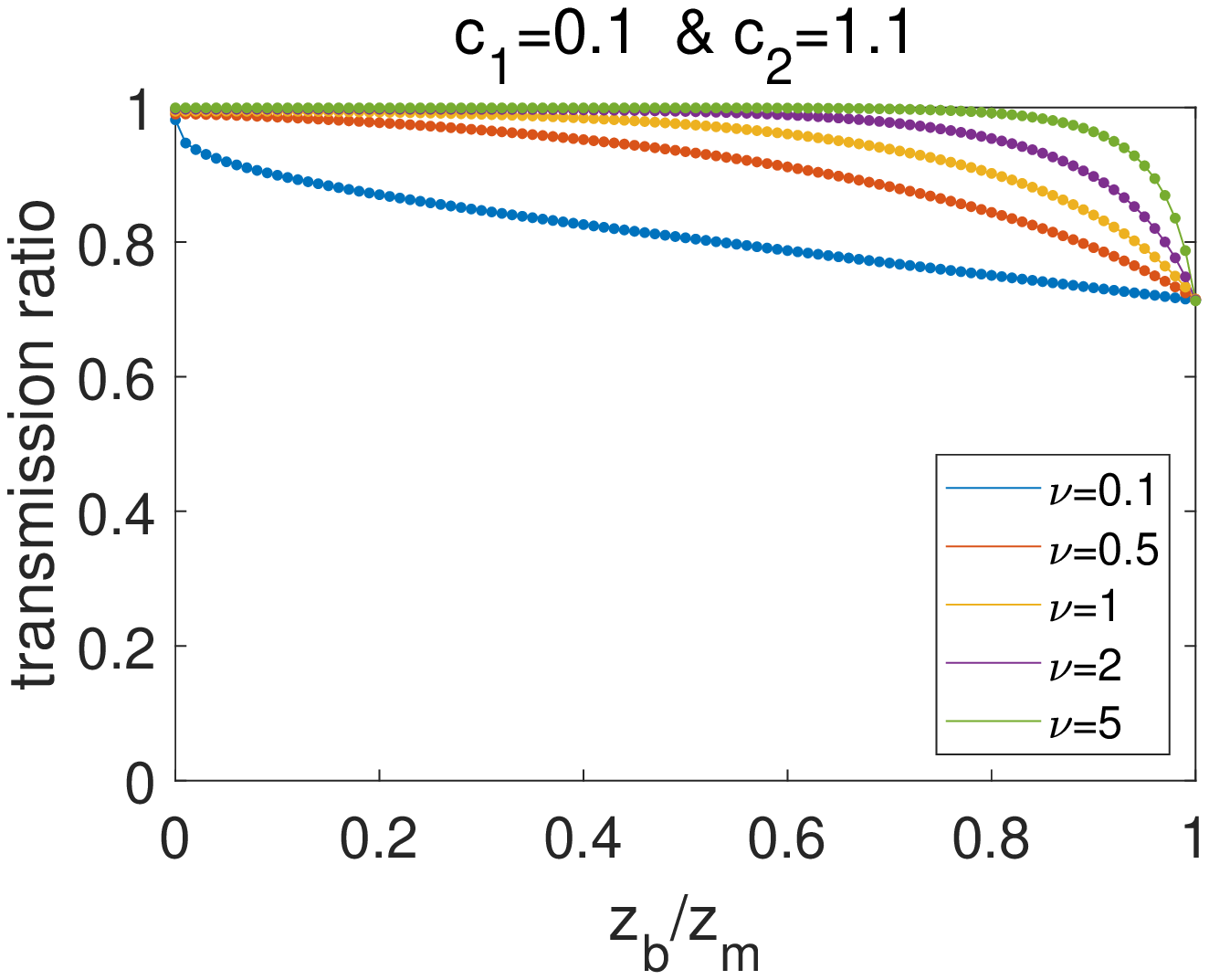}
\caption{}
\end{subfigure}
\begin{subfigure}{0.32\textwidth}
\includegraphics[width=\linewidth]{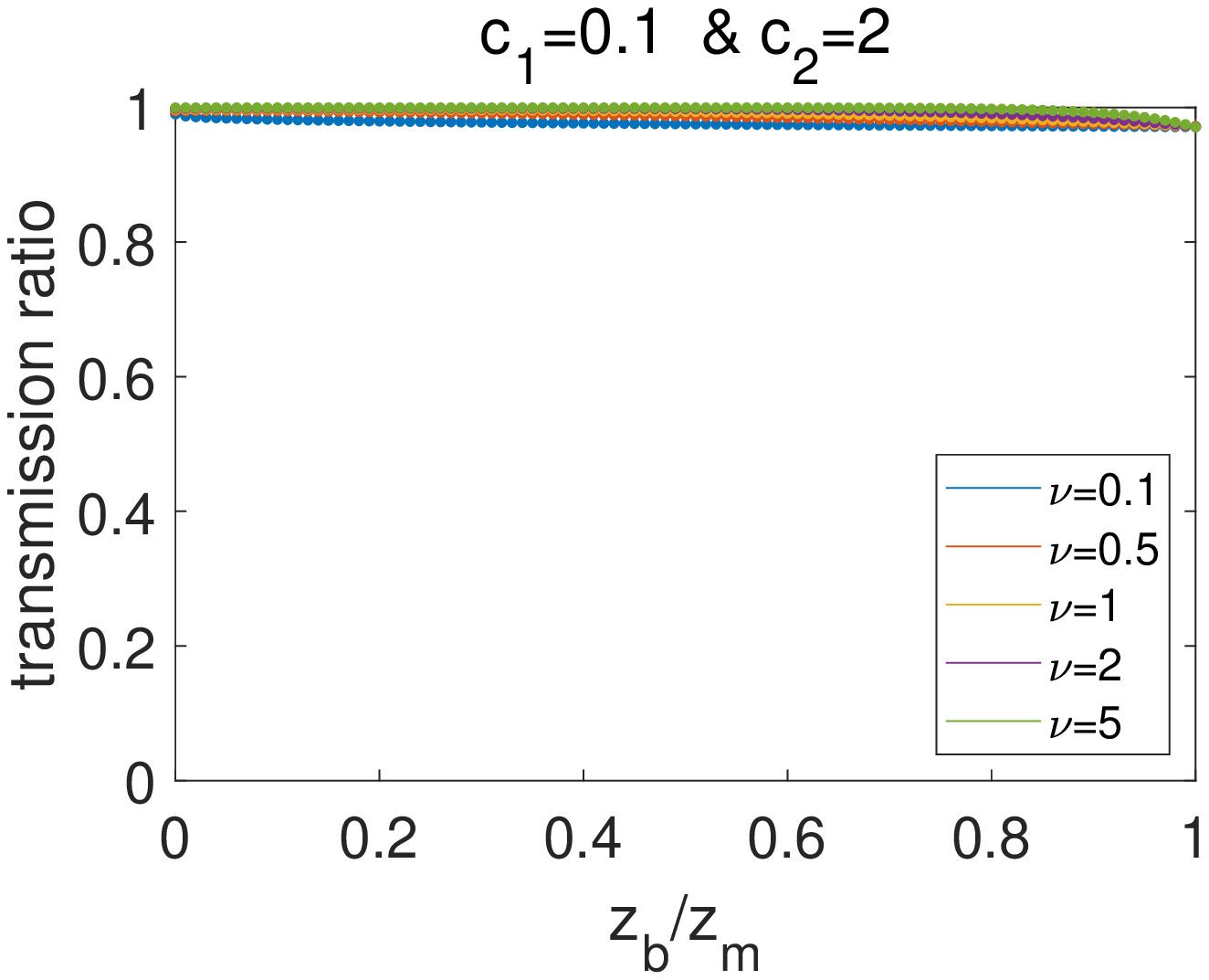}
\caption{}
\end{subfigure}
\begin{subfigure}{0.32\textwidth}
\includegraphics[width=\linewidth]{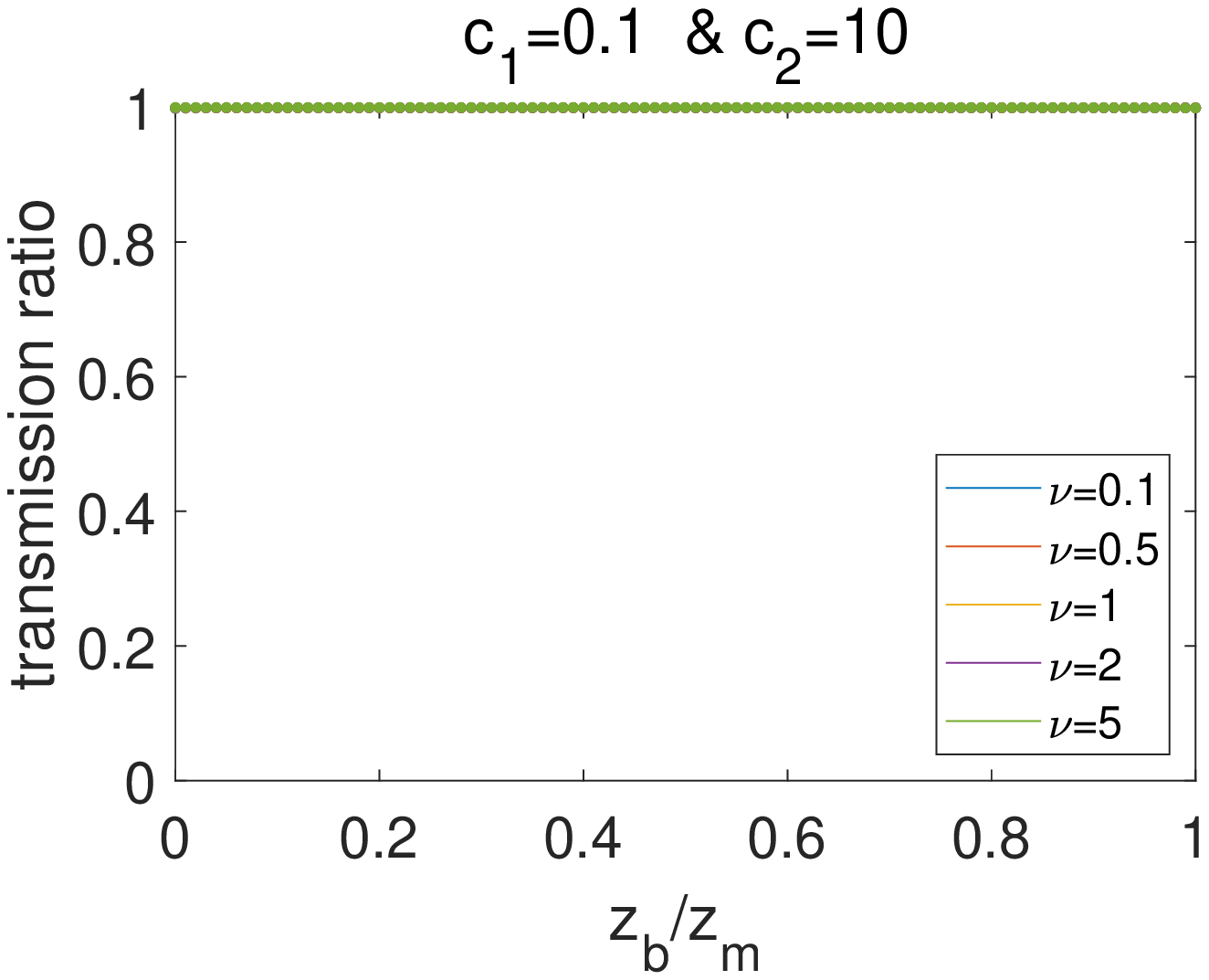}
\caption{}
\end{subfigure}

\medskip

\begin{subfigure}{0.32\textwidth}
\includegraphics[width=\linewidth]{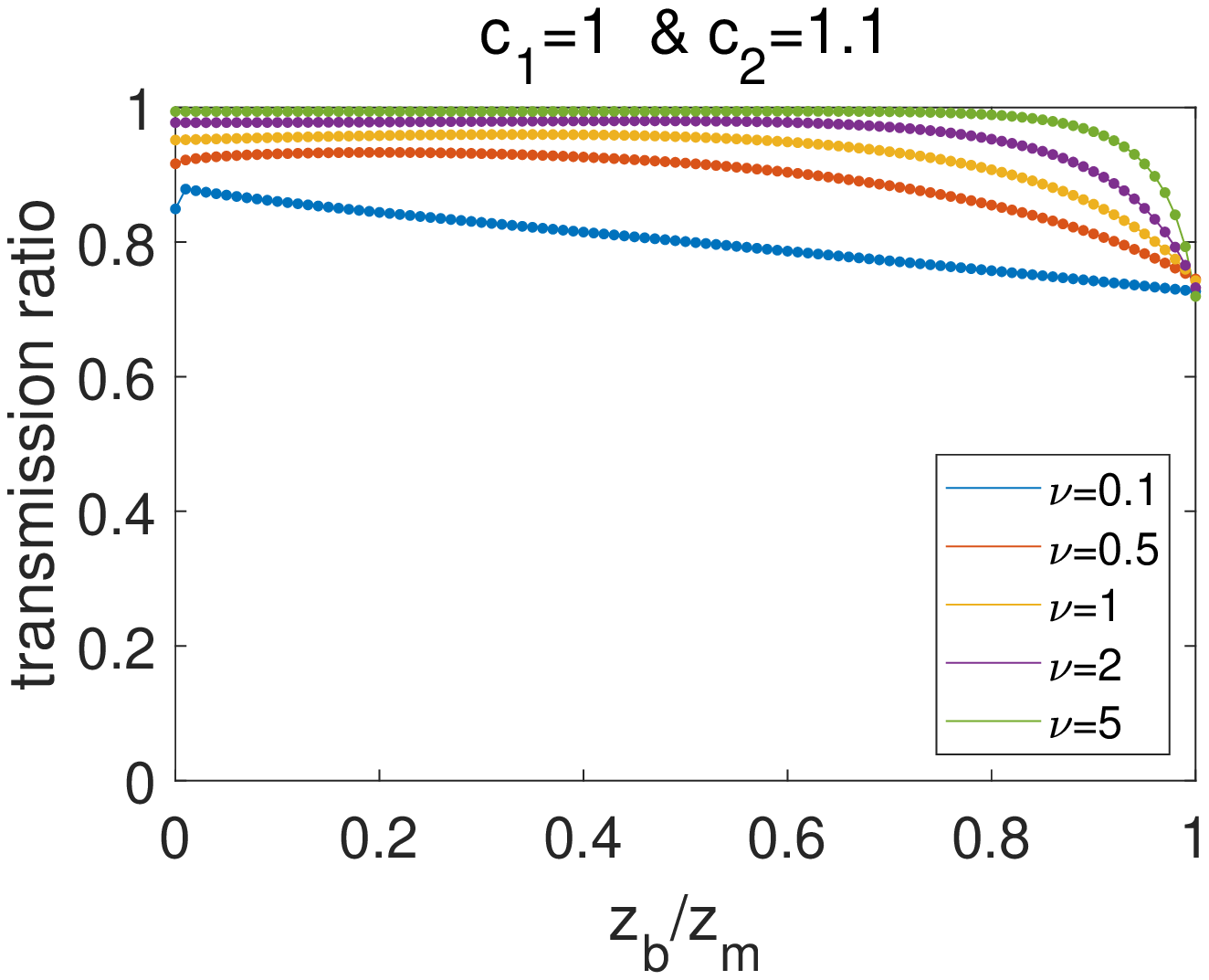}
\caption{}
\end{subfigure}
\begin{subfigure}{0.32\textwidth}
\includegraphics[width=\linewidth]{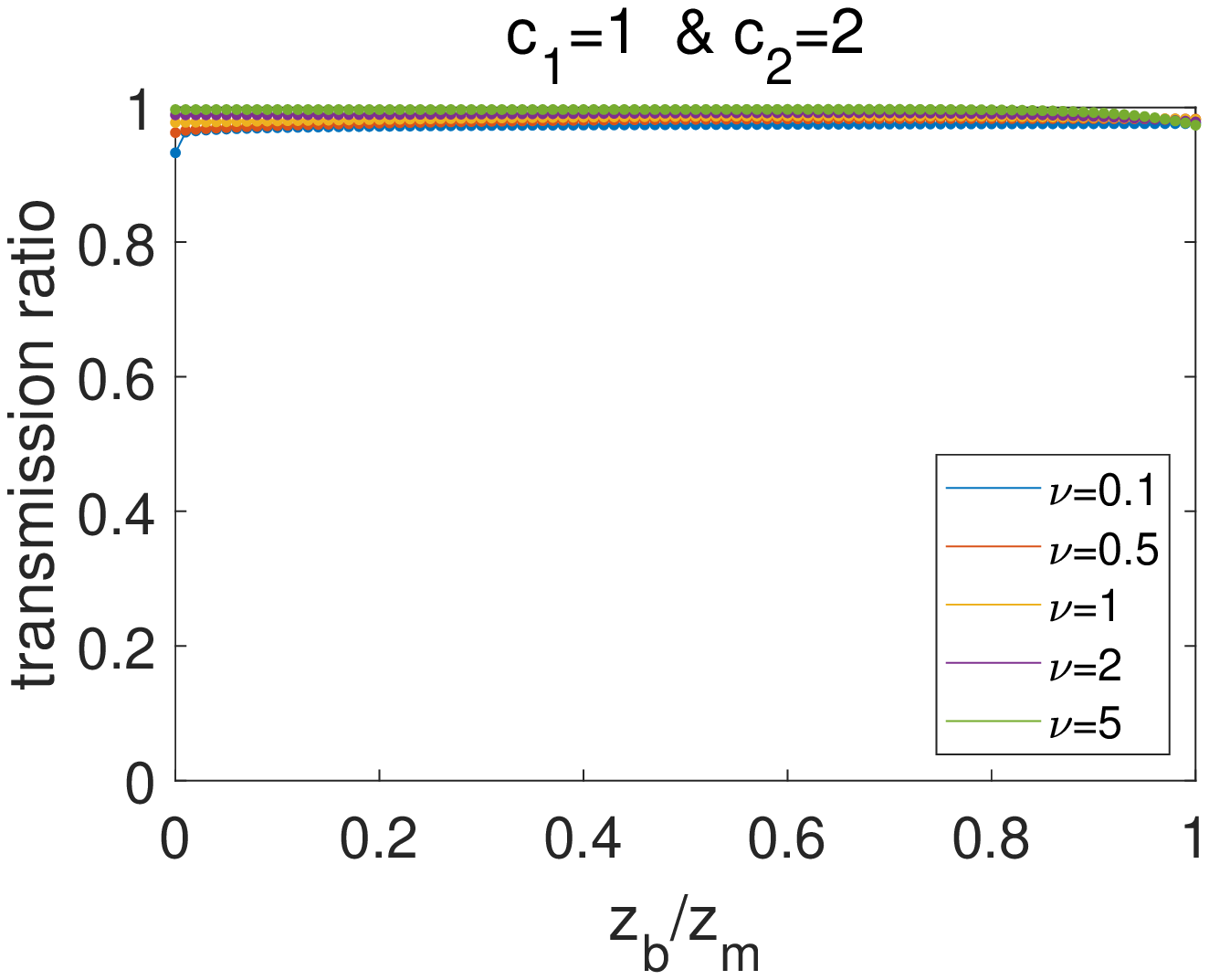}
\caption{}
\end{subfigure}
\begin{subfigure}{0.32\textwidth}
\includegraphics[width=\linewidth]{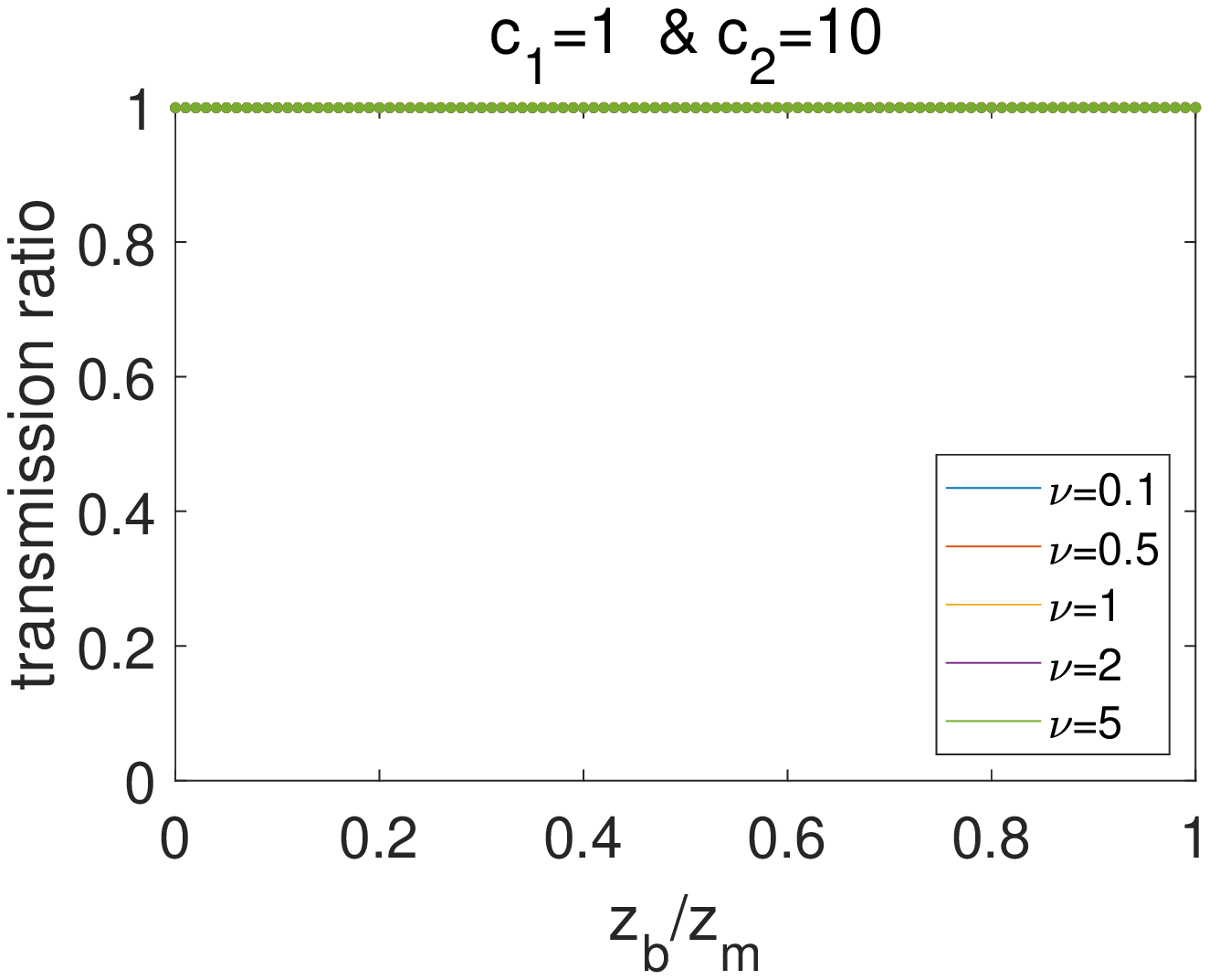}
\caption{}
\end{subfigure}

\medskip

\begin{subfigure}{0.32\textwidth}
\includegraphics[width=\linewidth]{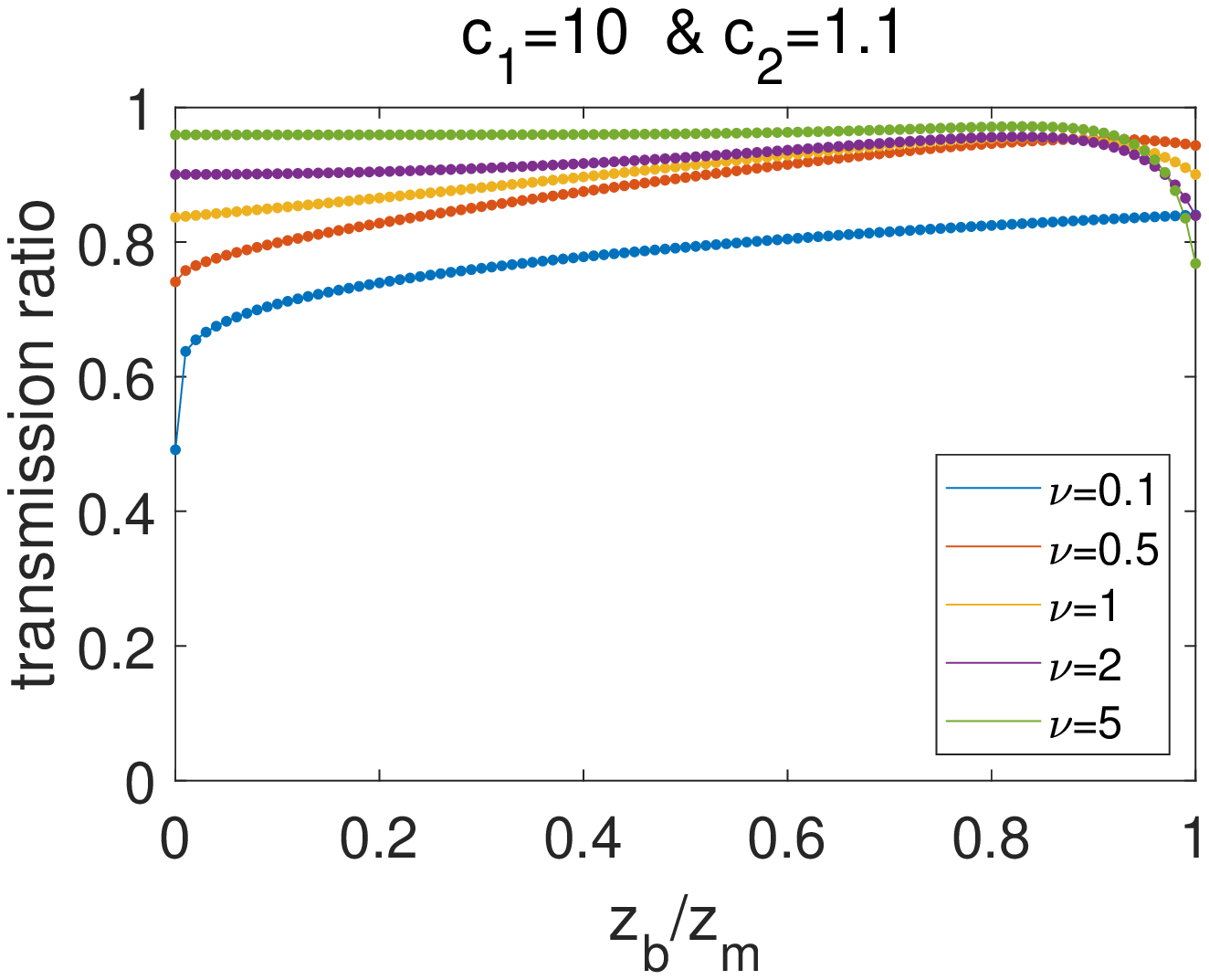}
\caption{}
\end{subfigure}
\begin{subfigure}{0.32\textwidth}
\includegraphics[width=\linewidth]{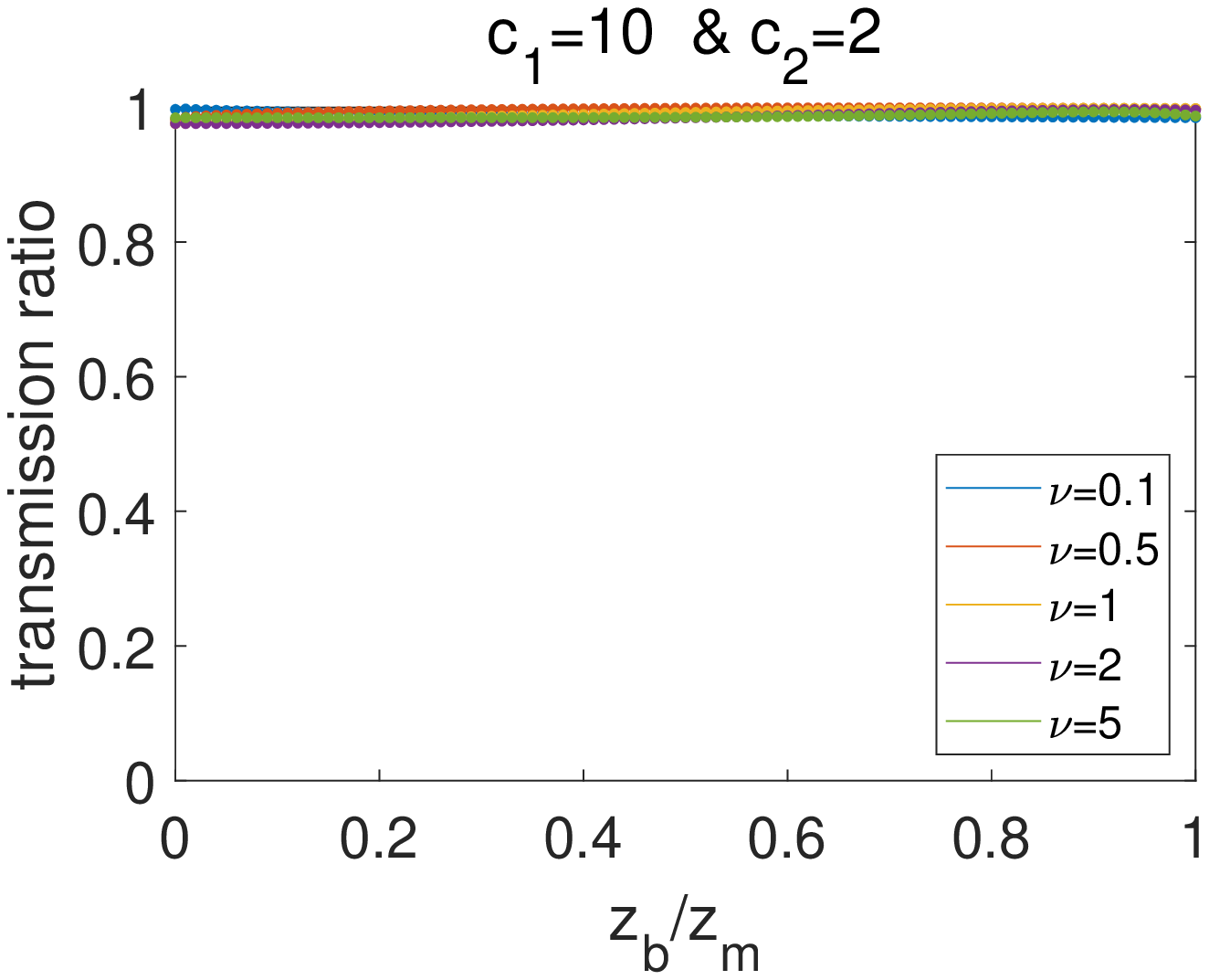}
\caption{}
\end{subfigure}
\begin{subfigure}{0.32\textwidth}
\includegraphics[width=\linewidth]{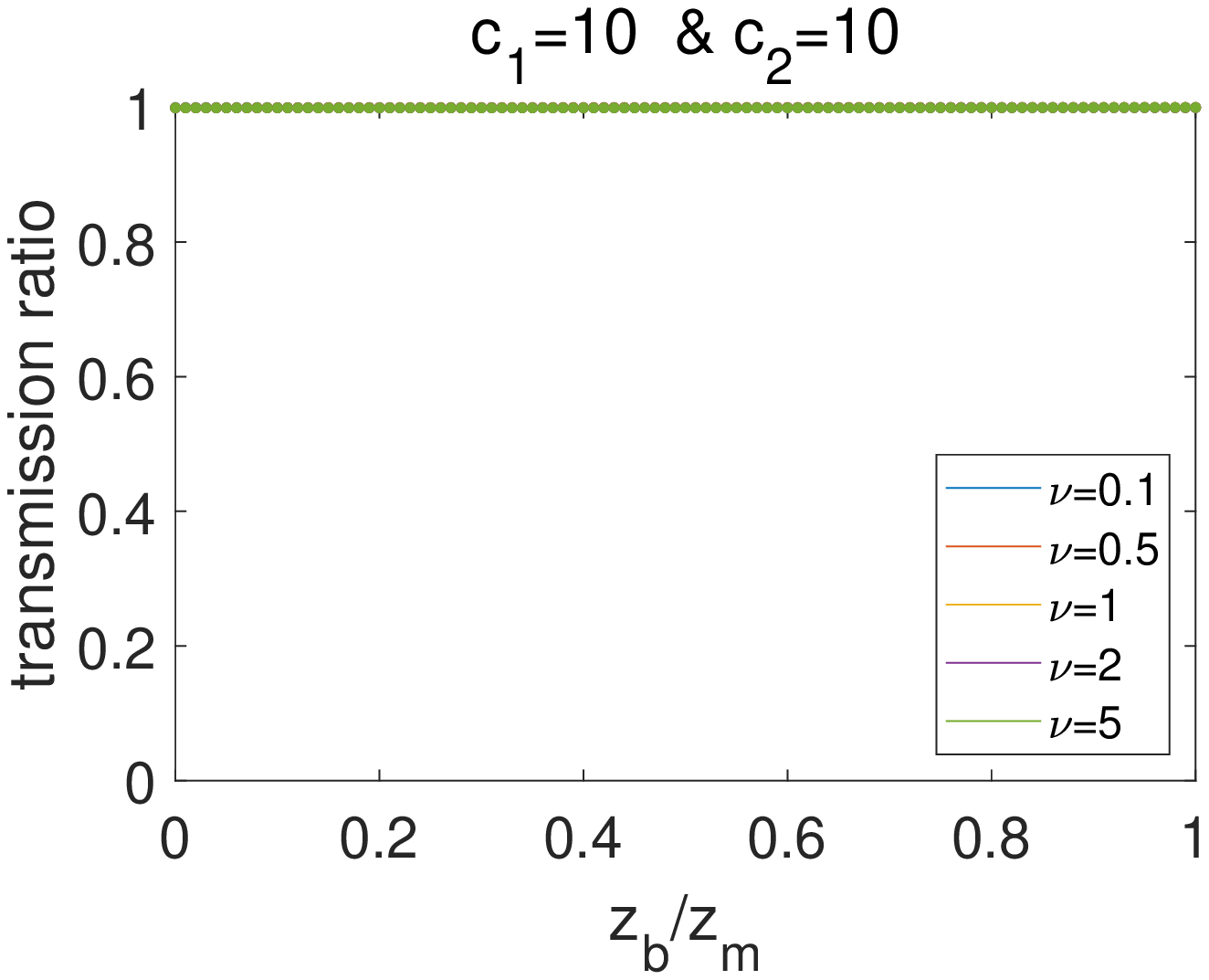}
\caption{}
\end{subfigure}

\medskip

\caption{Transmission ratio as a function of the relative position of the chosen boundary for sub-inertial waves. Different combinations of $c_{1}=0.1,1,10$ and $c_{2}=1.1,2,10$ are shown in panels. In each panel, transmission ratios at different $\nu=0.1,0.5,1,2,5$ are shown with colored lines.\label{fig:f5}}
\end{figure}

\begin{figure}
\centering

\begin{subfigure}{0.32\textwidth}
\includegraphics[width=\linewidth]{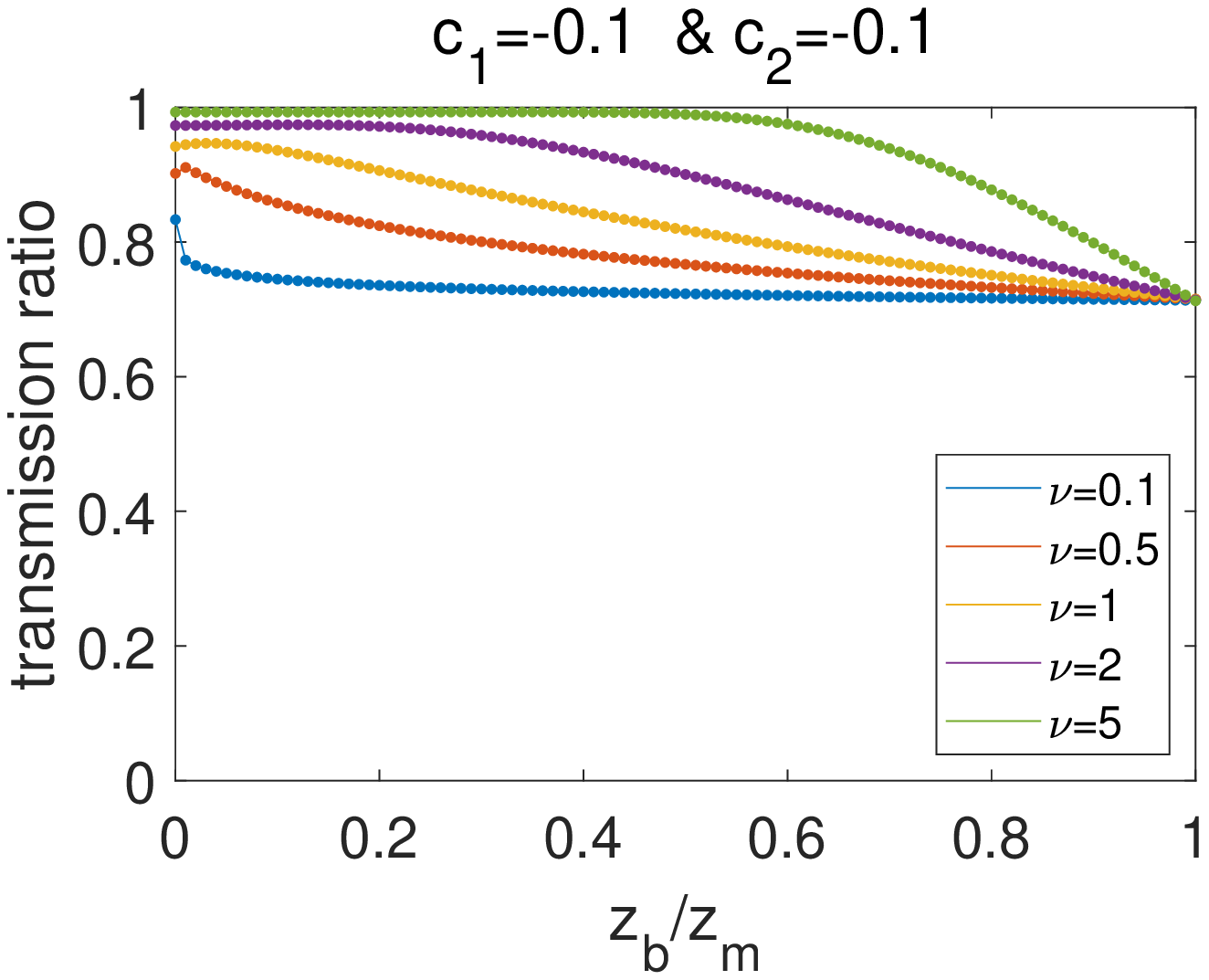}
\caption{}
\end{subfigure}
\begin{subfigure}{0.32\textwidth}
\includegraphics[width=\linewidth]{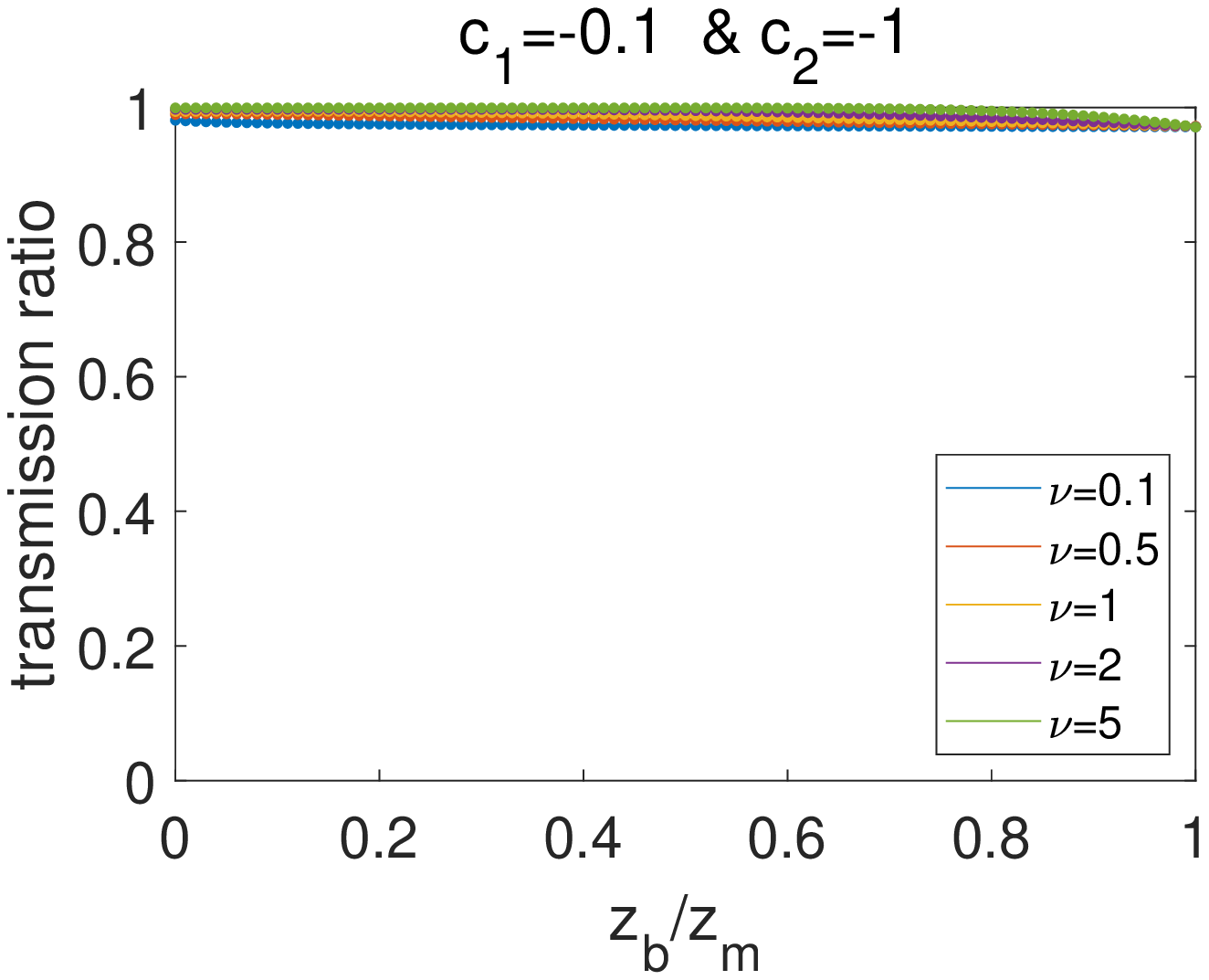}
\caption{}
\end{subfigure}
\begin{subfigure}{0.32\textwidth}
\includegraphics[width=\linewidth]{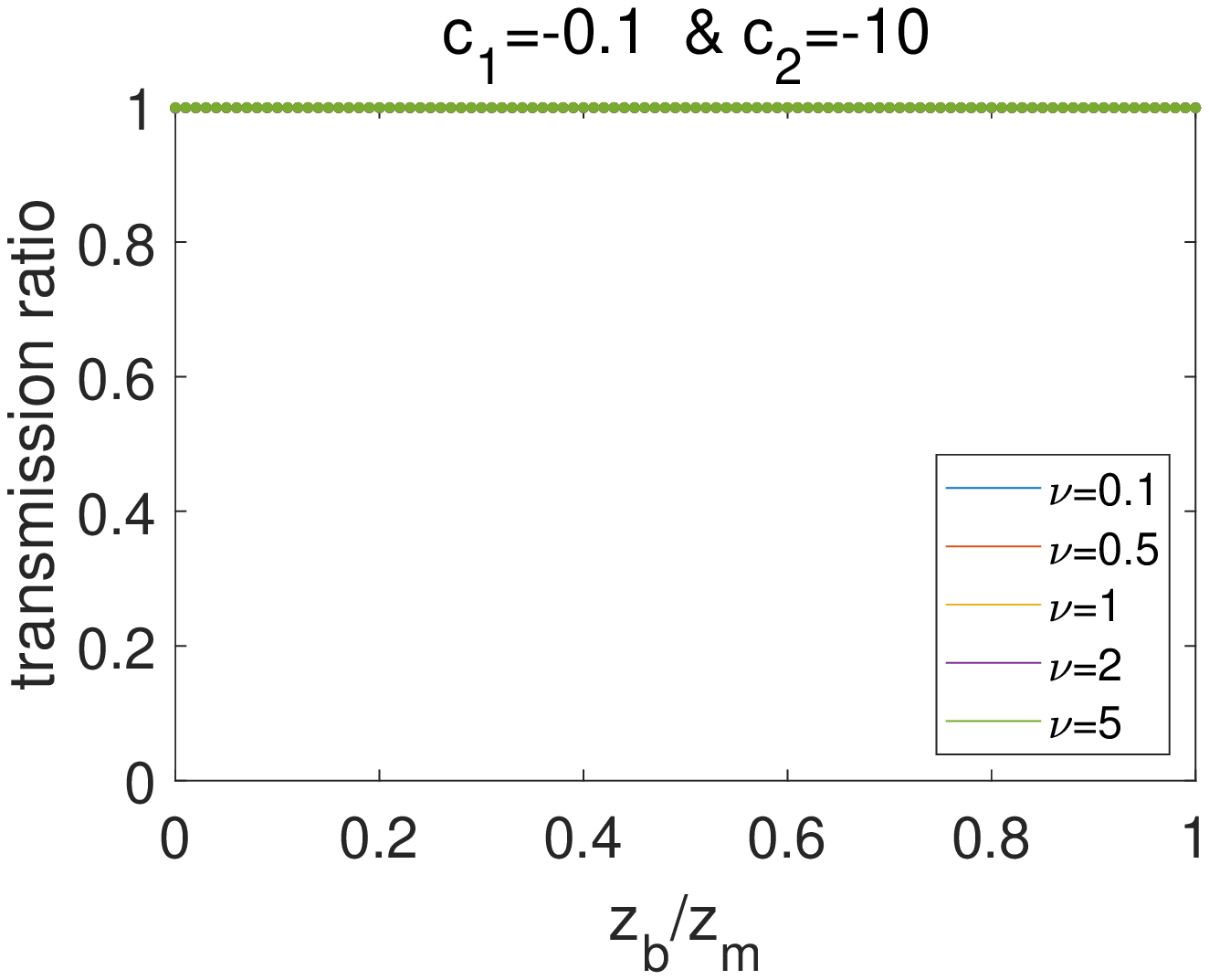}
\caption{}
\end{subfigure}

\medskip

\begin{subfigure}{0.32\textwidth}
\includegraphics[width=\linewidth]{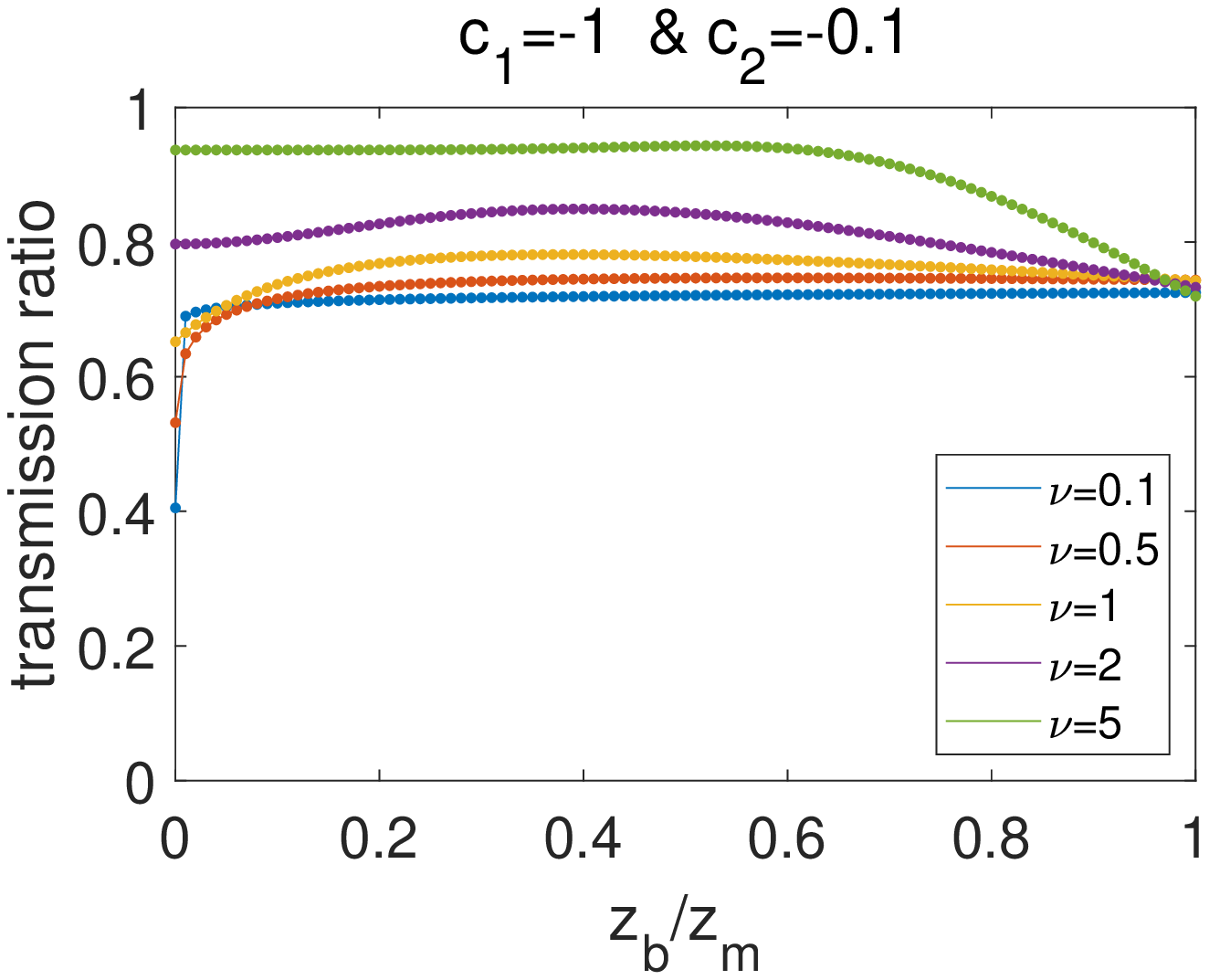}
\caption{}
\end{subfigure}
\begin{subfigure}{0.32\textwidth}
\includegraphics[width=\linewidth]{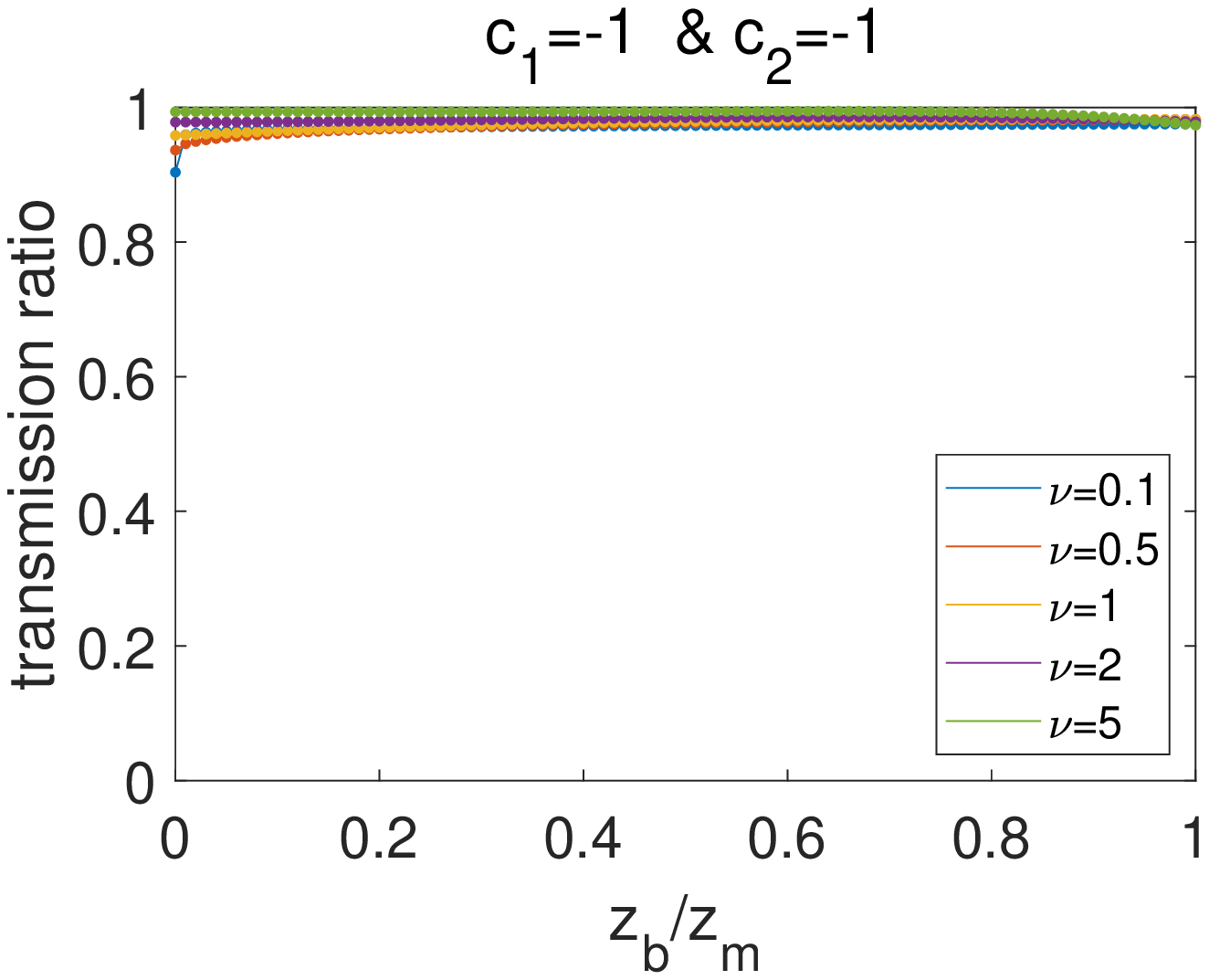}
\caption{}
\end{subfigure}
\begin{subfigure}{0.32\textwidth}
\includegraphics[width=\linewidth]{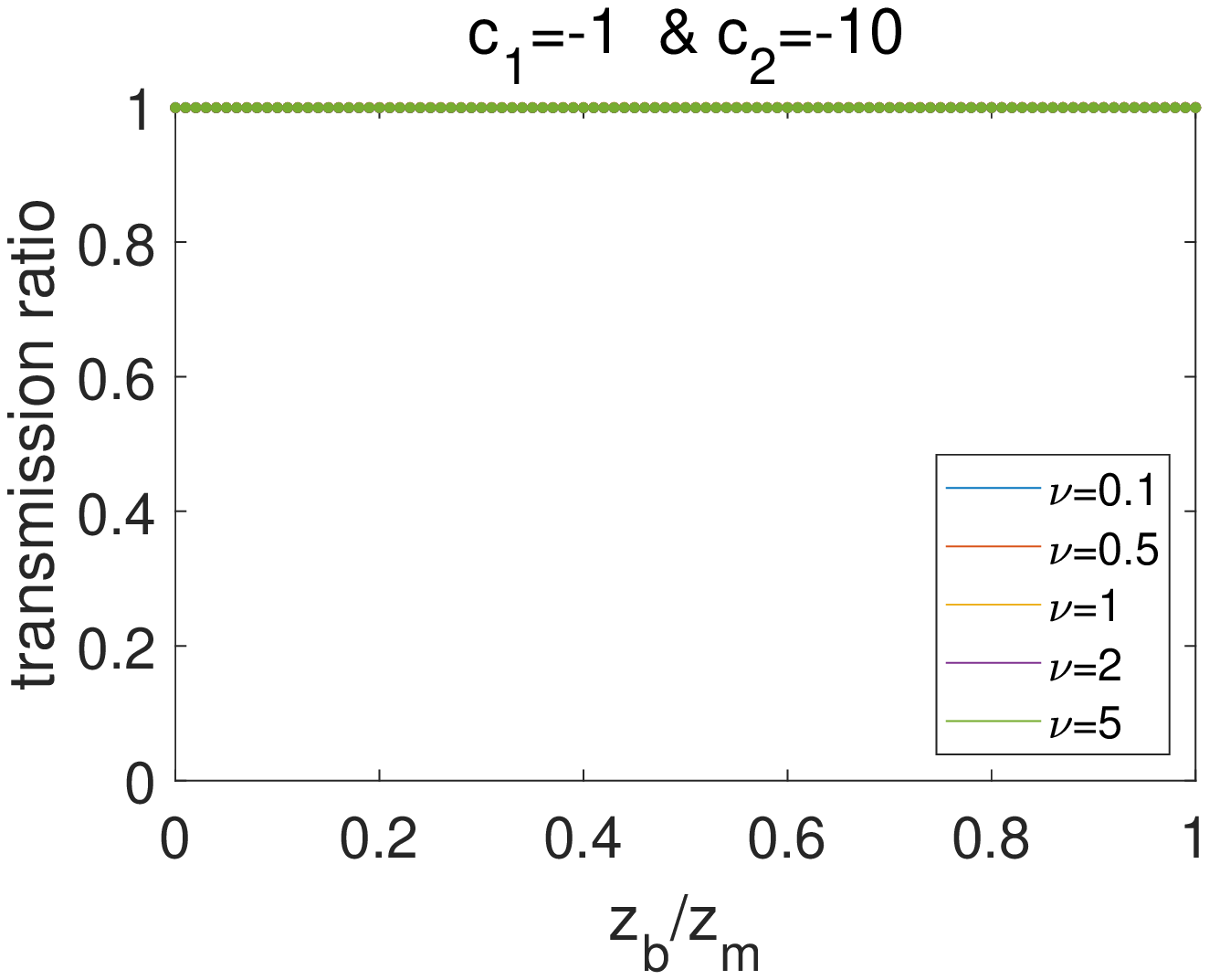}
\caption{}
\end{subfigure}

\medskip

\begin{subfigure}{0.32\textwidth}
\includegraphics[width=\linewidth]{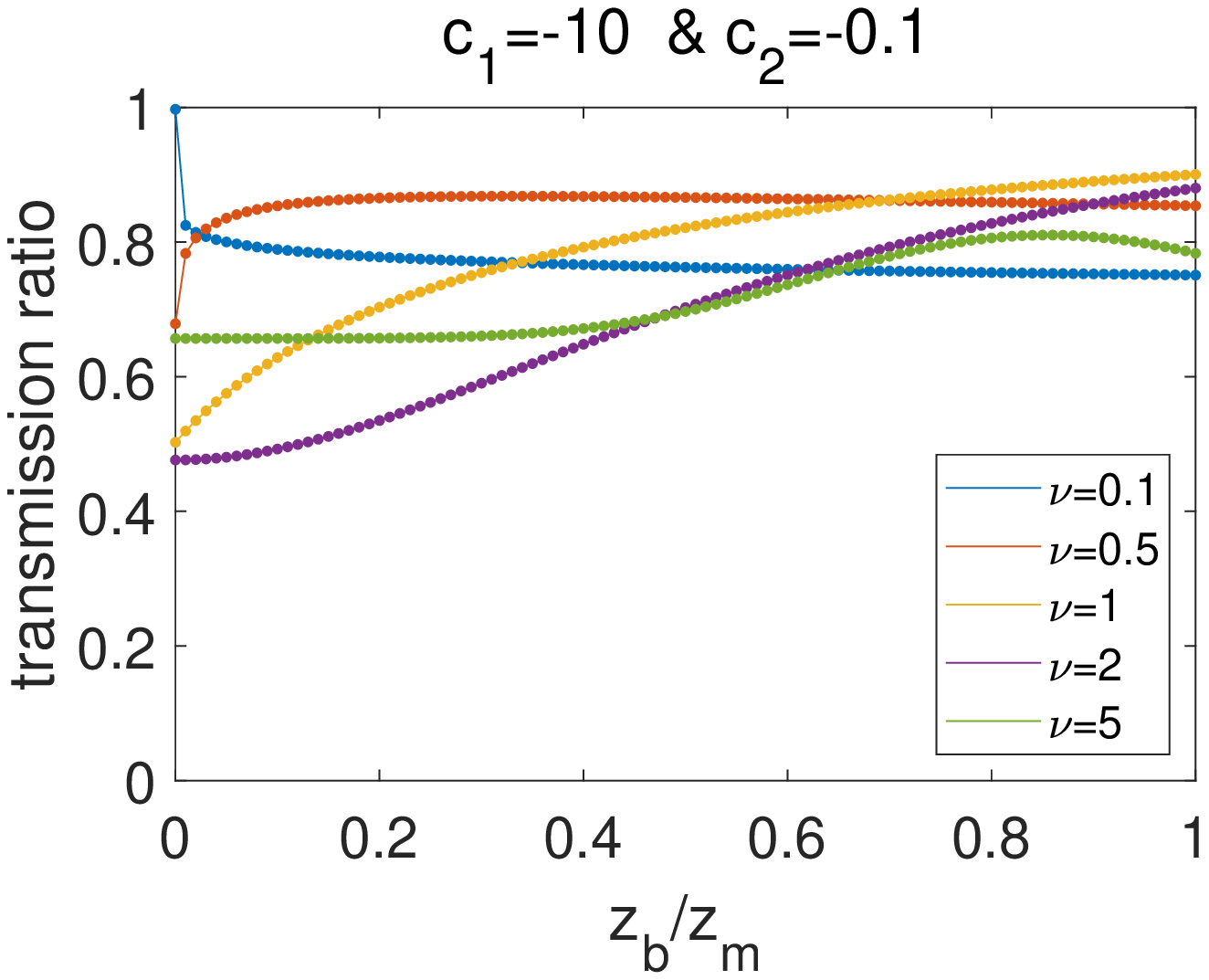}
\caption{}
\end{subfigure}
\begin{subfigure}{0.32\textwidth}
\includegraphics[width=\linewidth]{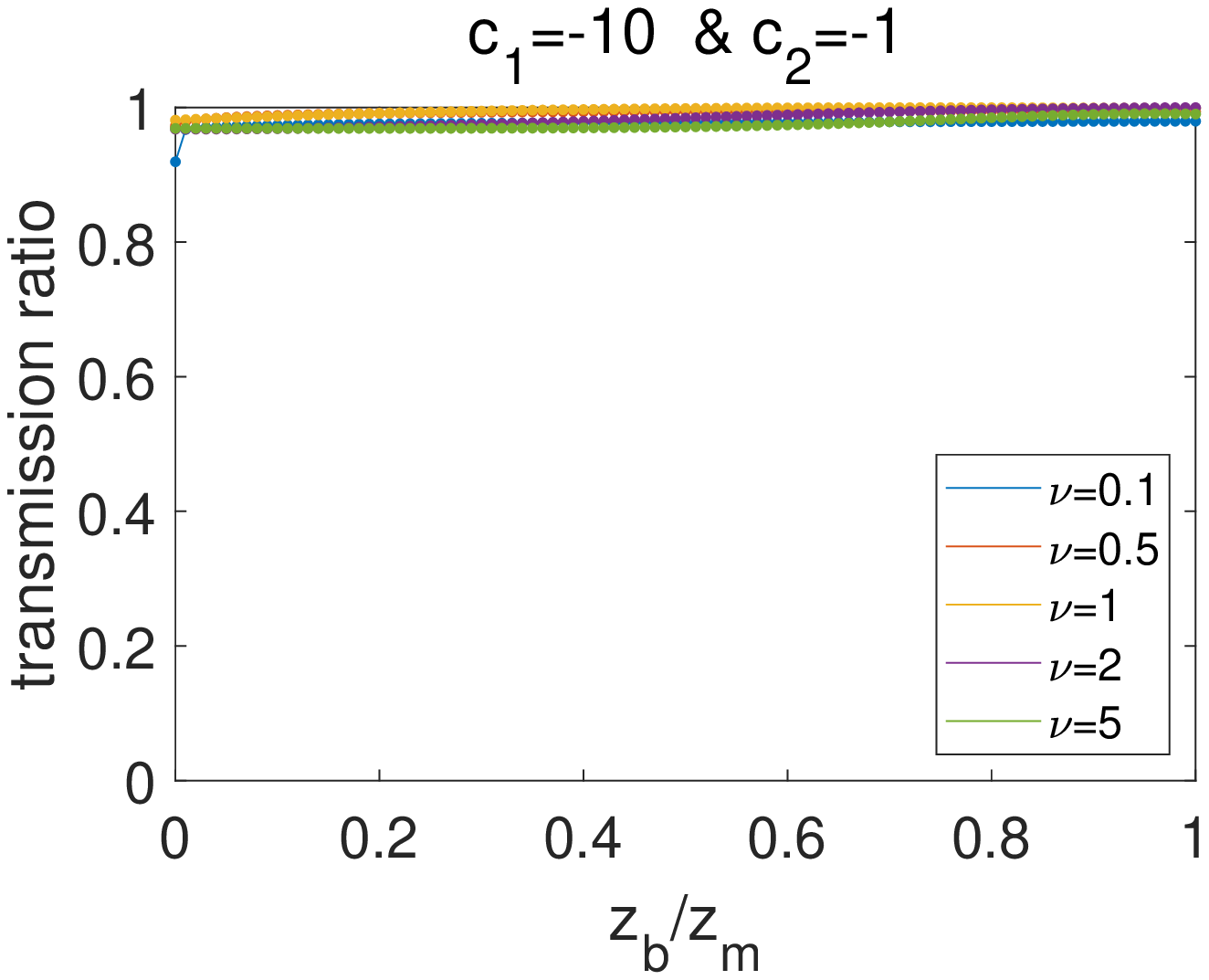}
\caption{}
\end{subfigure}
\begin{subfigure}{0.32\textwidth}
\includegraphics[width=\linewidth]{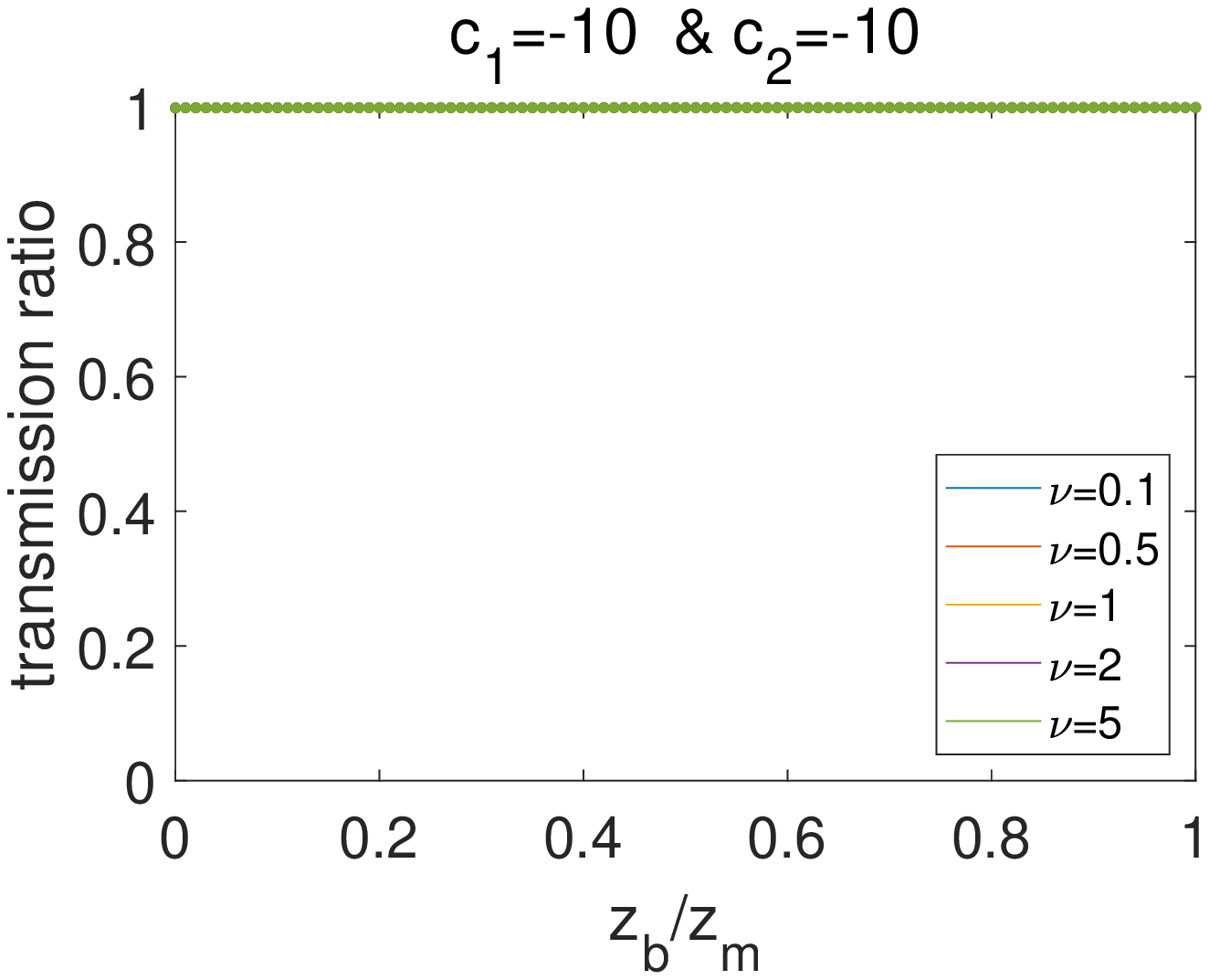}
\caption{}
\end{subfigure}

\medskip

\caption{Transmission ratio as a function of the relative position of the chosen boundary for super-inertial waves. Different combinations of $c_{1}=-0.1,-1,-10$ and $c_{2}=-0.1,-1,-10$ are shown in panels. In each panel, transmission ratios at different $\nu=0.1,0.5,1,2,5$ are shown with colored lines.\label{fig:f6}}
\end{figure}

Here we show transmission ratios for sub-inertial ($\sigma^2<f^2$) and super-inertial waves ($\sigma^2>f^2$), respectively. For sub-inertial waves, both $c_{1}$ and $c_{2}$ are positive values; for super-inertial waves, however, both $c_{1}$ and $c_{2}$ are negative values (see the section \ref{sec:s4} for further discussions). Fig.\ref{fig:f5} shows transmission ratios at different values of $\nu$, $c_{1}$, and $c_{2}$ for sub-inertial waves. The x-axis is the relative position of the boundary. As the wavenumber of the transmitted wave is approximated, we note that numerical results do not have to be completely consistent with our previous WKB and non-WKB analysis. Our intension here is to gain some insights by comparative studies. From the figure, we see the following trends. First, transmissions are always efficient when $c_{2}$ is large. A large value of $c_{2}$ can be achieved when $N_{max}^2/(2\Omega)^2$ or $f^2-\sigma^2$ is small, which corresponds to situations at fast rotation or at critical latitudes. Second, the far-field transmission ratio (the transmission ratio at the location $z_{b}/z_{m}=1$) increases with $z_{m}k$ for any $\nu$. The effect of $z_{m}k$ on far-filed transmission ratio can be investigated by keeping $c_{2}$ a constant and varying $c_{1}$ (that is, keeping $z_{m}$ a constant and varying $k$). From fig.~\ref{fig:f5}, we see that the far-field transmission ratio has an increasing trend with increasing $z_{m}k$ for any $\nu$.
The results for super-inertial waves are quite similar to those obtained in sub-inertial waves (see fig.~\ref{fig:f6}), and we shall not repeat the discussion here. Therefore, for a given wave (fixed $k$ and $\sigma$), the variation rate of stratification ($N_{max}^2/z_{m}$ for a fixed $N_{max}^2$) has important effect on the efficiency of transmission. The far-field transmission is more efficient when the variation of stratification is slower.

\subsection{Non-traditional effects}\label{sec:s4}
Traditional approximation, in which the horizontal Coriolis parameter $\tilde{f}=0$, is often used in geophysical fluids.
Under the traditional approximation, the frequency range for wave propagation is
\begin{eqnarray}
\min(f^2,N_{max}^2)<\sigma^2<\max(f^2,N_{min}^2)~.
\end{eqnarray}
The minimum value of $N_{min}=0$ is achieved in the convection zone, thus the wave frequency must be sub-inertial $N_{max}^2<\sigma^2<f^2$. Obviously, no inertial or gravito-inertia wave can survive in both convective and stable layers if the stable layer is strongly stratified with $N_{max}^2>f^2$. However, the case is different when non-traditional effects present. With non-traditional effects, Figs.\ref{fig:f3}(a-c) clearly show that waves can propagate in both convective and stable layers, although the stable layer is strongly stratified.

In the traditional approximation, when buoyancy frequency profile $N^2(z)=\gamma_{2}z^{\nu}$ is assumed, super-inertial waves in the stable layer are always kept away at distances $z>(\sigma^2/\gamma)^{1/\nu}$ from the interface, so as to ensure $N^2>\sigma^2 >f^2$. Wave rays propagating toward the interface are reflected back to the stable layer, without ever reaching the interface. This situation is similar to the antiwaveguide sound propagation adjacent to water surface \citep{brekhovskikh2003fundamentals}. On the other hand, sub-inertial waves are confined next to the interface at distances $z<(\sigma^2/\gamma)^{1/\nu}$, so as to ensure $N^2<\sigma^2 <f^2$.

When non-traditional effects present, the condition for wave propagation becomes (\ref{eq6}), which can be reorganized as
\begin{eqnarray}
N^2<\frac{\sigma^2(f^2+\tilde{f}_{s}^2-\sigma^2)}{f^2-\sigma^2}~,
\end{eqnarray}
for sub-inertial waves $\sigma^2<f^2$; and
\begin{eqnarray}
N^2>\frac{\sigma^2(\sigma^2-f^2-\tilde{f}_{s}^2)}{\sigma^2-f^2}~,
\end{eqnarray}
for super-inertial waves $\sigma^2>f^2$. Sub-inertial waves will be confined next to the interface at distances
\begin{eqnarray}
z<\left(\frac{\sigma^2(f^2+\tilde{f}_{s}^2-\sigma^2)}{\gamma_{2}(f^2-\sigma^2)}\right)^{1/\nu}~,
\end{eqnarray}
which is similar to that obtained in traditional approximation. However, the situation for super-inertial waves is different. For super-inertial waves with $\sigma^2>f^2+\tilde{f}_{s}^2$, waves are always kept away from the interface at distances
\begin{eqnarray}
z>\left(\frac{\sigma^2(\sigma^2-f^2-\tilde{f}_{s}^2)}{\gamma_{2}(\sigma^2-f^2)}\right)^{1/\nu}~,
\end{eqnarray}
while for super-inertial waves with $f^2<\sigma^2<f^2+\tilde{f}_{s}^2$, no similar restriction is obtained and waves can freely propagate in the stable layer. Thus, sub-inertial waves can only propagate a limited distance, but super-inertial waves are possible to propagate far away from the interface.

\begin{figure}
\centering

\begin{subfigure}{0.32\textwidth}
\includegraphics[width=\linewidth]{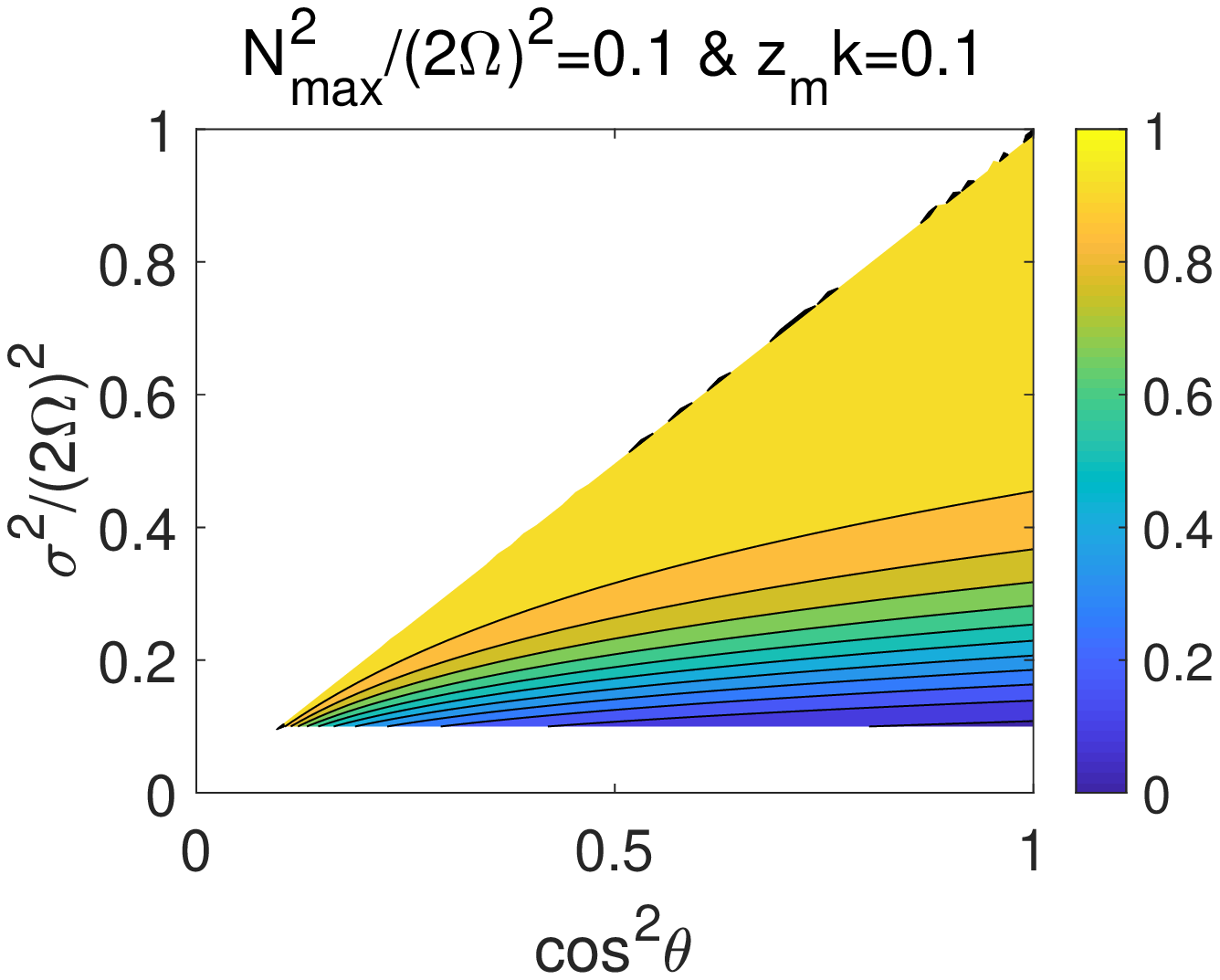}
\caption{}
\end{subfigure}
\begin{subfigure}{0.32\textwidth}
\includegraphics[width=\linewidth]{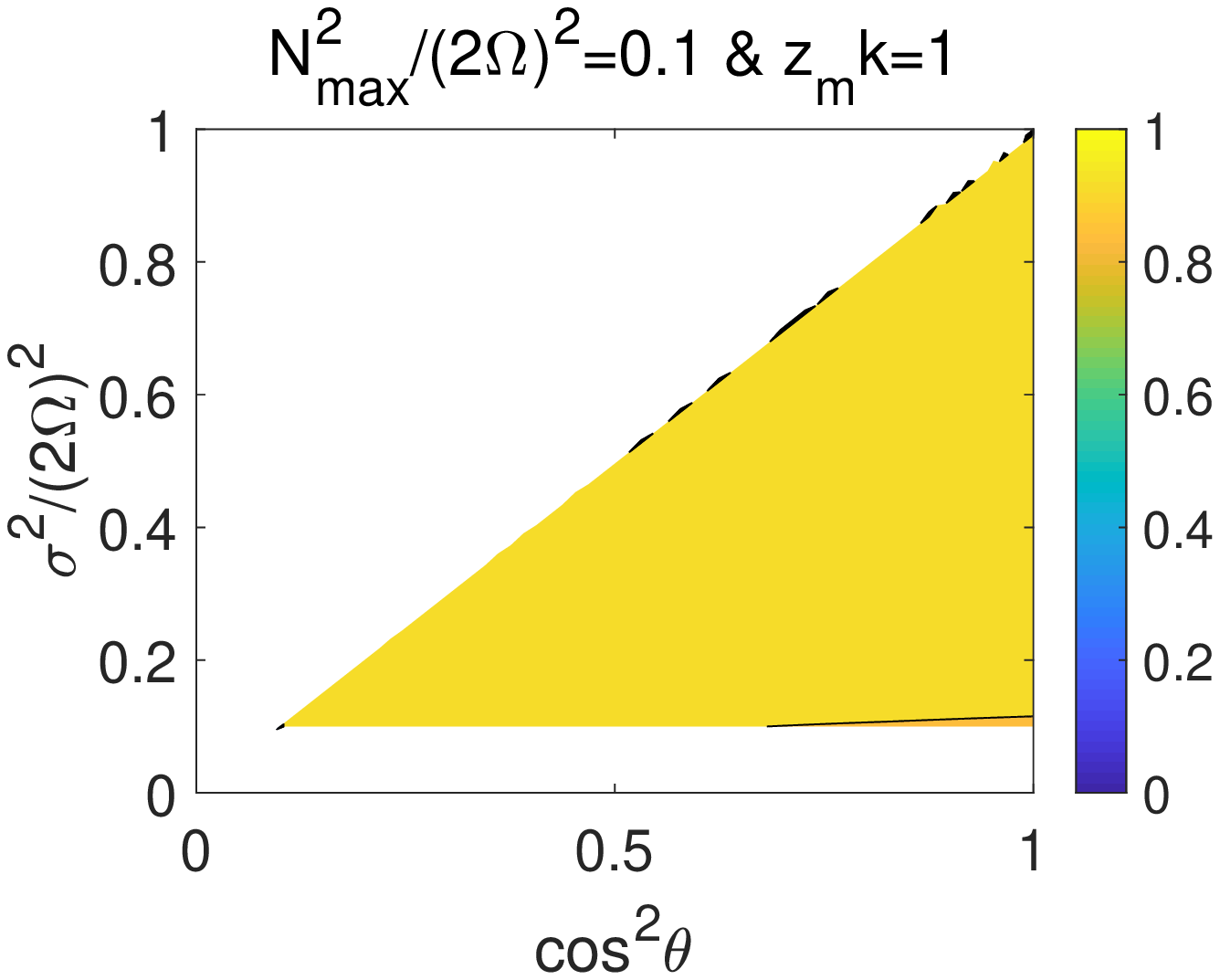}
\caption{}
\end{subfigure}
\begin{subfigure}{0.32\textwidth}
\includegraphics[width=\linewidth]{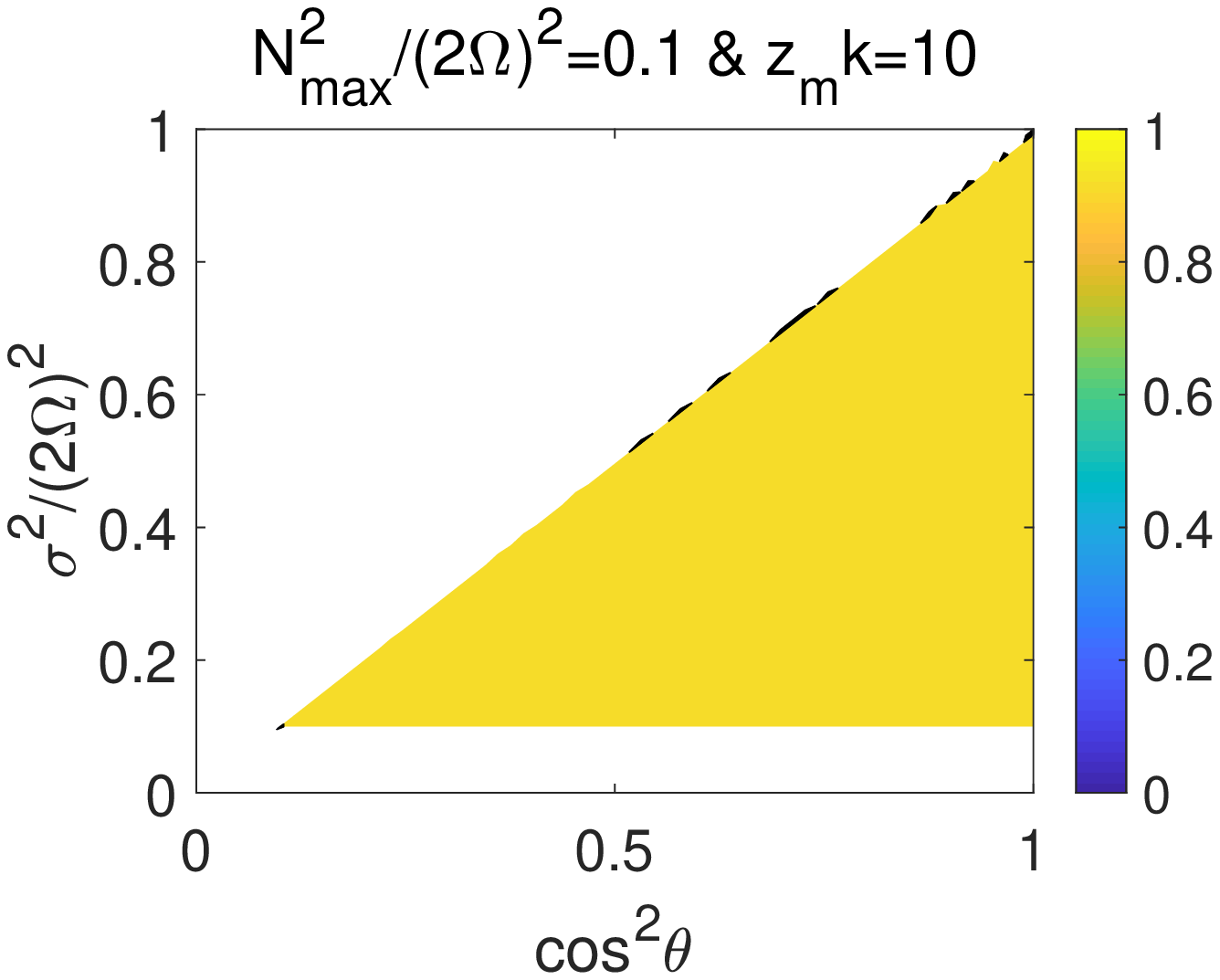}
\caption{}
\end{subfigure}

\caption{Wave transmissions under the traditional approximation. Transmission ratios computed at different $z_{m}k$ (=0.1,1,10 from the left to right) when the stratification is linearly varying $N^2=\gamma_{2}z$. In all the three cases, the parameter $N_{max}^2/(2\Omega)^2=0.1$. Waves can only survive in the colored regions. The regions are left white in the contour plots if waves cannot survive. \label{fig:f7}}
\end{figure}

To further illustrate the differences, we show wave transmission ratios at the interface in the traditional approximation for $N^2=\gamma_{2}z$ in Fig.~\ref{fig:f7}. In the traditional approximation, only frequencies in $(2\Omega)^2\geq f^2>\sigma^2>N^2$ can survive in both convective and stable layers. Therefore, on consideration of the transmission problem, wave frequencies must be sub-rotational with $\sigma^2<(2\Omega)^2$ (see also Fig.11 and their explanations in \citet{gerkema2008geophysical}). As $N^2$ must be smaller than $f^2$, here we only show the weakly stratified cases with $N_{max}^2/(2\Omega)^2=0.1$. It is obvious that frequencies are confined in the region $f^2>\sigma^2>N^2$. One consequence is that waves could hardly propagate at low latitudes. However, when non-traditional effects are included, from Fig.~\ref{fig:f4}, we see that wave propagation frequency ranges are much wider. This is especially true for waves propagating at low latitudes.

In the study of the tidal response of a massive rotating star, \citet{savonije1995non} found that subrotational responses with traditional approximation are confined in a poleward region, while without traditional approximation they can be found from equator to pole. The result is consistent with our analysis. This phenomena is also reviewed in \citet{gerkema2008geophysical}.
In the study of tidal oscillations of a rotating star with 1.5 solar masses, \citet{dintrans2000oscillations} used a stratification profile which is assumed to be proportional to radius. This setting is similar to the linearly varying stratification discussed in our paper. In their cases, the star has a strongly stratified middle stable layer surrounded by inner and outer convective layers. In a subrotational case ($\sigma/(2\Omega)=2/3$), they found that characteristic trajectories associated with $H_{2}$-mode ($f>\sigma>0$) propagate both in the convective and radiative layers, while those associated with $E_{1}$-mode ($N_{max}>\sigma>f$) remain trapped in the convective layer (see fig.6a in \citet{dintrans2000oscillations}). Their figure also shows that $H_{2}$-mode waves are mainly trapped at low latitudes with the colatitude $\theta \gtrsim \arccos (2/3)\approx 48^{\circ}$. Since $H_{2}$-mode can survive both in the convective and radiative layers, their result can be readily compared with our calculation in Figs.~\ref{fig:f4}(a-c). As shown in the figure, when $z_{m}k$ is large (see Fig.~\ref{fig:f4}(c)), wave transmissions can be efficiently transmitted at low latitudes $\theta<\theta_{c}$ (the diagonal line from the lower left to the upper right in the figure is $\theta=\theta_{c}$). Our calculation shows good agreement with their result. If traditional approximation is made, however, $H_{2}$-mode can never survive both in convective and radiative layers when $N_{max}^2>f^2$.

\section{Conclusion and Discussion}
We have discussed the wave propagation in a partially stratified rotating fluid with two configurations. In configuration 1, the wave propagates from the convective layer to the stable layer. In configuration 2, the wave propagates from the stable layer to the convective layer. First, we have discussed the width of the frequency range on the existence of wave solution. We have the following major findings:
\begin{itemize}
\setlength{\itemsep}{0pt}
\setlength{\parsep}{0pt}
\setlength{\parskip}{0pt}
\item[(1)] When the stable layer is strongly stratified (slow rotation), waves prefer to survive at low latitudes rather than high latitudes.
\item[(2)] When the stable layer is weakly stratified (rapid rotation), waves can survive at any latitude if the meridional wavenumber is large.
\end{itemize}

Second, we have discussed the relationship between the vertical phase and group velocities of a wave in the rotating fluid. We have the following major findings:
\begin{itemize}
\setlength{\itemsep}{0pt}
\setlength{\parsep}{0pt}
\setlength{\parskip}{0pt}
\item[(3)] On a general {\it f}-plane, the vertical group velocity and the modified vertical phase velocity have same directions for a sub-inertial wave, and opposite directions for a super-inertial wave.
\item[(4)] On a non-tilted plane, the vertical group and phase velocities of a gravito-inertia wave are always in opposite directions in strongly stratified fluid, and same directions in weakly stratified fluid.
\end{itemize}

Third, we have investigated the efficiency of wave transmission in these configurations. We have not observed significant difference on the efficiency of wave transmission between these two configurations. We have considered different stratification structures in the stable layer: the uniform stratification, and the continuously varying stratification. For the continuously varying stratification when the square of buoyancy frequency $N^2$ varies as an algebraic function $N^2 \propto z^{\nu}$ ($\nu \geq 1$), we have discussed the transmission ratio  both analytically (for $\nu\geq 1$) and numerically (for any $\nu>0$). We have the following major findings:
\begin{itemize}
\setlength{\itemsep}{0pt}
\setlength{\parsep}{0pt}
\setlength{\parskip}{0pt}
\item[(5)] For the uniform stratification, the transmission at the interface is efficient when the stable layer is weakly stratified (rapid rotation), or when the wave is at the critical latitude.
\item[(6)] For the continuously varying stratification ($N^2 \propto z^{\nu}$ with $\nu>0$), the far-field transmission is efficient when the stable layer is weakly stratified (rapid rotation), or when the wave is at the critical latitude, or when the thickness of the stratification layer is far greater than the horizontal wavelength (it means that $N^2$ varies slowly from 0 to $N_{max}^2$ at a distance much greater than the horizonal wavelength).
\end{itemize}

\begin{figure}
\centering
\includegraphics[width=\linewidth]{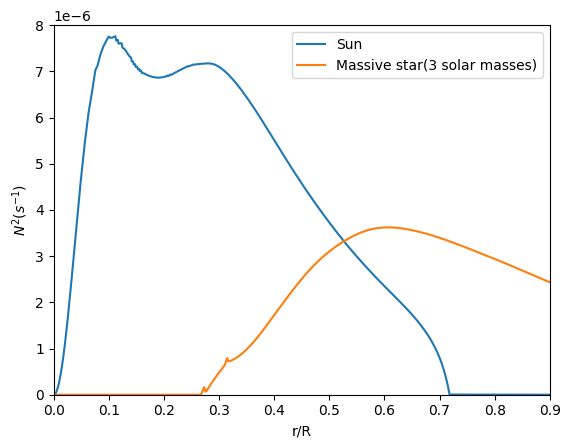}
\caption{Brunt-V\"ais\"al\"a frequency as a function of radius. The blue curve shows the solar case, and the orange curve shows the case of a main-sequency massive star with three times solar masses. The structures are computed with the MESA code \citep{paxton2010modules}. \label{fig:f8}}
\end{figure}

For each stratification structure, it is also interesting to note that the wave transmission ratio at the interface is identical for both configurations, no matter what direction the wave propagates. It only depends on the characteristics of the wave (wave frequency and wavenumber) and the fluid (degree of stratification).
Our findings have useful applications in the prediction of wave transmission in rotating stars or planets. Our calculation shows that wave transmission largely depends on the stratification structure. Wave transmissions are more efficient when the transition layer of radiative-convective varies slower. Fig.\ref{fig:f8} shows typical structures of buoyancy frequency squared in two main sequence stars. The blue curve denotes the Sun, and the orange curve is a massive star with three times solar masses. For this massive star, it can be seen that the transition of buoyancy frequency squared is almost linear; and for the Sun, the transition is more abrupt. Thus we expect that the wave transmission is more efficient in this massive star.

Linear instability analysis of rotating viscous flow has shown that teleconvection (fluid motion is vigorous in the stable layer) can be driven when the stable layer is weakly stratified in spherical geometry \citep{zhang02}. This phenomenon has also been verified in the analysis of rapidly rotating viscous flow on {\it f}-planes \citep{cai2020penetrative}. These results have borne some similarities with our conclusions. So probably our arguments still hold in other geometries. It would be interesting to verify them in different geometries. It has to be mentioned that our analysis is based on local {\it f}-planes in Cartesian coordinates, in which the curvature effect has not been included. New types of waves, such as Rossby and Poincar\'e waves, can propagate if the variation of the Coriolis parameter on latitude is taken into consideration. Also, {\it f}-plane only focuses on local waves with short wavelengths. In a global sphere, transmission of global waves may differ significantly from that of local waves \citep{pontin2020wave}. Future extensions may include considering wave transmissions in beta planes or a global sphere. Many stars and planets have multiple layer structures. \citet{sutherland2010internal} discussed transmission of gravity waves into neighbouring neutrally stable regions. \citet{andre2017layered} discussed transmission of internal waves in layered structures. \citet{gerkema05} considered multiple waveguides for waves in the ocean. Wave transmissions with varying stratification in a layered structure would be an interesting question for future research.

\acknowledgements
We thank three anonymous reviewers for their insightful comments and suggestions on this manuscript.
T.C. has been supported by NSFC (No.11503097), the Guangdong Basic and Applied Basic Research Foundation (No.2019A1515011625), the Science and Technology Program of Guangzhou (No.201707010006), the Science and Technology Development Fund, Macau SAR (Nos.0045/2018/AFJ, 0156/2019/A3), and the China Space Agency Project (No.D020303). C.Y. has been supported by the National Natural Science Foundation of China (grants 11373064, 11521303, 11733010, 11873103), Yun-nan National Science Foundation (grant 2014HB048), and Yunnan Province (2017HC018). X.W. has been supported by National Natural Science Foundation of China (grant no.11872246) and Beijing Natural Science Foundation (grant no. 1202015). This work is partially supported by Open Projects Funding of the State Key Laboratory of Lunar and Planetary Sciences.\\

\noindent{\bf Declaration of Interests\bf{.}} The authors report no conflict of interest.\\

\noindent{\bf  Author ORCID\bf{.}} T. Cai, https://orcid.org/0000-0003-3431-8570; Y. Cong, https://orcid.org/0000-0003-0454-7890; X. Wei, https://orcid.org/0000-0002-8033-2974 \\

\appendix

\section{}\label{appendixa}
In this appendix, we analyze the dependence of the width of frequency range $\sigma_{max}^{2}-\sigma_{min}^2$ on the colatitude $\theta$, the plane wave propagating direction $\alpha$, and the degree of stratification $N_{max}^2/(2\Omega)^2$.
The width of the frequency range (\ref{eq9}) can be simplified as
\begin{eqnarray}
G(\theta,\mu_{1},\mu_{2})=G_{1}(\theta,\mu_{1},\mu_{2})+\sqrt{G_{1}^2 (\theta,\mu_{1},\mu_{2})+4G_{2}(\theta,\mu_{1},\mu_{2})}~,
\end{eqnarray}
where $G(\theta,\mu_{1},\mu_{2})=\sigma_{max}^2-\sigma_{min}^2$, $G_{1}(\theta,\mu_{1},\mu_{2})=\cos^2\theta +\mu_{1}\sin^2\theta-\mu_{2}$, $G_{2}(\theta,\mu_{1},\mu_{2})=\mu_{1}\mu_{2}\sin^2\theta$, $\mu_{1}=\sin^2 \alpha$, and $\mu_{2}=N^2_{max}/(2\Omega)^2$.

\begin{figure}
\centering
\begin{subfigure}{0.45\textwidth}
\includegraphics[width=\linewidth]{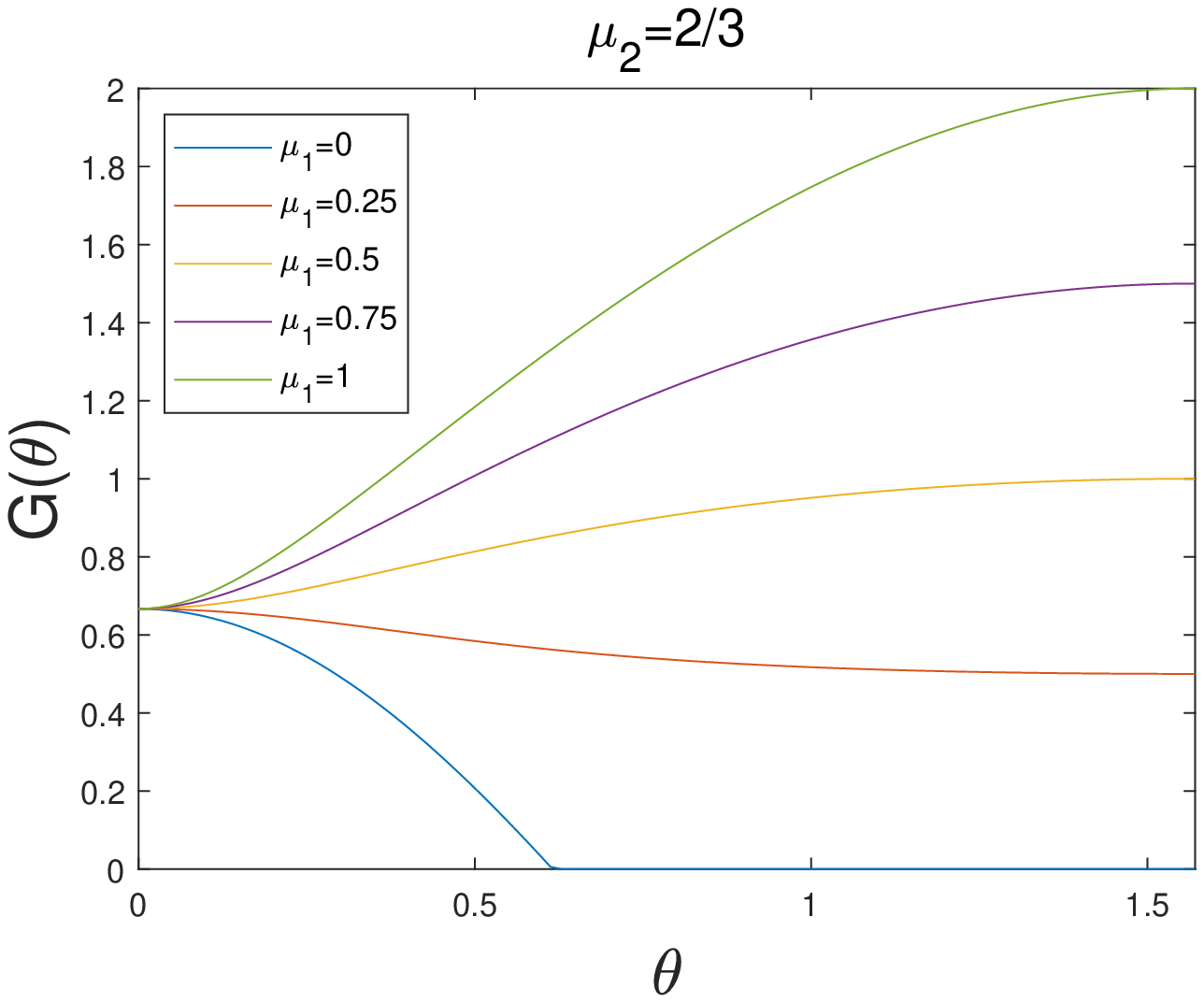}
\caption{}
\end{subfigure}
\begin{subfigure}{0.45\textwidth}
\includegraphics[width=\linewidth]{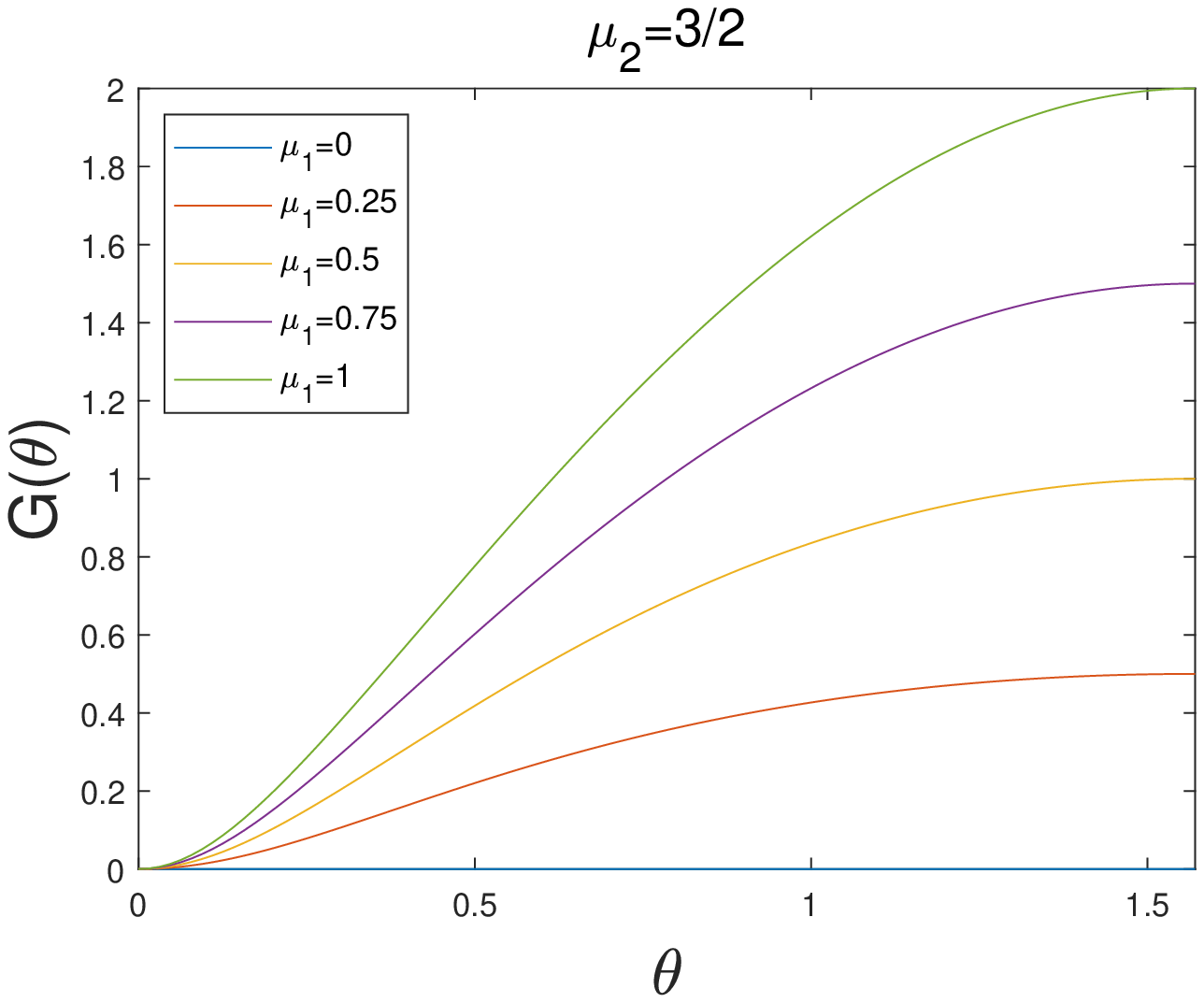}
\caption{}
\end{subfigure}
\caption{(a)$G(\theta)$ as a function of $\theta$ when $\mu_{2}=2/3$. In this case, $\mu_{2}<1$ and $\mu_{1c}=1/3$. (b)$G(\theta)$ as a function of $\theta$ when $\mu_{2}=3/2$. In this case, $\mu_{2}>1$. \label{fig:f9}}
\end{figure}

We first investigate the monotonicity of $\sigma_{max}^2-\sigma_{min}^2$ on $\theta$. Taking the first derivative with respect to $\theta$ and setting $G_{\theta}(\theta)=0$, we obtain
\begin{eqnarray}
G_{1\theta}^2 G_{2}=G_{1}G_{1\theta}G_{2\theta}+G_{2\theta}^2~.
\end{eqnarray}
From the above equation, we obtain that the critical point should satisfy
\begin{eqnarray}
\mu_{2}=1-\mu_{1}~,
\end{eqnarray}
and $G_{\theta}$ has the same sign as $\mu_{2}-(1-\mu_{1})$. If $\mu_{2}>1$, then the frequency width increases with $\theta$. If $\mu_{2}<1$, there exists a critical value $\mu_{1c}=1-\mu_{2}$, and the frequency width increases (decreases) with $\theta$ when $\mu_{1}>\mu_{1c}$ ($\mu_{1}<\mu_{1c}$). Figs.\ref{fig:f9}(a) and (b) presents two cases with $\mu_{2}<1$ and $\mu_{2}>1$, respectively. Fig.\ref{fig:f9}(b) clearly shows that $G(\theta)$ increases with $\theta$ when $\mu_{2}>1$. Fig.\ref{fig:f9}(a) shows that the trend of $G(\theta)$ depends on a critical value $\mu_{1c}$ when $\mu_{2}<1$.

Then we consider the monotonicity of $\sigma_{max}^2-\sigma_{min}^2$ on $\alpha$. Taking the first derivative of $G$ on $\mu_{1}$, we have
\begin{eqnarray}
G_{\mu_{1}}=(1+\frac{\cos^2\theta+\mu_{1}\sin^2\theta+\mu_{2}}{\sqrt{G_{1}^2 +4G_{2}}})\sin^2\theta\geq 0
\end{eqnarray}
Therefore, the frequency width increases with $\sin^2\alpha$.

Last, we consider the monotonicity of $\sigma_{max}^2-\sigma_{min}^2$ on the degree of stratification. The first derivative of G on $\mu_{2}$ gives
\begin{eqnarray}
G_{\mu_{2}}=\frac{-4\mu_{1}\sin^2\theta \cos^2\theta}{(\sqrt{G_{1}^2 +4G_{2}})(\mu_{1}\sin^2\theta+\mu_{2}-\cos^2\theta+\sqrt{G_{1}^2 +4G_{2}})}\leq 0
\end{eqnarray}
Therefore, the frequency width decreases with $N_{max}^2/(2\Omega)^2$.

\section{}\label{appendixb}
In this appendix, we deduce the averaged internal wave energy flux, which is needed in the discussion of wave transmission and reflection.
Applying the operation $\bm{\nabla} \times$ to (\ref{eq2}), we obtain the vorticity equation
\begin{eqnarray}
(\bm{\nabla} \bm{\times} \bm{u})_{t} +\bm{\nabla} \bm{\times} (\bm{f} \bm{\times} \bm{u})-\bm{\nabla} b \bm{\times} \hat{\bm{z}}=0~. \label{a1}
\end{eqnarray}
From this equation, we can obtain the vertical component of vorticity
\begin{eqnarray}
\bm{\hat{z}} \bm{\cdot} \bm{\nabla} \bm{\times} \bm{u}=-\frac{1}{i\sigma}(\bm{f} \bm{\cdot} \bm{\nabla}) w~. \label{a2}
\end{eqnarray}
Solving (\ref{eq3}) and (\ref{a2}), we get the horizontal velocities
\begin{eqnarray}
&&u=\frac{(i\sigma\cos\alpha-f\sin\alpha ) w_{z}-(ik\tilde{f}_{s}\sin\alpha)w }{k \sigma}~,\label{a5}\\
&&v=\frac{(i\sigma\sin\alpha+f\cos\alpha ) w_{z}+(ik\tilde{f}_{s}\cos\alpha)w}{k \sigma}~,\label{a6}
\end{eqnarray}
from which, we can also obtain the modified pressure $p$
\begin{eqnarray}
p&&=\frac{1}{ik\sin\alpha}(i\sigma v-fu)~\nonumber\\
&&=\frac{[(f^2-\sigma^2) \Imag (w_{z}/w)+k\tilde{f_{s}}f]+i[k\tilde{f}\sigma\cos\alpha-(f^2-\sigma^2)\Re(w_{z}/w)]}{k^2\sigma}w~. \label{a7}
\end{eqnarray}
(\ref{a7}) indicates that continuous pressure at the interface requires that the first derivative of vertical velocity to be continuous.
Now we consider the energy equation. Computing $\bm{u}\cdot(\ref{eq2})+b/N^2 \times (\ref{eq3})$, we can obtain the energy equation
\begin{eqnarray}
(\frac{|\bm{u}|^2}{2}+\frac{b^2}{2N^2})_{t}+\nabla\cdot(p\bm{u})=0~, \label{a9}
\end{eqnarray}
where $E_{kin}=|\bm{u}|^2/2$ is the kinetic energy, and $E_{pe}=b^2/(2N^2)$ is the potential energy, and $E_{wave}=E_{kin}+E_{pe}$ is the wave energy.
From (\ref{a5}-\ref{a6}) and $b=-N^2 w/(i\sigma)$, we can calculate the averaged wave energy
\begin{eqnarray}
\left<E_{wave}\right>&&=\left<E_{kin}\right>+\left<E_{pe}\right>\nonumber\\
&&=\frac{1}{4\sigma^2 k^2}[|f\Psi_{z}/\Psi+ik\tilde{f}_{s}|^2+\sigma^2(|\Psi_{z}/\Psi|^2+k^2)+N^2 k^2]|\psi|^2~.\label{a13}
\end{eqnarray}
and the vertical component of the averaged internal wave energy flux
\begin{eqnarray}
\left<F\right>&&=\left<\Real(p)\Real(w)\right>\nonumber\\
&&=\left<\frac{[(f^2-\sigma^2) \Imag(w_{z}/w)+k\tilde{f_{s}}f]\Real(w)\Real(w)}{k^2\sigma}\right>~\nonumber\\
&&=\frac{(f^2-\sigma^2) \Imag(\Psi_{z}/\Psi)+k\tilde{f_{s}}f}{2k^2\sigma}|\Psi|^2~\nonumber\\
&&=\frac{C \delta + C\Imag(\psi_{z}/\psi)+kB}{2k^2\sigma}|\psi|^2~\nonumber\\
&&=\frac{C\Imag(\psi_{z}/\psi)}{2k^2\sigma}|\psi|^2~.
\end{eqnarray}

\section{}\label{appendixc}
We have considered the continuously varying stratification in the stable layer under the WKB approximations. For two special cases with $\nu=1$ and $\nu=2$, exact solutions can be found without additional approximations. For these two special cases, we attempt to solve (\ref{eq13}) directly to verify the validity of the WKB results.

\subsection{Solution for linearly varying $N^2$}
With a linear relation $N^2=\gamma_{2} z$, Eq.~(\ref{eq13}) can be written as
\begin{eqnarray}
\psi_{zz}+k^2\frac{B^2-A_{0}C-\gamma_{2}{C}z}{C^2}\psi=0~.
\end{eqnarray}
Making a substitution $z'=Sgn(C)\ell^{2/3}(z-z_{t})$, we can transform it into a standard form:
\begin{eqnarray}
\psi_{z'z'}=z'\psi~,
\end{eqnarray}
where $\ell^2=k^2\gamma_{2}/|C|$ and $z_{t}=(B^2-A_{0}C)/(\gamma_{2}C)$. This equation is a Sturm-Liouville problem and its solution can be found by using Airy functions:
\begin{eqnarray}
\psi(z)=b_{1}Ai(z')+b_{2}Bi(z')~,
\end{eqnarray}
where $Ai$ and $Bi$ are the first and second kinds of Airy functions. It has been found that the Airy functions have the following asymptotic expansions for $z'\ll -1$ \citep{olver2010nist,zhang2006wave};
\begin{eqnarray}
&&Ai(z') \sim \pi^{-1/2} (-z')^{-1/4}\sin X(z')=(2i)^{-1}\pi^{-1/2} (-z')^{-1/4}\left(e^{iX(z')}-e^{-iX(z')}\right)~,\\
&&Bi(z') \sim \pi^{-1/2} (-z')^{-1/4}\cos X(z')=2^{-1}\pi^{-1/2} (-z')^{-1/4}\left(e^{iX(z')}+e^{-iX(z')}\right)~,
\end{eqnarray}
where $X(z')=2/3(-z')^{3/2}+\pi/4$.
At the interface, we have
\begin{eqnarray}
&&\psi(0)\simeq \pi_{0}(\frac{b_{2}-ib_{1}}{2}e^{iX_{0}}+\frac{b_{2}+ib_{1}}{2}e^{-iX_{0}})~,\\
&&\psi_{z}(0) \simeq \frac{1}{4}z_{0}^{-1}\psi(0)-\pi_{0}q(\frac{b_{1}+ib_{2}}{2}e^{iX_{0}}+\frac{b_{1}-ib_{2}}{2}e^{-iX_{0}})~,
\end{eqnarray}
where $z_{0}=Sgn(C)z_{t}>0$, $\pi_{0}=\pi^{-1/2}\ell^{-1/6}z_{0}^{-1/4}$, $X_{0}=X(-\ell^{2/3}z_{0})$, and the relation $q=\ell z_{0}^{1/2}$ is used. After some trivial calculations, we find that the transmission ratio can be evaluated as
\begin{eqnarray}
\eta=(\frac{q}{q})(\frac{|\pi_{0}b_{2}|}{|a_{1}|})^2=\frac{64z_{0}^2q^2}{64z_{0}^2q^2+1}=1-\zeta~, \label{c17}
\end{eqnarray}
where
\begin{eqnarray}
z_{0}q=\frac{z_{m}k(B^2-A_{0}C)^{3/2}}{N_{max}^2 C^{2}}~.
\end{eqnarray}
Apparently, $\eta$ increases with $z_{0}q$.
Comparing (\ref{c17}) with (\ref{eq70}), we can see that the following relation holds
\begin{eqnarray}
\frac{1}{8z_{0}q}=\beta=\frac{N_{max}^2 C^{2}}{8z_{m}k(B^2-A_{0}C)^{3/2}}~.
\end{eqnarray}

\subsection{Solution for quadratically varying $N^2$}
For the quadratic stratification in the stable layer, we assume the profile is $N^2=\gamma_{2}z^2$. Equation (\ref{eq13}) is
\begin{eqnarray}
\psi_{zz}+k^2\frac{B^2-A_{0}C-\gamma_{2}{C}z^2}{C^2}\psi=0~.
\end{eqnarray}
Substituting $z'=(4\gamma_{2}k^2/|C|)^{1/4}z$, we can transform the equation to a standard form of Sturm-Liouville probelm:
\begin{eqnarray}
\psi_{z'z'}-\frac{1}{4}Sgn(C)z'^2\psi=-\lambda\psi~,
\end{eqnarray}
where $\lambda=k(B^2-A_{0}C)/(2\gamma_{2}^{1/2}|C|^{3/2})$. The solution of the above equation can be found by using the parabolic cylinder functions \citep{olver2010nist}.

For $C>0$, the problem is
\begin{eqnarray}
\psi_{z'z'}-\frac{1}{4}z'^2\psi=-\lambda\psi~,
\end{eqnarray}
and the solution is
\begin{eqnarray}
\psi(z)=b_{1}U(-\lambda,z')+b_{2}V(-\lambda,z')~,
\end{eqnarray}
where $U$ and $V$ are Weber parabolic cylinder functions.
Asymptotic expansions of $U(-\lambda,z')$ and $V(-\lambda,z')$ can be found when $\lambda$ is large and $0\leq z'<2\sqrt{\lambda}$ \citep{olver2010nist}:
\begin{eqnarray}
&&U(-\lambda,z')\sim 2(1-\tilde{z}^2)^{-1/4}\sin X(\tilde{z})~,\\
&&V(-\lambda,z')\sim 2\epsilon (1-\tilde{z}^2)^{-1/4}\cos X(\tilde{z})~.
\end{eqnarray}
where $\mu=\sqrt{2}\lambda^{1/2}$, $\tilde{z}=z'/(\sqrt{2}\mu)$, $\epsilon=1/\Gamma(\lambda+1/2)$, $\Gamma(\xi)$ is the Gamma function of $\xi$, and $X(\tilde{z})=\lambda(\arccos \tilde{z}-\tilde{z}\sqrt{1-\tilde{z}^2})+\frac{1}{4}\pi$.
At the interface, we have
\begin{eqnarray}
&&\psi(0)\simeq ({\epsilon b_{2}-ib_{1}})e^{iX_{0}}+({\epsilon b_{2}+ib_{1}})e^{-iX_{0}}~,\\
&&\psi_{z}(0) \simeq -iq[({\epsilon b_{2}-ib_{1}})e^{iX_{0}}-(\epsilon b_{2}+ib_{1})e^{-iX_{0}}]~,
\end{eqnarray}
where $X_{0}=(2\lambda+1)\pi/4$.

For $C<0$, the problem is
\begin{eqnarray}
\psi_{z'z'}+\frac{1}{4}z'^2\psi=-\lambda\psi~,
\end{eqnarray}
and its solution is
\begin{eqnarray}
\psi(z)=b_{1}W(-\lambda,z')+b_{2}W(-\lambda,-z')~,
\end{eqnarray}
where $W$ is another type of Weber parabolic cylinder function \citep{olver2010nist}. The asymptotic solution of $W$ in $0\leq z'<2\sqrt{\lambda}$ can be found as
\begin{eqnarray}
&& W(-\lambda,z')\sim \lambda^{-1/4}(\tilde{z}^2+1)^{-1/4} \cos Y(\tilde{z})~,\\
&& W(-\lambda,-z')\sim \lambda^{-1/4}(\tilde{z}^2+1)^{-1/4} \sin Y(\tilde{z})~,
\end{eqnarray}
where $Y(\tilde{z})=\lambda[\tilde{z}\sqrt{\tilde{z}^2+1}+\ln(\tilde{z}+\sqrt{\tilde{z}^2+1})]+\pi/4$. 
Thus at the interface, we have
\begin{eqnarray}
&&\psi(0)\simeq 2^{-1}\lambda^{-1/4} [(b_{1}-ib_{2})e^{iY_{0}}+(b_{1}+ib_{2})e^{-iY_{0}}]~, \\
&&\psi_{z}(0)\simeq 2^{-1}\lambda^{-1/4}iq[(b_{1}-ib_{2})e^{iY_{0}}-(b_{1}+ib_{2})e^{-iY_{0}}]~,
\end{eqnarray}
where $Y_{0}=\pi/4$. 

After some trivial calculations, we find that the transmission ratio at the interface is always $\eta \equiv 1$.

\bibliographystyle{jfm}
\bibliography{wave}

\end{document}